\documentclass[fleqn,usenatbib]{mnras}
\usepackage{newtxtext,newtxmath}
\usepackage[T1]{fontenc}
\usepackage[normalem]{ulem}
\usepackage{mathtools}
\usepackage{cuted}

\title[The $H_2O$ and $CO_2$ production rates of Comet 67P/C--G]{Modelling the water and carbon dioxide production rates of Comet 67P/Churyumov--Gerasimenko}

\author[Bj\"{o}rn J. R. Davidsson et al.]{
Bj\"{o}rn J. R. Davidsson,$^{1}$\thanks{E-mail: bjorn.davidsson@jpl.nasa.gov}
Nalin H. Samarasinha,$^{2}$
Davide Farnocchia,$^{3}$
Pedro J. Guti\'{e}rrez$^{4}$
\\
$^{1}$Jet Propulsion Laboratory, California Institute of Technology,  M/S 183--401, 4800 Oak Grove Drive, Pasadena, CA 91109, USA\\
$^{2}$ Planetary Science Institute, 1700 E Ft Lowell Road, Suite 106, Tucson, AZ 85719, USA\\
$^{3}$ Jet Propulsion Laboratory, California Institute of Technology, M/S 301--121, 4800 Oak Grove Drive, Pasadena, CA 91109, USA\\
$^{4}$ Instituto  de Astrof\'{i}sica de Andaluc\'{i}a--CSIC, Camino bajo de Hu\'{e}tor, 50, 18008 Granada,  Spain
}

\date{Accepted 2021 October 23. Received 2021 October 12; in original form 2021 July 13.}

\pubyear{2021}

\begin{document}
\label{firstpage}
\pagerange{\pageref{firstpage}--\pageref{lastpage}}
\maketitle

\begin{abstract}
The European Space Agency \emph{Rosetta/Philae} mission to Comet 67P/Churyumov--Gerasimenko in 2014--2016 is the most complete and diverse investigation of a comet carried out thus far. 
Yet, many physical and chemical properties of the comet remain uncertain or unknown, and cometary activity is still not a well--understood phenomenon. 
We here attempt to place constraints on the nucleus abundances and sublimation front depths of $\mathrm{H_2O}$ and $\mathrm{CO_2}$ ice, and to 
reconstruct how the nucleus evolved throughout the perihelion passage. We employ the thermophysical modelling code `Numerical Icy Minor Body evolUtion Simulator', 
or \textsc{nimbus}, to search for conditions under which the observed $\mathrm{H_2O}$ and $\mathrm{CO_2}$ production rates are simultaneously reproduced  
before and after perihelion. We find that the refractories to water--ice mass ratio of relatively pristine nucleus material is $\mu\approx 1$, that airfall material has $\mu\approx 2$, 
and that the molar abundance of $\mathrm{CO_2}$ relative $\mathrm{H_2O}$ is near $30$ per cent. The dust mantle thickness is typically $\stackrel{<}{_{\sim}}2\,\mathrm{cm}$. 
The average $\mathrm{CO_2}$ sublimation front depths near aphelion were $\sim 3.8\,\mathrm{m}$ and $\sim 1.9\,\mathrm{m}$ on the northern and southern hemispheres, 
respectively, but varied substantially with time. We propose that airfall material is subjected to substantial fragmentation and pulverisation due to thermal fatigue 
during the aphelion passage. Sub--surface compaction of material due to $\mathrm{CO_2}$ activity near perihelion seems to have reduced the diffusivity 
in a measurable way. 
\end{abstract}


\section{Introduction} \label{sec_intro}

One of the main goals of the European Space Agency \emph{Rosetta/Philae} mission \citep{glassmeieretal07} to Comet 67P/Churyumov--Gerasimenko (hereafter, 67P/C--G) 
is to understand comet activity. Specifically, this goal was formulated as the `Study of the development of cometary activity and the processes in 
the surface layer of the nucleus and inner coma (dust/gas interaction)' \citep{schwehmandschulz99}. In essence, this problem includes 
understanding: 1) the composition of the nucleus (i.~e., the relative abundances of refractories and different types of volatiles) and how it relates 
to that of the coma; 2) the location of different volatiles underneath the surface, e.~g., the thickness of the dust mantle and depths where super-- 
and hyper--volatiles are being released; 3) the physical parameters that govern the diffusion of heat and mass in the near--surface layer; 
4) the mechanisms that govern comet outgassing, including the reasons for perihelion outgassing asymmetries; 5) the mechanisms that govern 
dust mantle formation, erosion, and the dust coma particle size--frequency distribution; 6) the effect of comet outgassing on the spin properties and orbit of the nucleus. 

The purpose of this paper is to contribute to this \emph{Rosetta} goal by placing 
novel constraints on the nucleus refractories/water--ice mass ratio, the nucleus molar $\mathrm{CO_2}$ abundance relative $\mathrm{H_2O}$, the depths of 
$\mathrm{H_2O}$ and $\mathrm{CO_2}$ sublimation fronts, and the vapour diffusivity (indicative of the size--scale of near--surface macro porosity) in Comet 67P/C--G. 
To achieve these goals we use the thermophysical model \textsc{nimbus} \citep[Numerical Icy Minor Body evolUtion Simulator;][]{davidsson21} to investigate 
under what conditions the model reproduces the global production rates of $\mathrm{H_2O}$ and $\mathrm{CO_2}$ measured by ROSINA throughout most 
of the \emph{Rosetta} mission \citep{fougereetal16b}\footnote{Revised ROSINA production rates were published when most of our work had been 
completed \citep{combietal20,lauteretal20}. However, differences with respect to \citet{fougereetal16b} are small and do not materially affect the result presented in the paper.}. Furthermore, we calculate the corresponding forces acting on the nucleus due to outgassing, to investigate the conditions 
under which the observed net non--gravitational changes during one orbit are being reproduced.

The longer--term goal of this work is to apply the currently developed nucleus outgassing torques to the spin state 
evolution model developed by \citet{samarasinhaandbelton95}, \citet{samarasinhaandmueller02}, and \citet{samarasinhaetal11}, to investigate 
the conditions under which the measured changes to the spin period, spin axis orientation, and gradual non--gravitational changes 
to the orbit are being reproduced. This may constrain the moments of inertia of the nucleus, and further our understanding of the  
interior and the activity of Comet 67P/C--G.

Various attempts to estimate the refractories/water--vapour mass ratio in the 67P/C--G coma, as well as the refractories/ices mass ratio 
of the 67P/C--G nucleus have been reviewed by \citet{choukrounetal20}. Unfortunately, they are not yet well--constrained. The \emph{lower limits} 
of the refractories/water--vapour mass ratio range $0.01$--$5$ for different methods and authors, while the \emph{upper limits} range 
from 1.7 to at least 10 \citep{rotundietal15,fulleetal16b,biveretal19,combietal20,choukrounetal20}. The \emph{lower limits} on 
the refractories/ices mass ratio of the nucleus range $0.2$--$3.1$, while the \emph{upper limits} range from 3 to 99 
\citep{heriqueetal16,fulleetal16,fulleetal17,blumetal17,huetal17,patzoldetal19,choukrounetal20}. Most of these values rely on: 1) estimates of the coma dust mass obtained by 
retrieving the dust size--frequency distribution from OSIRIS images and determining the bulk densities of individual particles collected by the 
GIADA and COSIMA instruments; 2) estimates of the coma vapour mass by ROSINA, VIRTIS, and MIRO; 3) interpretations in terms of composition 
of the measured nucleus bulk density, permittivity, total nucleus net mass loss, and integrated water vapour loss. These estimates are complicated 
by the fact that a disputed amount of material with unknown ice abundance is ejected into the coma, where some of the ice is sublimated, before the 
solids reunite with the nucleus as airfall material \citep[e.~g.,][]{thomasetal15b,kelleretal15,kelleretal17,davidssonetal21}. 

The nucleus refractories/water--ice mass ratio was estimated as 9--99 by \citet{huetal17} and as 19 by \citet{blumetal17}, based on nucleus thermophysical 
modelling that fitted the pre--perihelion water production rate. These models assumed a fixed dust mantle thickness (no erosion and a static water sublimation front). 
\citet{skorovetal20} demonstrated the importance of considering a variable dust mantle thickness in order to reach the steep observed  dependence of water production 
on heliocentric distance. We here extend these studies by considering thermophysical modelling that includes mantle erosion and a moving sublimation front, leading to a 
dynamically changing dust mantle thickness, which evolves independently on different latitudes. The coupled heat conduction and gas diffusion processes are calculated 
more rigorously than in these previous works, and we perform  a detailed study of the diffusivity of the nucleus material, needed in order to reproduce the observations. 
Furthermore, we consider both the pre-- and post--perihelion branches, which have significantly different water production rate dependencies on heliocentric distance 
\citep{hansenetal16,fougereetal16b}, in order to investigate the cause of this perihelion water production asymmetry. Doing so, our goal is to obtain a more 
reliable estimate of the nucleus refractories/water--ice mass ratio, by considering a more realistic thermophysical model.

\citet{hoangetal20} applied a thermophysical model with the goal of simultaneously reproducing the $\mathrm{H_2O}$, $\mathrm{CO_2}$, and $\mathrm{CO}$ 
production rates of Comet 67P/C--G for two short pre--perihelion time intervals (Sep~17 through Oct~13, 2014, and Jan 11 through Feb 5, 2015). They considered different combinations of 
abundances and initial depths of the sublimation fronts. Their models that best reproduced the $\mathrm{H_2O}$ production did not perform well for $\mathrm{CO_2}$, 
and \emph{vice versa}. No conclusion was drawn regarding the depth below the surface where $\mathrm{CO_2}$ is located.

\citet{hernyetal21} used thermophysical modelling to reproduce the $\mathrm{CO_2/H_2O}$, $\mathrm{CO/H_2O}$, and $\mathrm{CO/CO_2}$ abundance ratios 
observed by ROSINA. Unfortunately, their model produced 5 times more water vapour than observed, therefore they scaled down all modelled production rates by 
that factor. The overproduction occurred because the modelled southern hemisphere lacked a dust mantle, which led to an extreme erosion of $17\,\mathrm{m}$ with 
both $\mathrm{H_2O}$ and $\mathrm{CO_2}$ exposed at the surface. \citet{hernyetal21} acknowledged that neither ice exposure nor level of erosion were consistent with the observed nucleus. 
Because of the scaling they introduced, the resulting abundances and northern $\mathrm{CO_2}$ sublimation front depth must be considered highly uncertain. Therefore, we 
are not aware of any study of the \emph{Rosetta} data that convincingly determined the depth of the $\mathrm{CO_2}$ sublimation front, or estimated the nucleus $\mathrm{CO_2}$ abundance. 
We here attempt such determinations, by striving to simultaneously fit the measured production rates of both $\mathrm{H_2O}$ and $\mathrm{CO_2}$, considering both the pre-- and post--perihelion branches. 

In this paper, a novel approach is taken towards dust mantle erosion. We first describe the need for such an approach, before outlining the approach itself. 
Classically, the removal of dust grains at the upper boundary relied on calculating the 
net force resulting from nucleus gravity, the centrifugal force from nucleus rotation, and the drag force on grains due to nucleus outgassing 
\citep[e.~g.,][]{shulman72,fanaleandsalvail84,rickmanetal90,espinasseetal93,oroseietal95}. It was assumed that the dust grains would become cohesionless as soon as the surrounding 
ice sublimated. This approach was questioned by some authors \citep[e.~g.][]{kuehrtandkeller94,mohlmann95}, who emphasised the potential role of dust mantle cohesion. 
Laboratory measurements of the cohesion of dust--mantle analogues (i.~e., determining the net effect of inter--grain van der Waals forces) show that uniform but highly porous 
aggregates of $\mathrm{\mu m}$--sized silica grains have tensile strengths $\sim 10^3\,\mathrm{Pa}$ \citep{guettlereta09}, but that hierarchically arranged aggregates of such grains may have effective 
tensile strengths as low as $\sim 1\,\mathrm{Pa}$ \citep{skorovandblum12}. 

Based on a comparison between the expected mantle tensile strengths and the calculated local peak gas pressure within the mantles, \citet{skorovandblum12} concluded that 
$\mathrm{H_2O}$ sublimation is not capable of driving cometary dust activity, and that $\mathrm{CO_2}$ sublimation only does so at heliocentric distances 
$r_{\rm h}\stackrel{<}{_{\sim}} 3\,\mathrm{au}$. Using similar lines of argumentation, \citet{gundlachetal15} concluded that the dust mantle of 67P/C--G necessarily must be very weak, 
thus strongly hierarchical, consisting of weakly attached `pebbles' in the mm--cm--size range. This has been taken as evidence \citep[e.~g.,][]{gundlachetal15,blumetal17} that comet nuclei 
formed through the gentle gravitational collapse of pebble swarms \citep[e.~g.][]{nesvornyetal10,wahlbergjanssonjohansen14}, in turn formed by streaming instabilities in the Solar Nebula 
\citep[e.~g.,][]{youdingoodman05,johansenetal07}. However, for CO--driven activity, \citet{jewittetal19} demonstrate that the size of pebbles that is needed in order to bring cohesion 
below the peak gas pressure, is too large for such pebbles to be dragged into the coma, unless the heliocentric distance is $r_{\rm h}\stackrel{<}{_{\sim}} 7\,\mathrm{au}$. This was
 dubbed the `cohesion bottleneck' by \citet{jewittetal19}. Because strong dust activity is common among comets and Centaurs in the $7$--$12\,\mathrm{au}$ region 
\citep{jewitt09,sarneczkyetal16,kulyketal18}, and sometimes takes place as far out as $\sim 25\,\mathrm{au}$ \citep{jewittetal17,huietal18}, the existence of a cohesion bottleneck 
(as currently formulated) does not seem to be empirically supported. Resolving this issue is important, particularly because it has consequences for the debate on comet formation.

Because of the difficulties surrounding the cohesion bottleneck problem, and the orders--of--magnitude uncertainties in the actual strengths of cometary mantles, 
we here start the development of a novel approach to dust mantle erosion. In this paper, we do not attempt to apply a dust mantle erosion rate based on 
first principles, because those principles (as evident from the description above) are not well--understood. Instead, we will enforce an erosion rate based on 
our best understanding of the empirical dust production rate of 67P/C--G, inferred from various types of Rosetta observations. 
We then adjust the nucleus model parameters (primarily the refractories/water--ice mass ratio and the diffusivity) until 
the model matches the observed $\mathrm{H_2O}$ and $\mathrm{CO_2}$ production rates (and by construction, the dust production rate). By doing so, we build an archive that 
contains the temperature and the vapour pressures of $\mathrm{H_2O}$ and $\mathrm{CO_2}$ as functions of depth, that arise as natural consequences of energy and mass conservation, 
at given mantle erosion rates. In a forthcoming publication, this numerical information will be used to `reverse--engineer' the dust erosion process. Specifically, correlations between 
vapour pressures, temperatures, their amplitudes and frequencies of oscillation, and associated mantle erosion rates will be established for different heliocentric distances, 
nucleus latitudes, and time of day. By considering various fatigue processes and crack propagation mechanisms, this may allow for a better understanding of dust production, and 
estimates of the tensile strength of the 67P/C--G mantle. The final goal of that exercise will be to define an algorithm that can be applied in thermophysical models, that calculates 
the appropriate dust mantle erosion rate based on the instantaneous temperature and vapour pressure profiles with depth, that arise during simulations. This algorithm, that is 
benchmarked against 67P/C--G, can then be tested for other comets. 

This paper is organised as follows. The thermophysical model is briefly recapitulated in Section~\ref{sec_model} \citep[for a detailed description, see ][]{davidsson21}. 
Section~\ref{sec_rates} discusses the production rates of water and dust of 67P/C--G. Specifically, Section~\ref{sec_rates_water} discusses the problem of isolating 
the contribution from the nucleus itself to the observed water production (the thermophysical model water production rate should be compared to that of the nucleus, and 
not to the total observed rate that potentially also have contributions from an extended source). Section~\ref{sec_rates_erosion} defines the time--dependent mantle 
erosion rate, based on the inferred empirical dust production, that here is used as \emph{input} to the thermophysical model, for the reasons described above. 
Section~\ref{sec_results} contains our results and focuses on four topics: 1) the pre--perihelion $\mathrm{H_2O}$ and $\mathrm{CO_2}$ production rates (Section~\ref{sec_results_preper}); 
2) the post--perihelion $\mathrm{H_2O}$ and $\mathrm{CO_2}$ production rates (Section~\ref{sec_results_postper}); 3) the resulting erosion of the nucleus, the depths of the 
$\mathrm{H_2O}$ and $\mathrm{CO_2}$ sublimation fronts, and their temperatures (Section~\ref{sec_results_fronts}); 4) the forces on the nucleus due to outgassing and the 
resulting non--gravitational change of the orbit (section~\ref{sec_results_forces}). These results are discussed in Section~\ref{sec_discussion} and our conclusions are summarised in Section~\ref{sec_conclusions}.

\section{The thermophysical model} \label{sec_model}

In order to calculate the thermophysical evolution of the nucleus, we use \textsc{nimbus} developed by \citet{davidsson21}. \textsc{nimbus} models the nucleus as a 
$1932\,\mathrm{m}$ radius spherical body \citep[surface--area equivalent to the real nucleus;][]{jordaetal16} 
consisting of a porous mixture of refractory dust and ices (in the current manuscript only crystalline water ice and $\mathrm{CO_2}$ ice are considered). 
The model body is divided into 18 equal--angle latitudinal slabs, and 100 radial cells per latitude (growing in geometric progression from $4\,\mathrm{mm}$ thickness 
at the surface to $200\,\mathrm{m}$ at the core). The model nucleus is given the spin axis orientation (equatorial system $\{\alpha,\,\delta\}=\{69.37^{\circ},\,64.132^{\circ}\}$) from 
the shape reconstruction \citep{jordaetal16}, orbital elements from the Minor Planet Center, and \textsc{nimbus} calculates local illumination conditions at each latitude for any given orbital position 
and rotational phase. Nominally, \textsc{nimbus} tracks the solid--state and radiative conduction of absorbed solar heat radially and latitudinally. Sub--surface $\mathrm{H_2O}$ and $\mathrm{CO_2}$ 
ice sublimate and consume energy when sufficiently warm, and the vapours are diffusing radially and latitudinally along local temperature and gas pressure gradients (transporting energy through advection). 
Longitudinal flows are ignored because the rotation speed of the nuclear surface vastly exceeds the diffusion speed. The vapour either finds its way across the outer nucleus boundary and enters the coma, 
or recondenses internally if reaching a sufficiently cold region (releasing latent energy). 

However, the current application uses an alternative implementation called \textsc{nimbusd} (with \textsc{d} for `dust'). Here, the latitudinal heat and gas diffusion 
are switched off for technical reasons in order to allow for erosion of dust (and other solids, if present) from the upper surface and shrinking of the nucleus. For the short time scales 
considered here (one orbit) these flows are negligible. If the uppermost locations of $\mathrm{H_2O}$ and $\mathrm{CO_2}$ ice at a given latitudinal slab 
(called the `sublimation fronts' of respective species) withdraw underground because the front moves faster than the erosion rate, an ice--free dust mantle is formed along with a 
$\mathrm{CO_2}$--free region that only contains $\mathrm{H_2O}$ ice and refractories. The gas diffusion rate, and hence the total comet outgassing, is sensitive to the depths of the sublimation fronts 
and the near--surface temperature and gas pressure gradients. Furthermore, the removal of ice and corresponding increase of the porosity reduces the heat conductivity (all these 
processes are being modelled). Because all simulations in this paper are using \textsc{nimbusd}, there is no risk of confusion between the variants, thus the thermophysical model is simply 
referred to as \textsc{nimbus} in the following.

A detailed description of the differential equations for energy and mass conservation that are being solved by \textsc{nimbus}, as well as 
many auxiliary physical functions for heat conduction, heat capacity, saturation pressures, mass flux rates \emph{et cetera}, are provided by \citet{davidsson21} and 
are not repeated here. We emphasise that the applied conductivity and heat capacity are constantly evolving quantities that are functions of the particular composition, 
porosity, and temperature that prevail at a given time and location. However, the resulting value of the thermal inertia near the surface is typically 
$\sim 40$--$60\,\mathrm{MKS}$ ($1\,\mathrm{MKS}=1\,\mathrm{J\,m^{-2}\,K^{-1}\,s^{-1/2}}$), similar to what has been measured for 67P/C--G \citep{schloerbetal15,spohnetal15,marshalletal18}. 

By relying on laboratory measurements of most material parameters applied in \textsc{nimbus}, the number of free parameters is small. The free parameters to be 
discussed in the current manuscript are limited to the erosion rate of solids from the surface (see Section~\ref{sec_rates}), the initial depths of the sublimation fronts applied at aphelion, 
the refractories/water--ice mass ratio $\mu$, the molar $\mathrm{CO_2}$ abundance relative to water, and the gas diffusivity that is being parameterised by the length $L$, radius $r_{\rm p}$, and 
tortuosity $\xi$ of tubes in the Clausing formula \citep[see equation~46 in ][]{davidsson21}. Different values may be assigned to the northern and southern hemispheres when needed, as 
described in Section~\ref{sec_results}.

The \textsc{nimbus} calculations provide internal temperatures and abundances, as well as the nucleus production rates for $\mathrm{H_2O}$ and 
$\mathrm{CO_2}$ for each latitude slab, at a temporal resolution that corresponds to a dynamical time step (typically varying between a few seconds and 
a few hundreds of seconds). Saving that amount of data to file for simulations that cover years in the simulated world would create unmanageable data volumes. 
Therefore, \textsc{nimbus} saves selected rotational periods at a given resolution. In this paper, every $12^{\rm th}$ rotation period was stored (roughly once per 
week given the $\sim 12\,\mathrm{h}$ rotation period), with a $10^{\circ}$ resolution in terms of rotational phase. In order to produce continuous outgassing curves, 
every stored rotation is copied to the following eleven ones. The errors introduced by this approach are small, but result in a certain level of discontinuity seen in some 
of the figures. 

The escape of gases across the upper surface gives rise to a force acting on the nucleus because of linear momentum conservation. 
Calculating this reaction force acting on a surface sublimating into vacuum is not trivial. \citet{davidssonandskorov04} used a 
Monte Carlo approach to calculate the velocity distribution function of water molecules emerging from a highly porous medium 
with a near--surface temperature gradient. They used the velocity distribution as a source function to a Direct Simulation Monte Carlo (DSMC) model 
that calculates the evolution of the velocity distribution function due to molecular collisions within the base of the comet coma (i.~e., a numerical 
solution to the Boltzmann equation containing a non--zero collision integral). This 
type of calculations are necessary in environments where the gas is far from being in thermodynamic equilibrium (i.~e., the velocity 
distribution function is strongly non--Maxwellian and the hydrodynamic Euler equations for conservation of mass, momentum, and energy 
do not apply). The model by \citet{davidssonandskorov04} evaluates the force acting on the surface as the sum of momentum transfer 
during molecular ejection from the nucleus because of sublimation, and inelastic molecule--surface collisions due to coma gas that 
has been backscattered toward the nucleus within the Knudsen layer. Such a complex treatment is out of scope in the current paper. Also, we cannot 
directly apply the existing simulations by \citet{davidssonandskorov04} for two reasons: 1) those are only valid when ice is intimately mixed 
with dust up to and including the surface, while the current \textsc{nimbus} simulations result in the formation of ice--free dust mantles; 
2) those only considered $\mathrm{H_2O}$ ice, while the current work considers both $\mathrm{H_2O}$ and $\mathrm{CO_2}$.

Therefore, we apply the classical approach \citep[e.~g.,][]{rickman89} of defining a force that is proportional to the 
local outgassing rate $Z\,\mathrm{(molec\,m^{-2}\,s^{-1})}$ and to the mean molecular speed of a Maxwellian gas $\langle V\rangle$ 
(regardless of direction of travel), multiplied to a momentum transfer coefficient $\eta$ that corrects for the fact that the mean molecular velocity along the surface 
normal does not equal $\langle V\rangle$,
\begin{equation} \label{eq:00a}
\mathbfit{F}=-\eta\sum_i\sum_j Z_{i,j}m_j\langle V\rangle_{i,j}s_i\mathbfit{n}_i.
\end{equation}
Here, the summation is made over all nucleus facets $i$ having outward surface normals $\mathbfit{n}_i$ and areas $s_i$, and over the two species $j=1$ ($\mathrm{H_2O}$) 
and $j=2$ ($\mathrm{CO_2}$). The mean molecular speed is given by 
\begin{equation} \label{eq:00b}
\langle V\rangle_{i,j}=\sqrt{\frac{8k_{\rm B}T_i}{\upi m_j}}
\end{equation}
where $k_{\rm B}$ is the Boltzmann constant, $T_i$ is the surface temperature of facet $i$, and $m_j$ is the mass of a $\mathrm{H_2O}$ ($j=1$) or 
$\mathrm{CO_2}$ ($j=2$) molecule.

The evaluation of $\eta$ is a topic of debate, with values typically falling in the range $0.4\leq\eta\leq 1$ 
\citep[see][and references therein]{rickman89,davidssonandskorov04}. We nominally ran the simulations with $\eta=1$, but re--normalise 
in Section~\ref{sec_results_forces} in order to comply with measured non--gravitational changes of the orbit. Although \textsc{nimbus} works with 
a spherical model nucleus rotating at a fixed $12.4043\,\mathrm{h}$ period \citep{mottolaetal14}, we apply a mapping procedure that takes into account 
the actual nucleus shape and spin period changes over time, as described in Section~\ref{sec_results_preper}.

\section{Nucleus water production and erosion rates} \label{sec_rates}

The comet nucleus emits water vapour directly from its surface, along with fine dust ($\stackrel{<}{_{\sim}}\,1\,\mathrm{mm}$) that may be 
considered fully refractory, and larger chunks \citep[$1\,\mathrm{mm}$--$1\,\mathrm{m}$; e.g.][]{rotundietal15, davidssonetal15b, agarwaletal16} 
that potentially contain water ice. The latter might constitute an extended (distributed) source of water vapour in 
the coma, before the majority of these heavy particles reunite with the nucleus as airfall material \citep[e.g.][]{thomasetal15b, kelleretal15, kelleretal17, davidssonetal21}. 
The transition size of $\sim 1\,\mathrm{mm}$ between dust and chunks is determined by the diurnal thermal skin depth: dust grains are isothermal and unable to carry ice, 
while chunks are thermally heterogeneous. As demonstrated by \citet{davidssonetal21} for a $1\,\mathrm{cm}$ chunk, such particles may have a dust mantle, an icy interior, 
strongly different daytime and nighttime temperatures, and activity lifetimes on the order of days. The empirical information available is the total observed water production $Q_{\rm H_2O}$ from 
nucleus and chunks (for symbols used in Section~\ref{sec_rates}, see Table~\ref{tab1}). However, the contribution $Q_{\rm nuc}$ from the nucleus itself needs to be isolated, 
for direct comparison with water production rates calculated by \textsc{nimbus}. Furthermore, in the current application, \textsc{nimbus} needs the dust 
mantle erosion rate $E$ as input. We here use simple methods to retrieve those rates from available empirical data. We make the following assumptions:

\begin{trivlist}
\item A\#1. The dayside area elements on the nucleus and on coma chunks have the same average water production rates $q_{\rm w}$, 
and production rates of fine dust $q_{\rm d}$ (both measured in units of $\mathrm{kg\,m^{-2}\,s^{-1}}$). Thus, chunks are 
considered `miniature comets'.
\item A\#2. The size distribution of chunks is given by a power--law. It can be used to let all chunks be represented by 
an average radius $r_{\rm c}$ (in terms of surface area).
\item A\#3. The ejection rate of chunks into the coma is proportional to the total water production rate 
$Q_{\rm H_2O}$ of the comet, i.e., $n_{\rm c}\,\propto\,Q_{\rm H_2O}\,\mathrm{(chunks\,s^{-1})}$. This accounts 
for a nucleus source of chunks ($\propto Q_{\rm nuc}$), but also that large chunks in the $\mathrm{dm}$--$\mathrm{m}$ class 
will produce smaller chunks as well. 
\item A\#4. The production rate of fine dust is proportional to that of water vapour, or $q_{\rm d}\propto q_{\rm w}$. 
\end{trivlist} 

By making these assumptions, the dust production rate takes the same functional form as the water production rate. This can be motivated 
by considering the brightness of the comet, that is proportional to the amount of fine dust in the coma. \citet{hansenetal16} point out 
that the coma brightness determined by groundbased telescopes is highly correlated with the water production rate that has been inferred from the ROSINA measurements. 
A careful reconstruction of both the dust production rate and the water production rate was made by \citet{marschalletal20}. They find that the dust--to--gas mass 
ratio changed over time (i.~e., A\#4 is not strictly valid). However, during the period 200 days pre--perihelion to 100 days post--perihelion (when most of the gas and dust are 
being produced), the dust--to--gas mass ratio is still rather stable: \citet{marschalletal20} find that it is $0.5\pm 0.25$, except for a brief period around 100 days 
pre--perihelion when the dust--to--gas mass ratio increased to $\sim 1.3$. Considering that the production rate changed by orders of magnitude in this period, the fact that their ratio 
remained quasi--constant to within $\pm 50$ per cent for most of the time shows that A\#4 is a reasonable approximation.

\begin{table}
\begin{center}
\begin{tabular}{||l|l|l||}
\hline
\hline
Symbol & Description & Unit\\
\hline
$A_{\rm c}$ & Fraction of \underline{nucleus} area emitting chunks &\\
 & per nucleus rotation & \\
$A_{\rm tot}$ & Total \underline{comet} surface area & $\mathrm{m^2}$\\
$C$ & Correction factor in equation~(\ref{eq:07}) & \\
$E$ & Dust mantle erosion rate & $\mathrm{kg\,m^{-2}\,s^{-1}}$\\
$F(t)$ & Total \underline{comet} fine dust production rate &\\
 & to \underline{comet} $\mathrm{H_2O}$ production rate ratio &\\
$G(t)$ & Total \underline{nucleus} dust to \underline{comet} $\mathrm{H_2O}$\\
 &  production rate ratio &\\
$h$ & Thickness of airfall layer deposited\\
 & per apparition & $\mathrm{m}$\\
$k_{\rm c}$ & Proportionality constant in equation~\ref{eq:04} & $\mathrm{kg^{-1}}$\\
$M_{\rm H_2O}$ & Total comet $\mathrm{H_2O}$ mass loss per apparition & $\mathrm{kg}$\\
$M_{\rm L}$ & Total comet mass loss per apparition & $\mathrm{kg}$\\
$N_{\rm c}$ & Number of coma chunks &\\
$\tilde{N}_{\rm c}$ & Number of chunks ejected into coma\\
 & per apparition &\\
$n_{\rm c}$ & Ejection rate of chunks into the coma & $\mathrm{s^{-1}}$\\
$q$ & Differential size frequency distribution & \\
 & power--law index\\
$q_{\rm d}(t)$ & Fine dust production flux & $\mathrm{kg\,m^{-2}\,s^{-1}}$\\
$q_{\rm w}(t)$ & $\mathrm{H_2O}$ production flux & $\mathrm{kg\,m^{-2}\,s^{-1}}$\\
$Q_{\rm H_2O}(t)$ & Total \underline{comet} $\mathrm{H_2O}$ production rate & $\mathrm{kg\,s^{-1}}$\\
$Q_{\rm nuc}(t)$ & Total \underline{nucleus} $\mathrm{H_2O}$ production rate & $\mathrm{kg\,s^{-1}}$\\
$r_{\rm c}$ & Coma chunk radius & $\mathrm{m}$\\
$R_{\rm nuc}$ & Effective nucleus radius & $\mathrm{m}$\\
$\Delta t_{\rm f}$ & Dynamical life--time of coma chunks & $\mathrm{s}$\\
$\rho_{\rm c}$ & Airfall layer bulk density & $\mathrm{kg\,m^{-3}}$\\
$\psi$ & Airfall layer macro porosity & \\
\hline 
\hline
\end{tabular}
\caption{Summary of the functions and parameters used for the dust production model. Note that areas, production rates \emph{et cetera} 
of the `comet' refers to the sum of contributions from the nucleus and chunks in the coma. When only the nucleus 
contribution is relevant, that is noted specifically.}
\label{tab1}
\end{center}
\end{table}

\subsection{The nucleus water production rate} \label{sec_rates_water}

In order to calculate the nucleus water outgassing rate,
\begin{equation} \label{eq:01}
Q_{\rm nuc}(t)=2\upi R_{\rm nuc}^2q_{\rm w}(t),
\end{equation}
we need $q_{\rm w}(t)$, which is defined by
\begin{equation} \label{eq:02}
q_{\rm w}(t)=\frac{Q_{\rm H_2O}(t)}{A_{\rm tot}(t)},
\end{equation}
according to A\#1, where $A_{\rm tot}$ is the total sublimating area. $A_{\rm tot}$ is ultimately constrained by the amount of 
airfall accumulated during one apparition, as described in the following. If the number of coma chunks at a given moment is $N_{\rm  c}$ then (according to A\#2)
\begin{equation} \label{eq:03}
A_{\rm tot}(t)=2\upi r_{\rm c}^2N_{\rm c}(t)+2\upi R_{\rm nuc}^2.
\end{equation}
We assumed (A\#3) that 
\begin{equation} \label{eq:04}
n_{\rm c}(t)=k_{\rm c}Q_{\rm H_2O}(t),
\end{equation}
therefore $N_{\rm c}$ is given by 
\begin{equation} \label{eq:05}
N_{\rm c}(t)=k_{\rm c}Q_{\rm H_2O}(t)\Delta t_{\rm f},
\end{equation}
where $\Delta t_{\rm f}$ is the average dynamical life--time of chunks in the coma (the flight time from source to airfall target).
In order to evaluate the proportionality constant $k_{\rm c}$, we integrate equation~(\ref{eq:04}) over one orbit,
\begin{equation} \label{eq:06}
\tilde{N}_{\rm c}=\int_0^Pn_{\rm c}(t)\,dt=k_{\rm c}\int_0^PQ_{\rm H2O}(t)\,dt=k_{\rm c}M_{\rm H_2O}
\end{equation}
(where $P$ is the orbital period, and $M_{\rm H_2O}$ is the total amount of water vapour produced during the apparition), 
and use the fact that the total number of chunks $\tilde{N}_{\rm c}$ produced during the perihelion passage is related to the 
thickness of the airfall deposition layer accumulated on the northern hemisphere of the comet,
\begin{equation} \label{eq:07}
2\upi R_{\rm nuc}^2 h\rho_{\rm c}(1-\psi)=C\left(\frac{4\upi}{3}r_{\rm c}^3\rho_{\rm c}\right)\tilde{N}_{\rm c}.
\end{equation}
The left hand side of equation~(\ref{eq:07}) is the total mass of accumulated airfall material in the northern hemisphere, assuming that it 
covers half the nucleus surface, has a layer thickness $h$, and that the chunks with bulk density $\rho_{\rm c}$ assemble in this 
layer with a macro--porosity of $\psi$. The right hand side of equation~(\ref{eq:07}) has the mass of a single coma chunk within parenthesis, 
which yields the total mass of coma chunks when multiplied with the total number of chunks $\tilde{N}_{\rm c}$. The correction factor $C$ is 
compensating for the fact that $r_{\rm c}$ and $\tilde{N}_{\rm c}$ are defined to represent the coma chunks in terms of their surface area, 
which is not necessarily a good measure of their volume. Note that $\rho_{\rm c}$ cancel out in equation~(\ref{eq:07}), i.e., the current derivation does 
not depend on the bulk density of the coma chunks.\\

We now proceed to evaluate $r_{\rm c}$ and $C$. For a differential size frequency distribution of chunks on power--law form, $Kr^{-q}$, the 
surface area distribution is $Kr^{-q}r^2=Kr^{2-q}$. With lower and upper truncation radii $\{r_{\rm min},\,r_{\rm max}\}$, we define 
the most typical radius $r_{\rm c}$ (in terms of surface area) as
\begin{equation} \label{eq:08}
\frac{\int_{r_{\rm min}}^{r_{\rm c}}(r')^{2-q}\,dr'}{\int_{r_{\rm min}}^{r_{\rm max}}(r')^{2-q}\,dr'}=\frac{r_{\rm c}^{3-q}-r_{\rm min}^{3-q}}{r_{\rm max}^{3-q}-r_{\rm min}^{3-q}}=\frac{1}{2},
\end{equation}
i.e., the radius for which all smaller chunks collectively have the same surface area as the combined surface area of all larger chunks. 

The total surface area of chunks is
\begin{equation} \label{eq:09}
S_{\rm c}=4\upi K\int_{r_{\rm min}}^{r_{\rm max}}(r')^{2-q}\,dr'= \frac{4\upi K}{3-q}\left(r_{\rm max}^{3-q}-r_{\rm min}^{3-q}\right),
\end{equation}
which means that the number of $r_{\rm c}$--sized chunks carrying this area is 
\begin{equation} \label{eq:10}
N_{\rm c}=\frac{S_{\rm c}}{4\upi r_{\rm c}^2}.
\end{equation}
The differential volume distribution is $Kr^{-q}r^3=Kr^{3-q}$, which means that the total volume of all chunks is 
\begin{equation} \label{eq:10b}
V_{\rm c}=\frac{4\upi K}{3}\int_{r_{\rm min}}^{r_{\rm max}}(r')^{3-q}\,dr'= \frac{4\upi K}{3(4-q)}\left(r_{\rm max}^{4-q}-r_{\rm min}^{4-q}\right).
\end{equation}
The correction factor should make sure that the right--hand side of equation~(\ref{eq:07}) equals the correct volume, thus
\begin{equation} \label{eq:11}
V_{\rm c}=C\left(\frac{4\upi}{3}r_{\rm c}^3\right)N_{\rm c}.
\end{equation}
Insertion of equations~(\ref{eq:09})--(\ref{eq:10b}) into equation~(\ref{eq:11}) yields
\begin{equation} \label{eq:12}
C=\frac{3-q}{r_{\rm c}(4-q)}\frac{r_{\rm max}^{4-q}-r_{\rm min}^{4-q}}{r_{\rm max}^{3-q}-r_{\rm min}^{3-q}}.
\end{equation}

\begin{table}
\begin{center}
\begin{tabular}{||l|l|l||}
\hline
\hline
Symbol & Value & Unit\\
\hline
Independent constants & &\\
$h$ & 0.87 & $\mathrm{m}$\\
$M_{\rm L}$ & $(1.05\pm 0.34)\cdot 10^{10}$ & $\mathrm{kg}$\\
$q$ & 3.8 & \\
$R_{\rm nuc}$ & 1932 & $\mathrm{m}$\\
$r_{\rm min}$ &  0.0081 & $\mathrm{m}$\\
$r_{\rm max}$ & 1 & $\mathrm{m}$\\
$\Delta t_{\rm f}$ & 12 & $\mathrm{h}$\\
$\psi$ & 0.4 & \\
\hline
Dependent constants & & \\
$C$ & 2.34 & \\
$k_{\rm c}$ & $38.4$ & $\mathrm{chunks\,kg^{-1}}$\\
$M_{\rm H_2O}$ & $5.13\cdot 10^9$ & $\mathrm{kg}$\\
$\tilde{N}_{\rm c}$ & $1.97\cdot 10^{11}$ & \\
$r_{\rm c}$ & 0.0184 & $\mathrm{m}$\\
\hline 
\hline
\end{tabular}
\caption{Numerical values for independent (input) and dependent (output) constants applied in the 
evaluation of equations~(\ref{eq:01})--(\ref{eq:12}).}
\label{tab2}
\end{center}
\end{table}

We now proceed to evaluate equations~(\ref{eq:01})--(\ref{eq:12}), for numerical values of independent and dependent parameters, see Table~\ref{tab2}. 
\emph{Philae}/ROLIS images of airfall material at Agilkia show that the differential size distribution of chunks has $q=3.8$ at radii 
$0.02\,\mathrm{m}\leq r\leq r_{\rm max}=0.5\,\mathrm{m}$ \citep{mottolaetal15}. We extend the distribution to $r_{\rm min}=0.0081\,\mathrm{m}$, 
where it starts to flatten, according to \emph{Philae}/CIVA images \citep{pouletetal17}. That yields $r_{\rm c}=0.0184\,\mathrm{m}$ (equation~\ref{eq:08}) 
and $C=2.34$ (equation~\ref{eq:12}). When evaluating equation~(\ref{eq:07}) we use $R_{\rm nuc}=1932\,\mathrm{m}$, for which a sphere has the 
same surface area as the nucleus of 67P/C--G \citep{jordaetal16}, the average thickness of seasonal airfall deposits $h=0.87\,\mathrm{m}$ 
according to \citet{davidssonetal21}, and a macro--porosity $\psi=0.4$ for airfall material, similar to that of gravitationally reaccumulated 
low--mass rubble--pile asteroids \citep{abeetal06}. This yields $\tilde{N}_{\rm c}=1.97\cdot 10^{11}$.

In order to define $Q_{\rm H_2O}(t)$ (here, $\mathrm{molec\,s^{-1}}$, which should be converted to $\mathrm{kg\,s^{-1}}$ when applied 
in equations~\ref{eq:02}--\ref{eq:06} and \ref{eq:14}) we fitted the following functions to \emph{Rosetta}/ROSINA measurements by \citet{fougereetal16b},
\begin{equation} \label{eq:13}
Q_{\rm H_2O}(t)=\left\{\begin{array}{l}
\displaystyle \Big[2.131\cdot 10^{28}r_{\rm h}^{-4.6821}\\
\\
\displaystyle +1.093\cdot 10^{28}\exp(-50(r_{\rm h}-q_{\rm orb}))\Big]\,\,\,\mathrm{inbound}\\
\\
\displaystyle 6.347\cdot 10^{28}r_{\rm h}^{-5.6326}\,\,\,\mathrm{outbound}
\end{array}\right.
\end{equation}
where the first+second and third rows correspond to the pre-- and post--perihelion branches, respectively, $r_{\rm h}=r_{\rm h}(t)$ is the 
heliocentric distance of the comet ($\mathrm{au}$), and $q_{\rm orb}=1.24\,\mathrm{au}$ is the perihelion distance. The 
two branches have distinctively different slopes far from perihelion \citep[also see][]{hansenetal16}, and the pre--perihelion 
exponential term in equation~(\ref{eq:13}) is needed to reproduce a near--perihelion surge in water production and to yield a seamless 
transition to the post--perihelion branch. Integrating equation~(\ref{eq:13}) yields a total water loss $M_{\rm H_2O}=5.13\cdot 10^9\,\mathrm{kg}$, 
which is consistent with other ROSINA--based estimates by \citet{lauteretal20} of $4.0\pm 0.6\cdot 10^9\,\mathrm{kg}$ and by \citet{combietal20} of $4.9\pm 1.5\cdot 10^9\,\mathrm{kg}$, 
but somewhat higher than $2.4\pm 0.1\cdot 10^9\,\mathrm{kg}$ based on \emph{Rosetta}/MIRO \citep{biveretal19} and losses ranging $(2.3$--$3.1)\cdot 10^9\,\mathrm{kg}$ 
recorded by SWAN/SOHO during three previous apparitions \citep{bertaux15}. This yields $k_{\rm c}=38.4\,\mathrm{kg^{-1}}$ from equation~(\ref{eq:06}). Applying 
$\Delta t_{\rm f}=12\,\mathrm{h}$ \citep[a typical flight--time of coma chunks;][]{davidssonetal21}, equations~(\ref{eq:01})--(\ref{eq:03}) can be evaluated. 

\begin{figure}
\scalebox{0.45}{\includegraphics{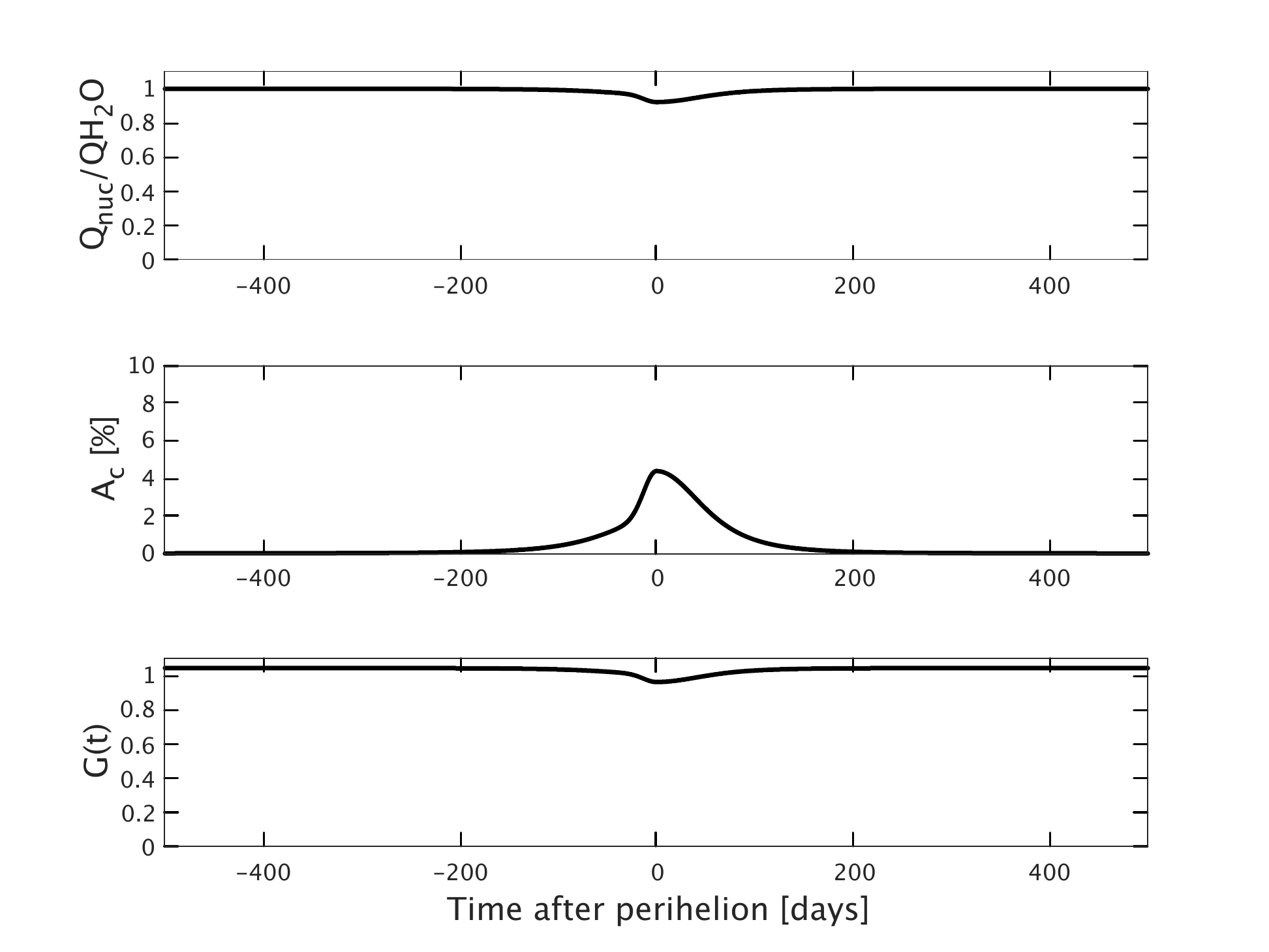}}
     \caption{\emph{Upper panel:} The inferred water production rate of the nucleus $Q_{\rm nuc}$ compared to the total observed water production rate $Q_{\rm H_2O}$ 
that also contains minor contributions from icy chunks in the coma. \emph{Middle panel:} The percentage of the nucleus illuminated surface (during one nucleus revolution) that 
ejects chunks in the $\mathrm{mm}$--$\mathrm{m}$ size range. \emph{Lower panel:} The production rate of solids eroding from the nucleus, expressed in units of the 
total observed water production rate.}
     \label{fig_prod_erosion}
\end{figure}

Figure~\ref{fig_prod_erosion} (upper panel) shows that $0.92\leq Q_{\rm nuc}/Q_{\rm H_2O}\leq 1$, i.e., the extended source of water contributes at 
most $8$ per cent of the water vapour near perihelion. Because this deviation is substantially smaller than the scatter in the measured water production rate data 
(see Figs.~\ref{fig_preper_H2O_CO2} and \ref{fig_postper_H2O_CO2}), we decided to compare \textsc{nimbus} water production rates directly with the observed $Q_{\rm H_2O}$ instead of $Q_{\rm nuc}$.

\subsection{The nucleus dust erosion rate} \label{sec_rates_erosion}

Equations~(\ref{eq:06})--(\ref{eq:07}) provide the constant $k_{\rm c}$ that allows us to estimate the fraction of the nucleus surface area responsible for chunk ejection during 
one nucleus rotation period $P_{\rm rot}=12.4\times 3600\,\mathrm{s}$ \citep{mottolaetal14},

\begin{equation} \label{eq:14}
A_{\rm c}=\frac{r_{\rm c}^2  k_{\rm c}Q_{\rm H2O}P_{\rm rot}}{2R_{\rm nuc}^2}.
\end{equation}

With $A_{\rm c}$ not exceeding $\sim 4$ per cent at any time (Fig.~\ref{fig_prod_erosion}, middle panel), the vast majority of the surface is eroding because it ejects fine  ($\stackrel{<}{_{\sim}}\,1\,\mathrm{mm}$) dust. 
The water production of the nucleus is therefore controlled by the thickening or thinning of the dust mantle resulting from gradual small--scale erosion and water 
sublimation--front withdrawal. Chunk ejection will be relatively rare and is a local phenomenon. Therefore, we restrict the erosion rate (to be applied in \textsc{nimbus}) to $q_{\rm d}$ 
itself. We express the total nucleus dust erosion rate $E$ in units of the total water production rate,
\begin{equation} \label{eq:15}
E=G(t)Q_{\rm H_2O}=G(t)q_{\rm w}A_{\rm tot}(t).
\end{equation}
But
\begin{equation} \label{eq:16}
E=2\upi R_{\rm nuc}^2q_{\rm d}
\end{equation}
which means that 
\begin{equation} \label{eq:17}
  G(t)=\frac{2\upi R_{\rm nuc}^2q_{\rm d}}{q_{\rm w}A_{\rm tot}(t)}.
\end{equation}
According to A\#4,
\begin{equation} \label{eq:18}
q_{\rm d}=Fq_{\rm w},
\end{equation}
thus
\begin{equation} \label{eq:19}
  G(t)=\frac{2\upi R_{\rm nuc}^2F}{A_{\rm tot}(t)}.
\end{equation}

\begin{figure*}
\centering
\begin{tabular}{cc}
\scalebox{0.45}{\includegraphics{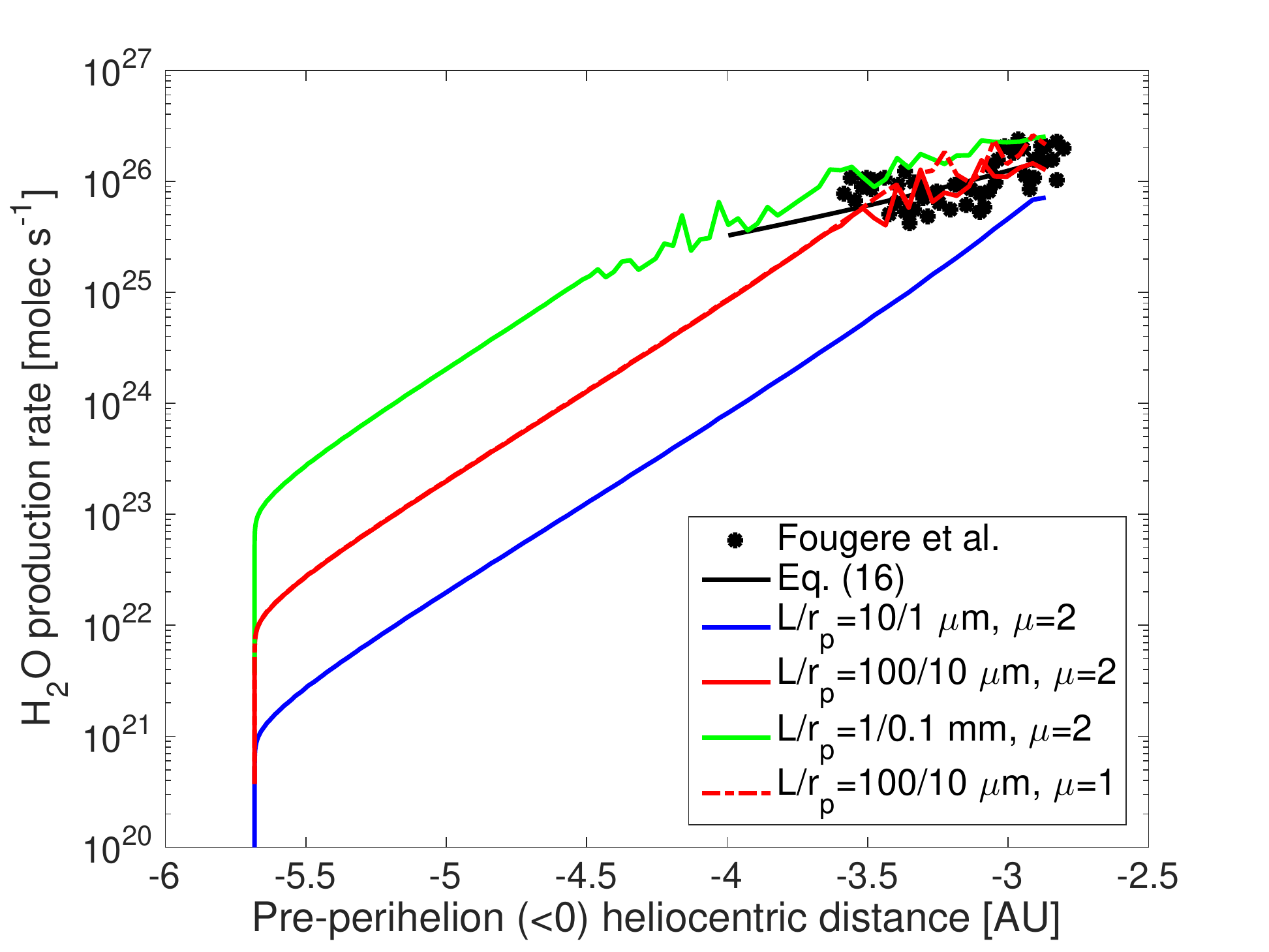}} & \scalebox{0.45}{\includegraphics{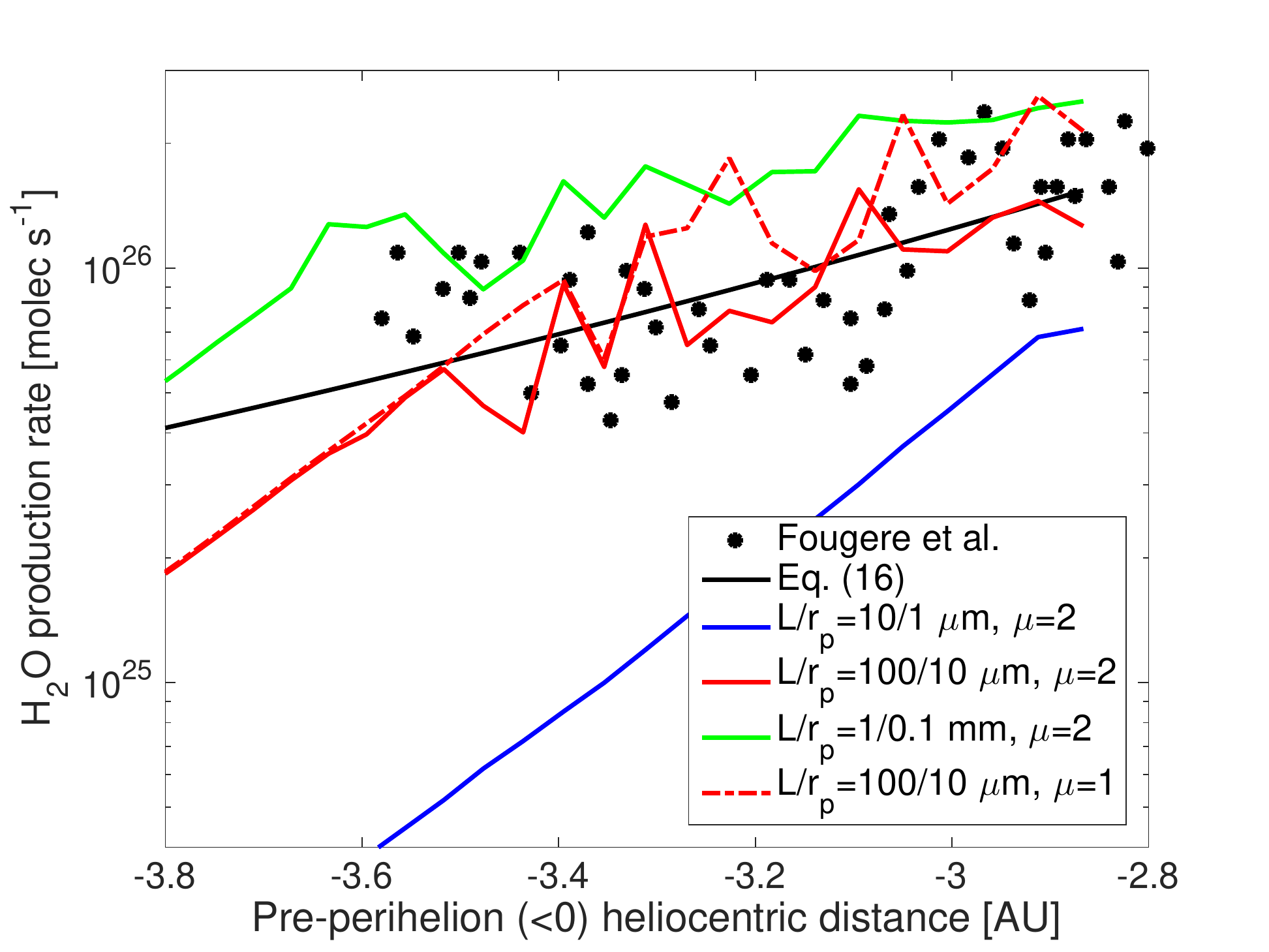}}\\
\end{tabular}
     \caption{\emph{Left:} The water production rate calculated from aphelion at $5.68\,\mathrm{au}$ to $2.9\,\mathrm{au}$ inbound, having a 
small overlap with the first available \emph{Rosetta}/ROSINA data. The arbitrary nucleus temperature initial condition is rapidly corrected. At large heliocentric 
distances, the water production rate is strongly sensitive to the diffusivity (via the tube length $L$, radius $r_{\rm p}$, and tortuosity $\xi$) but insensitive to the 
nucleus mass ratio $\mu$ between refractories and water ice. \emph{Right:} A close--up on the region where data and models overlap. The model 
with $\{L,\,r_{\rm p}\}=\{100,\,10\}\,\mathrm{\mu m}$, $\xi=1$, and $\mu=2$ provides the best fit. The effect on the water production rate by increasing the water 
ice mass fraction from 33 per cent ($\mu=2$) to 50 per cent ($\mu=1$) is smaller than the scatter in the measured data.}
     \label{fig_diffusivity}
\end{figure*}

The mass ratio $F$ between (escaping) fine dust and water vapour can be constrained through the total 
mass loss of the nucleus during the perihelion passage, $M_{\rm L}=(1.05\pm 0.34)\cdot 10^{10}\,\mathrm{kg}$, 
determined with \emph{Rosetta}/RSI \citep{patzoldetal19}. Specifically (equations~\ref{eq:02}, \ref{eq:06}),
\begin{equation} \label{eq:20}
M_{\rm L}=\int_0^P\left(q_{\rm w}+q_{\rm d}\right)A_{\rm tot}\,dt=M_{\rm H_2O}+FM_{\rm H_2O},
\end{equation}
or 
\begin{equation} \label{eq:21}
F=\frac{M_{\rm L}-M_{\rm H_2O}}{M_{\rm H_2O}}.
\end{equation}
Numerically, $F=1.05$, which means that the total amounts of refractories and water vapour lost to space by 67P/C--G 
was nearly equal. Equation~(\ref{eq:19}) yields $G(t)$, shown in the lower panel of Fig.~\ref{fig_prod_erosion}. 
With $0.96\leq G(t)\leq 1.05$, it is clear that the total erosion rate ($\mathrm{kg\,s^{-1}}$) of fine dust $E$ is very similar to the 
observed total water production rate $Q_{\rm H_2O}$, and the differences are smaller than the scatter in water production rate 
measurements (see Figs.~\ref{fig_diffusivity} and \ref{fig_preper_H2O}). Therefore, we decided to apply $Q_{\rm H_2O}$ itself, for the total dust erosion 
rate of the nucleus. Specifically, the local erosion rate ($\mathrm{kg\,m^{-2}\,s^{-1}}$) was set proportional to local illumination 
conditions (determined by latitude and time of day), such that the total amount of eroded dust per time unit equalled $Q_{\rm H_2O}$.

We note that $h=0.87\,\mathrm{m}$, estimated by \citet{davidssonetal21} through mass transfer calculations, is consistent 
with the deposition of $1.4\pm 0.8\,\mathrm{m}$ in Hapi \citep{cambianicaetal20} and $0.7\pm 0.3\,\mathrm{m}$ in Ma'at \citep{cambianicaetal21}, based on 
measurements of boulder shadow lengths and their temporal changes. \citet{marschalletal20} estimated a deposition of $0.14^{+0.22}_{-0.14}\,\mathrm{m}$. 
Applying $h=2.2\,\mathrm{m}$ \citep[the upper limit according to ][]{cambianicaetal20} instead of $h=0.87\,\mathrm{m}$ would increase the peak 
contribution from the extended source from 8 to 18 per cent. The fraction of the nucleus surface being responsible for chunk ejection would 
increase from 4 to 11 per cent. The erosion would be a factor $0.86\leq G(t)\leq 1.05$ times the total water production. With a larger number of 
chunks needed to produce a thicker airfall layer, the nucleus is responsible for a somewhat smaller fraction of the water production. Therefore, the erosion 
taking place because of the gradual removal of fine dust on $\sim 90$ per cent of the surface, proceeds at a somewhat lower rate. Given the factor 2--3 dispersion in 
measured water production rates, a factor 0.82 reduction of the nominal production rate that we try to match  (remembering that \textsc{nimbus} in principle 
should reproduce the contribution from the nucleus, not the full measured $Q_{\rm H2O}$ that also includes the extended source)  has a negligible impact on our results. 
A smaller $h$ value, as suggested by \citet{marschalletal20} would further motivate our decision of using $G(t)=1$.

\section{Results} \label{sec_results}

The spin axis orientation of 67P/C--G is such that the northern hemisphere is being illuminated pre--perihelion \citep[see, e.~g.;][]{kelleretal17}. 
The comet reaches the inbound equinox at $r_{\rm h}=1.67\,\mathrm{au}$ when the sub--solar point enters the southern hemisphere. 
At the $r_{\rm h}=1.24\,\mathrm{au}$ perihelion, the southern hemisphere is fully active while the northern hemisphere has polar night. 
When the outbound equinox is reached at $r_{\rm h}=2.6\,\mathrm{au}$ post--perihelion, the sub--solar point moves into northern 
latitudes again. It is crucially important to be aware of these changes when interpreting the behaviour of nucleus activity. We 
discuss the pre--perihelion branch in Section~\ref{sec_results_preper}, the post--perihelion branch in Section~\ref{sec_results_postper}, 
the overall properties of the solutions in Section~\ref{sec_results_fronts}, and the outgassing force properties in Section~\ref{sec_results_forces}.

\subsection{The pre--perihelion branch} \label{sec_results_preper}

The goal of the \textsc{nimbus} reproduction of the \emph{Rosetta}/ROSINA $\mathrm{H_2O}$ and $\mathrm{CO_2}$ production rate 
measurements \citep{fougereetal16b}
on the pre--perihelion branch is to constrain the physical and chemical properties of (primarily) the northern hemisphere of 67P/C--G under 
post--aphelion conditions. We first focus on the water production rate. 

A substantial number of short test simulations (i.~e., for limited ranges in heliocentric distance) were performed in order to: 1) understand 
the sensitivity of the solutions to different parameter values in different parts of the orbit; 2) to get a first feel for the relevant parameter ranges; 
3) work out a strategy for how to determine the diffusivity and nucleus mass ratio $\mu$ between refractories and water ice.

One set of tests showed that the influence of $\mu$ is dominating over that of diffusivity at perihelion. Specifically, going 
from $\mu=4$ to $\mu=1$ (20 to 50 per cent water ice by mass) led to a 250 per cent increase in water production rate (i.~e., the production 
rate depends linearly on the weight--percentage of water ice). However, increasing the diffusivity by a full three orders of magnitude 
(from $\{L,\,r_{\rm p}\}=\{10,\,1\}\,\mathrm{\mu m}$ to $\{10,\,1\}\,\mathrm{mm}$, using $\xi=1$) just resulted in a 50 per cent increase in water production. 
The reasons for the weak influence of diffusivity during strong sublimation were outlined by \citet{davidssonetal21}. In essence, diffusivity significantly changes 
the sub--surface temperature and vapour pressure distributions, but in such a way that the resulting outgassing rate remains quasi--constant. 
Furthermore, these preliminary tests indicated that $1\stackrel{<}{_{\sim}}\mu\stackrel{<}{_{\sim}} 2$ seems to provide the best reproduction of the gas production rate of 67P/C--G.

A second set of tests were performed far from perihelion (at $r_{\rm h}=3.5\,\mathrm{au}$). They revealed a strong dependence of 
the water production rate on diffusivity, but a comparably weak one on $\mu$. A three orders of magnitude increase in diffusivity leads to roughly 
the same factor of increase in the water production rate. However,  the water production rate still scales linearly with the weight percent of the 
water ice (changing from $\mu=2$ to $\mu=1$, or from 33 per cent to 50 per cent, increases the production rate by a factor 1.5). 

It is therefore clear that the diffusivity can be determined by considering an arbitrary (but realistic) $\mu$--value, and varying $\{L,\,r_{\rm p}\}$ until 
the synthetic water production rate matches the observed one at large heliocentric distances. This should not merely be done locally, but for a substantial 
piece of orbital arc prior to the test point, in order to allow for proper thermal adjustment of the nucleus. With the best--fit diffusivity at hand, it can be applied 
during local simulations at perihelion, in order to fit the actual $\mu$--value. The ultimate test of this combination of diffusivity and water abundance is to perform a full simulation 
from aphelion to perihelion and demonstrate that the entire empirical water production rate curve is fitted both in terms of shape and magnitude. 

In the following, we perform such an analysis, and illustrate the principles just described. We assumed $\xi=1$ and tested $\{L,\,r_{\rm p}\}$ combinations of $\{10,\,1\}\,\mathrm{\mu m}$, 
$\{100,\,10\}\,\mathrm{\mu m}$, and $\{1,\,0.1\}\,\mathrm{mm}$ (the second and third have 10 and 100 times higher diffusivity than the first, respectively). Note that the 
thermophysical solution is insensitive to the individual $\{L,\,r_{\rm p},\,\xi\}$ values (any combination of numerical values that yield the same diffusivity are equivalent).

These models were run for $\mu=2$ (an educated guess based on the previous test simulations), from the $r_{\rm h}=5.68\,\mathrm{au}$ aphelion, to the 
$3.0\leq r_{\rm h}\leq 3.6\,\mathrm{au}$ region where the first ROSINA data were acquired \citep{fougereetal16b}. Figure~\ref{fig_diffusivity} (left) shows that the 
best fit to the early ROSINA data was obtained for $\{L,\,r_{\rm p}\}=\{100,\,10\}\,\mathrm{\mu m}$ (red solid curve). The dependence of the water production on the 
diffusivity is strong, as previously mentioned. That best solution was also tested for a higher water ice abundance $\mu=1$ (red dashed--dotted curve). It is almost 
indistinguishable from the red curve, and Fig.~\ref{fig_diffusivity} (right) shows a close--up. The difference between the $\mu=1$ and $\mu=2$ solutions is smaller 
than the scatter in the empirical data.

As the comet approaches the Sun, the subsolar latitude gradually moves southward, so that the northern hemisphere has polar night at perihelion 
while the southern hemisphere is scorched by the Sun. The near--perihelion activity is therefore dominated by southern vapour and dust production. The challenge 
for \textsc{nimbus} is to properly reproduce this hand--over of prime responsibility for activity from the northern to the southern hemisphere. We decided to apply 
$\{L,\,r_{\rm p}\}=\{100,\,10\}\,\mathrm{\mu m}$ for the southern hemisphere of 67P/C--G as well. In a handful of simulations during one week centred on the perihelion passage, 
we attempted to constrain the refractory/water--ice mass ratio of the southern hemisphere, $\mu_{\rm S}$. One week ($\sim 14$ nucleus rotations) is sufficient to 
establish a balance between dust mantle erosion and $\mathrm{H_2O}$ sublimation--front motion, i.e., reaching a quasi--constant dust mantle thickness, and to 
establish a diurnal steady--state thermal cycling in the region responsible for water outgassing. The results of these simulations are summarised in Table~\ref{tab3}.

\begin{table}
\begin{center}
\begin{tabular}{||l|l||}
\hline
\hline
Refractories/water--ice  & Model perihelion $\mathrm{H_2O}$\\
mass ratio $\mu$ &  production rate\\
\hline
2.0 & $0.53Q_{\rm H_2O}$\\
1.5 & $0.64Q_{\rm H_2O}$\\
1.0 & $0.89Q_{\rm H_2O}$\\
0.5 & $1.69Q_{\rm H_2O}$\\
\hline 
\hline
\end{tabular}
\caption{\textsc{nimbus} was run for four different nucleus mass ratios between refractories and water ice $\mu$ during one week near perihelion. 
The resulting model water production rates are expressed with respect to the measured rate $Q_{\rm H_2O}=1.87\cdot 10^{28}\,\mathrm{molec\,s^{-1}}$ 
according to equation~(\ref{eq:13}).}
\label{tab3}
\end{center}
\end{table}

As can be seen, a reduction of the refractories/water--ice mass ratio from $\mu=2$ to $0.5$ makes the modelled water production grow 
from $0.53$ to $1.69$ times the observed rate $Q_{\rm H_2O}$ (as defined by equation~\ref{eq:13}). The best reproduction of the perihelion 
water production rate is obtained for $\mu=1$ with $0.89Q_{\rm H_2O}$ (particularly because the nucleus production rate $Q_{\rm nuc}$ in reality might be 
some $\sim 8$ per cent lower than the total observed water production rate according to Section~\ref{sec_rates_water}). 
.

\begin{figure}
\scalebox{0.45}{\includegraphics{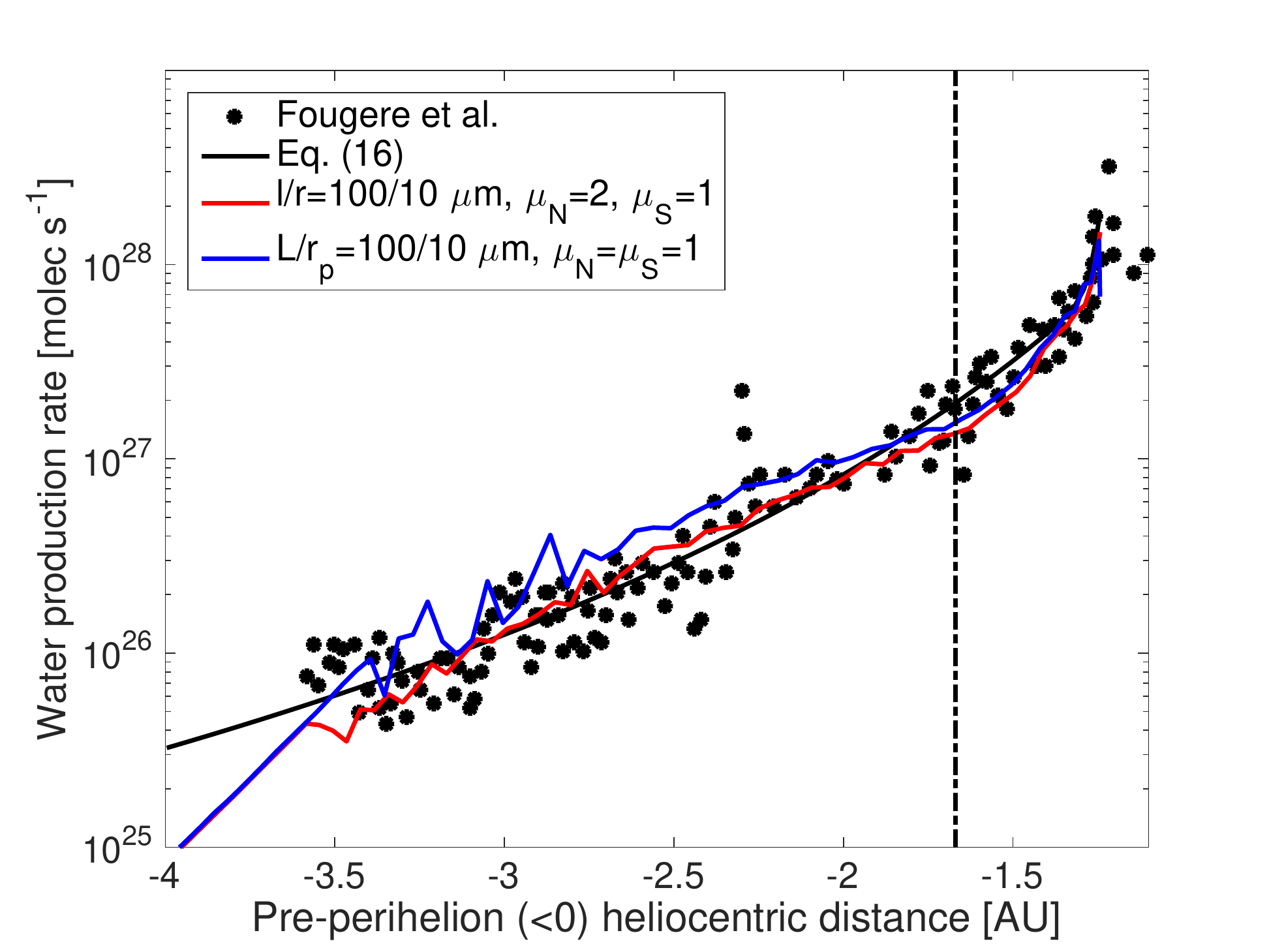}}
     \caption{\textsc{nimbus} models of the pre--perihelion water production rate closely follows the measured data 
when $\{L,\,r_{\rm p}\}=\{100,\,10\}\,\mathrm{\mu m}$, $\mu_{\rm N}=2$ in the north, and $\mu_{\rm S}=1$ in the south 
(and somewhat less so when $\mu_{\rm N}=\mu_{\rm S}=1$). The vertical line marks the inbound equinox.}
     \label{fig_preper_H2O}
\end{figure}

In order to test whether this solution ($\mu=1$ and $\{L,\,r_{\rm p}\}=\{100,\,10\}\,\mathrm{\mu m}$ on both hemispheres) is capable of reproducing the entire 
pre--perihelion water production rate branch, we ran a model from aphelion all the way up to perihelion with this parameter set. Figure~\ref{fig_preper_H2O} 
shows that the model (blue curve) provides a reasonable fit, although the production rate beyond $r_{\rm h}\approx 2\,\mathrm{au}$ tends to be on the high side. 
We therefore decided to introduce a hemispherical compositional dichotomy, with $\mu_{\rm S}=1$ in the south, but a somewhat larger $\mu_{\rm N}=2$ in the 
north (still using $\{L,\,r_{\rm p}\}=\{100,\,10\}\,\mathrm{\mu m}$ and $\xi=1$ everywhere). The resulting red curve in Figure~\ref{fig_preper_H2O} is somewhat 
more convincing, and we consider this our best solution. 

At this point, we turned the attention to the $\mathrm{CO_2}$ production. Observations in Aug--Sep 2014 by ROSINA indicated that the $\mathrm{CO_2}$ production 
predominantly emanated from the southern hemisphere \citep{hassigetal15}. The $\mathrm{CO_2/H_2O}$ production rate ratio was an order of magnitude higher in 
parts of the south compared to most of the north, though that number largely reflects a low southern water production due to the poor illumination conditions. 
The dominance of $\mathrm{CO_2}$ production in the south (although significant amounts where also produced from the Hapi region in the north) was confirmed 
by \citet{fougereetal16} for a longer Aug 2014 to June 2015 time--line. Initially, we therefore only considered models with $\mathrm{CO_2}$ ice on the 
southern hemisphere (using $\{L,\,r_{\rm p}\}=\{100,\,10\}\,\mathrm{\mu m}$, $\xi=1$ and $\mu_{\rm S}=1$, as determined earlier). 

In order to assign a molar $\mathrm{CO_2}$ nucleus abundance relative $\mathrm{H_2O}$, we first assumed that the abundance ratio would be close to that of the coma. 
The coma column density $\mathrm{CO_2/H_2O}$ ratio was $2.4\pm 0.6$ per cent over Aten/Babi, $3.0\pm 0.7$ per cent over Seth/Hapi, and $3.9\pm 1.0$ per cent over Imhotep in April 2015, 
according VIRTIS--M measurements analysed by \citet{migliorinietal16}. \citet{finketal16} analysed VIRTIS--M data from February and April 2015 and obtained column density 
$\mathrm{CO_2/H_2O}$ ratios ranging 3.3--8.5 per cent, from which they inferred a production rate $\mathrm{CO_2/H_2O}$ ratio of 2.2--5.6 per cent. \citet{hansenetal16} used 
ROSINA data to determine average gas mass losses of 83 per cent for $\mathrm{H_2O}$ and 10 per cent for $\mathrm{CO_2}$, corresponding to $\mathrm{CO_2/H_2O}=12$ per cent by mass, 
or $\mathrm{CO_2/H_2O}=4.9$ per cent by number. Based on these measurements, we first applied a nucleus $\mathrm{CO_2/H_2O}=5.5$ per cent molar ratio.

\begin{figure*}
\centering
\begin{tabular}{cc}
\scalebox{0.45}{\includegraphics{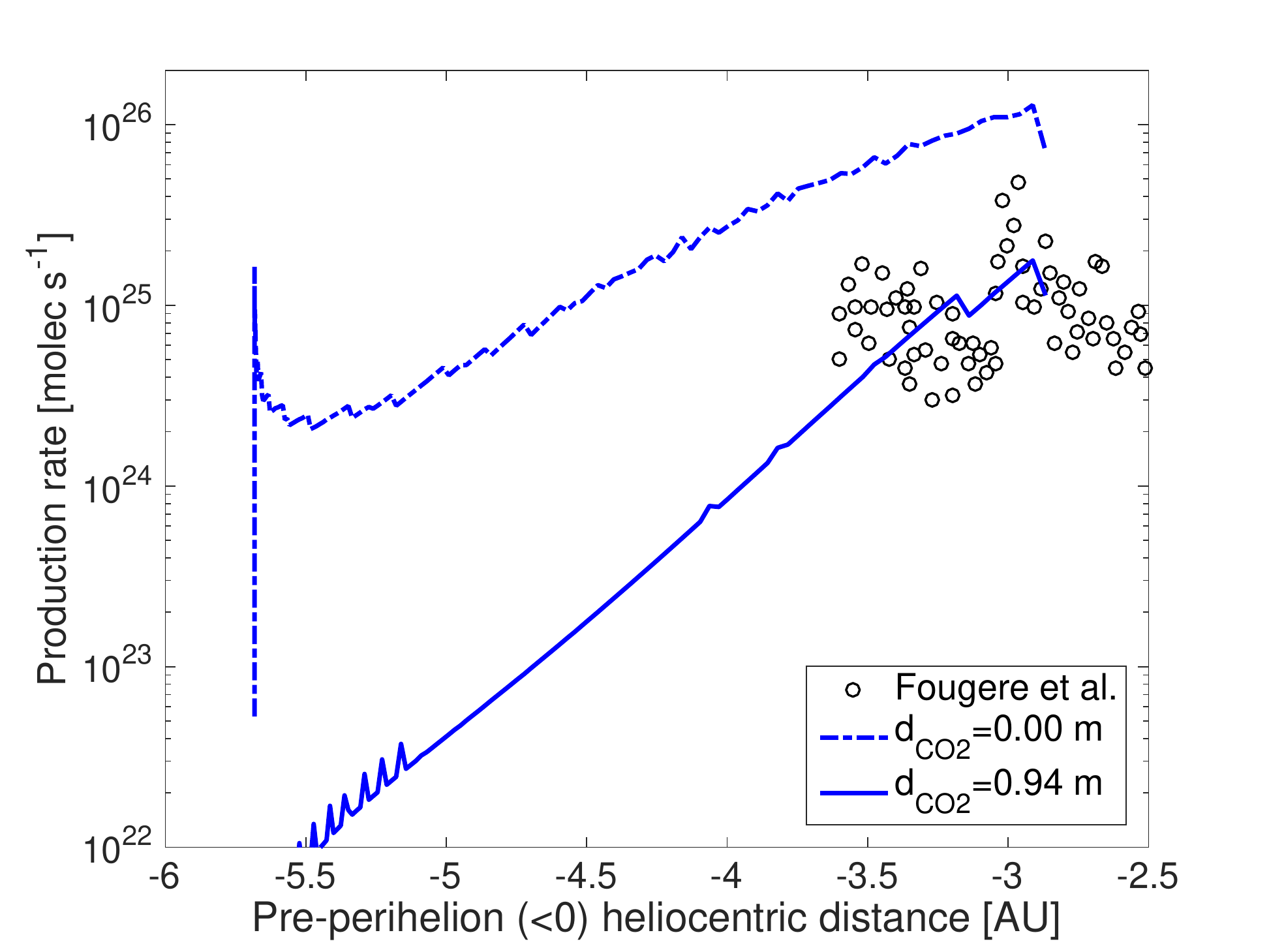}} & \scalebox{0.45}{\includegraphics{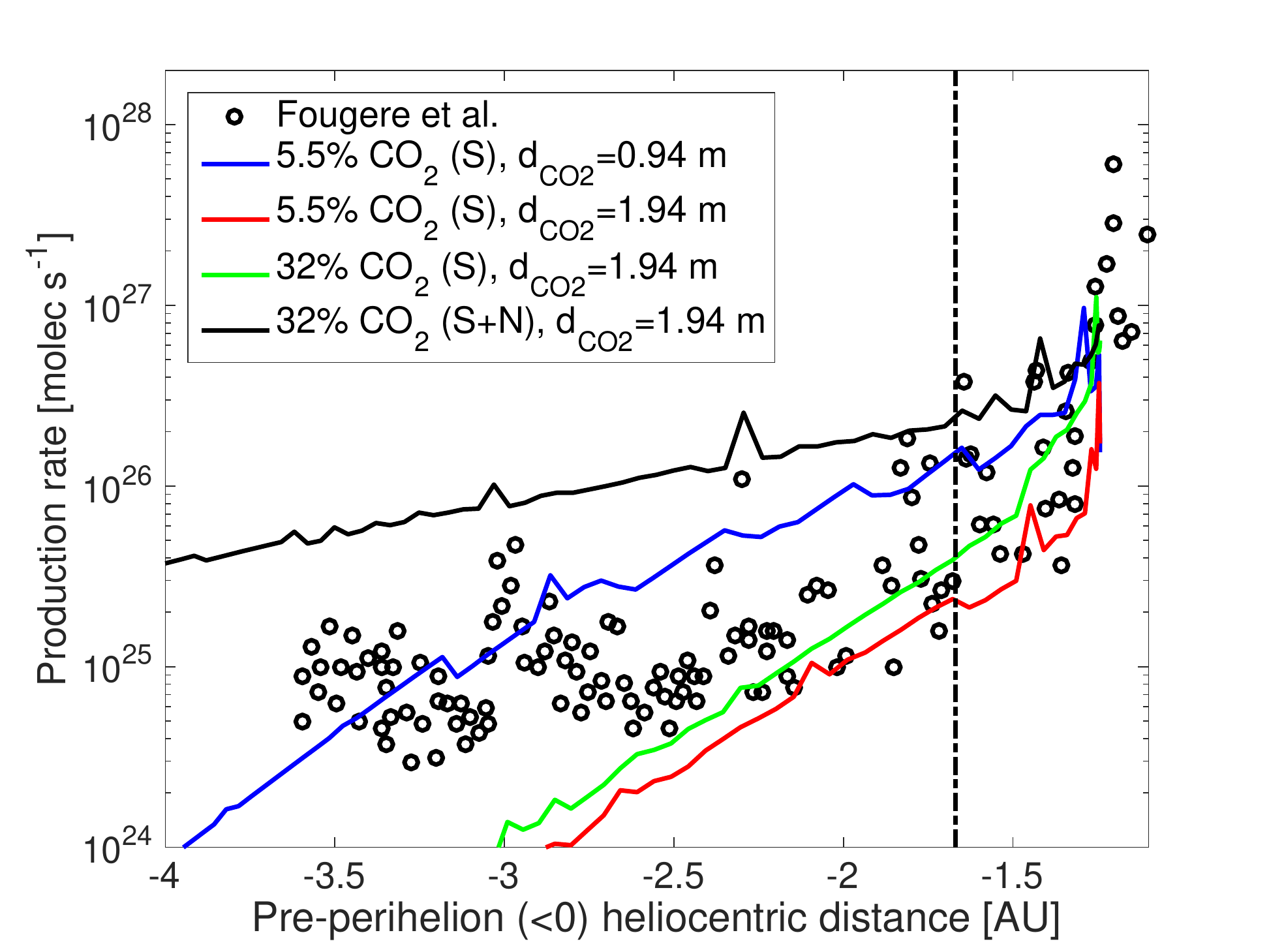}}\\
\end{tabular}
     \caption{\emph{Left:} Pre--perihelion $\mathrm{CO_2}$ production rates (until August 2014) resulting from having $\mathrm{CO_2}$ ice exclusively on the 
southern hemisphere, but at different initial depths $d_{\rm CO2}$. Both models have $\{L,\,r_{\rm p}\}=\{100,\,10\}\,\mathrm{\mu m}$, $\mu_{\rm S}=1$, and 
5.5 per cent $\mathrm{CO_2}$ relative to water by number. \emph{Right:} Pre--perihelion $\mathrm{CO_2}$ production rates (up to the August 2015 perihelion for models, but somewhat beyond 
for measurements to show that the $\mathrm{CO_2}$ production peaked post--perihelion) for different 
initial depths, and intrinsic nucleus $\mathrm{CO_2}$ abundances (relative $\mathrm{H_2O}$ by number). The model shown as a black curve has $\mathrm{CO_2}$ on 
both hemispheres, the others only have $\mathrm{CO_2}$ in the south. The vertical line marks the inbound equinox.}
     \label{fig_CO2A}
\end{figure*}

Figure~\ref{fig_CO2A} (left) shows one model for which $\mathrm{CO_2}$ was present up to the very surface on the southern hemisphere at  May 2012 aphelion. 
If that is the case, the $\mathrm{CO_2}$ sublimation front only has time to withdraw to a depth of at most $0.84\,\mathrm{m}$ by the time \emph{Rosetta} made its 
first observations in Aug 2014. That results in a $\mathrm{CO_2}$ production rate that is too high. The $\mathrm{CO_2}$ sublimation front must therefore be located 
deeper at the onset of simulations. Initial tests showed that $d_{\rm CO2}=0.94\,\mathrm{m}$ in the south resulted in a $\mathrm{CO_2}$ production rate that was 
about right in August 2014 (Fig.~\ref{fig_CO2A}, left).

When that model was propagated all the way to perihelion (blue curve in Fig.~\ref{fig_CO2A}, right) it tended to overshoot the observed production for a large 
fraction of the inbound orbit. It was therefore clear that the contribution from the southern hemisphere had to be smaller, in order to avoid the 
overshoot, and that some $\mathrm{CO_2}$ production from the north might be necessary, to still fit the large--distance data. We first increased the 
initial front depth in the south gradually, to $d_{\rm CO2}=1.94\,\mathrm{m}$, which produced a curve that followed the lower part of the 
data cloud at $r_{\rm h}\stackrel{<}{_{\sim}}\,2\,\mathrm{au}$ reasonably well (red curve in Fig.~\ref{fig_CO2A}, right).

At this point we also tested the sensitivity of the total $\mathrm{CO_2}$ production rate to the assumed nucleus $\mathrm{CO_2/H_2O}$ ratio. If a substantial fraction 
of cometary ices are presolar \citep[as suggested by the presence of $\mathrm{S}_2$ and by the xenon isotope composition;][]{calmonteetal16,martyetal17} 
the cometary $\mathrm{CO_2}$ composition may be close to that of protostars. Massive protostars have molar $\mathrm{CO_2}$ abundances relative $\mathrm{H_2O}$ 
of 10--23 per cent \citep{gerakinesetal99}, while low--mass protostars have $32\pm 2$ per cent \citep{pontoppidanetal08}. We here apply 32 per cent.

The dependence on the intrinsic abundance was rather weak at large heliocentric distance (see red and green curves in Fig.~\ref{fig_CO2A}, right). Even when applying almost a sixfold increase in abundance from 
$5.5$ per cent $\mathrm{CO_2}$ to 32 per cent $\mathrm{CO_2}$ relative to water, the resulting increase of the $\mathrm{CO_2}$ 
production rate was merely a factor $\sim 1.5$ at $r_{\rm h} \stackrel{>}{_{\sim}} 1.7\,\mathrm{au}$, substantially smaller than the scatter of the data. Observations of the 
$\mathrm{CO_2/H_2O}$ production rate ratio at large heliocentric distances therefore does not offer meaningful clues on the nucleus $\mathrm{CO_2}$ abundance. However, because of 
substantial water--driven erosion near the south pole, the $\mathrm{CO_2}$ fronts were locally brought very close to the surface. Near perihelion (within $r_{\rm h} \stackrel{<}{_{\sim}} 1.4\,\mathrm{au}$) 
the $\mathrm{CO_2}$ sublimation front depths stabilised because their propagation speeds matched that of nucleus erosion. This happened at $\sim 0.45\,\mathrm{m}$ depth when the 
abundance was 5.5 per cent, but at $\sim 0.15\,\mathrm{m}$ for 32 per cent abundance. At this point, the higher--abundance model produced at most $\sim 5$ times more 
$\mathrm{CO_2}$ vapour than the low--abundance model, somewhat short of the factor 5.8 intrinsic abundance difference. There are two likely contributing factors for this 
discrepancy: 1) downward diffusion and recondensation of vapour below the front during approach to the Sun has altered the $\mathrm{CO_2}$ abundance of the sublimation front at perihelion 
with respect to that of the deep interior; 2) there are smaller contributions from mid--southern latitudes that still are approaching steady state. 

The low--abundance model provided $2.6\cdot 10^{26}\,\mathrm{molec\,s^{-1}}$ just before perihelion (briefly spiking to $4\cdot 10^{26}\,\mathrm{molec\,s^{-1}}$), 
while the high--abundance model provided $1.4\cdot 10^{27}\,\mathrm{molec\,s^{-1}}$. The measured rates (within $0.1\,\mathrm{au}$ prior to perihelion) had a range 
$7.9\cdot 10^{25}$--$1.3\cdot 10^{27}\,\mathrm{molec\,s^{-1}}$ (average $4.8\cdot 10^{26}\,\mathrm{molec\,s^{-1}}$) 
before perihelion. However, the $\mathrm{CO_2}$ production peaked shortly after perihelion, with a range $6.4\cdot 10^{26}$--$6.0\cdot 10^{27}\,\mathrm{molec\,s^{-1}}$, and an average 
of $2.1\cdot 10^{27}\,\mathrm{molec\,s^{-1}}$. Because the 5.5 per cent--model would not 
be able to reach the observed range right after perihelion, we consider the 32 per cent--model more representative of the nucleus behaviour. Note, that it would not be possible to obtain a significantly higher 
$\mathrm{CO_2}$ production rate at perihelion simply by reducing the initial front depth at aphelion. Such a model would stabilise at a similar $\sim 0.45\,\mathrm{m}$ steady--state depth near perihelion, and 
provide similar amounts of $\mathrm{CO_2}$ vapour. We therefore think that a nucleus molar $\mathrm{CO_2/H_2O}$ abundance ratio of $\sim 0.3$ is a necessity to explain the observed data. 

Although the high--abundance model performed well at perihelion, it still grossly under--estimated the production at larger distances. We therefore introduced $\mathrm{CO_2}$ 
with $d_{\rm CO2}=1.94\,\mathrm{m}$ on the northern hemisphere as well (at 32 per cent abundance), but as seen in Fig.~\ref{fig_CO2A} (right), the resulting production rate (black curve) was too high.

\begin{figure}
\scalebox{0.45}{\includegraphics{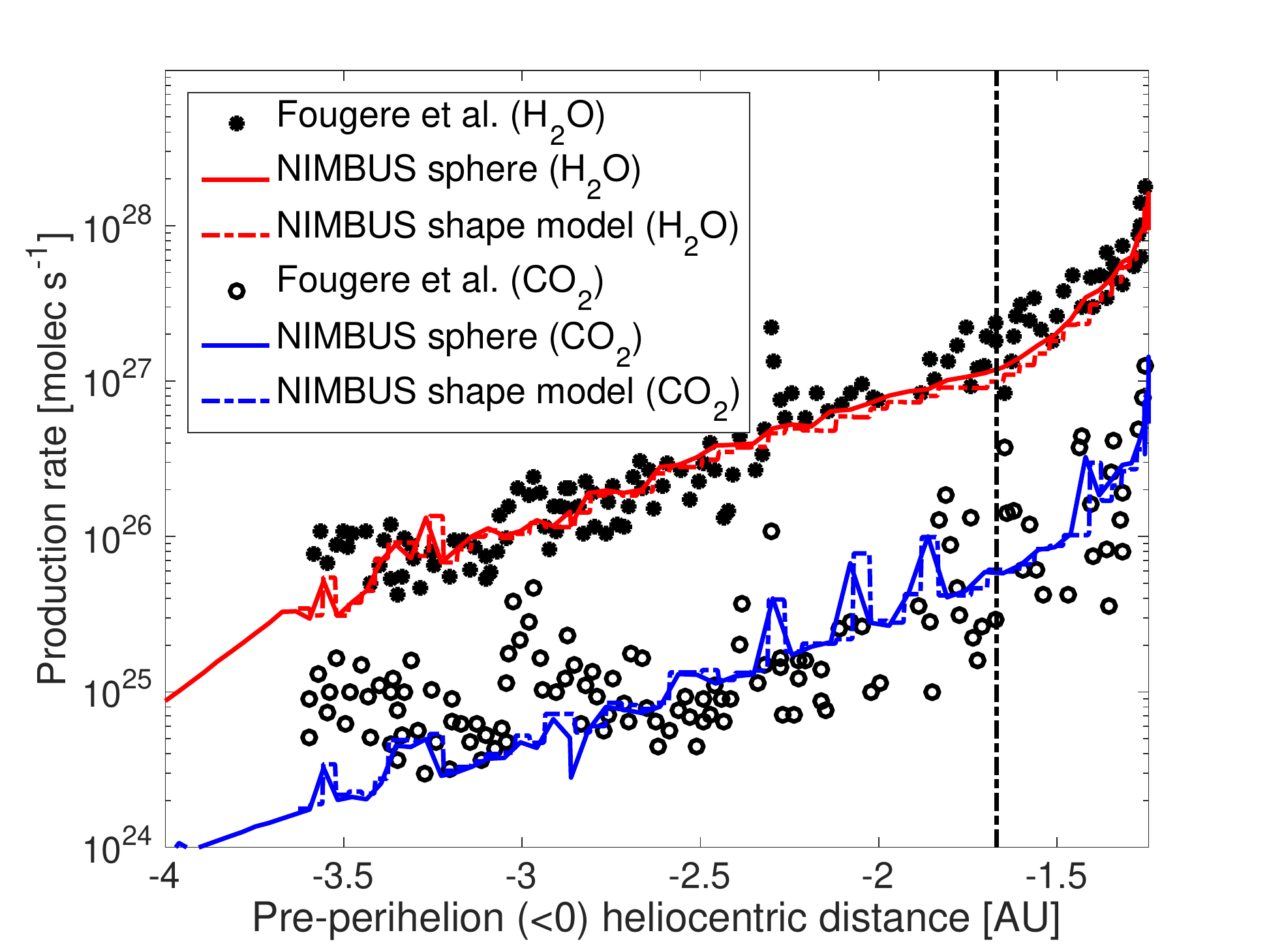}}
     \caption{\textsc{nimbus} models of the pre--perihelion $\mathrm{H_2O}$ and $\mathrm{CO_2}$ production rates for spherical and 67P/C--G--shaped 
model nuclei. The models have $\{L,\,r_{\rm p}\}=\{100,\,10\}\,\mathrm{\mu m}$, $\mu_{\rm N}=2$ in the north, and $\mu_{\rm S}=1$ in the south. The initial 
aphelion $\mathrm{CO_2}$ sublimation front depth was $d_{\rm CO2}=3.77\,\mathrm{m}$ in the north (11 per cent $\,\mathrm{CO_2}$) and $d_{\rm CO2}=1.94\,\mathrm{m}$ in the south (32 per cent $\,\mathrm{CO_2}$). 
The spikes in the $\mathrm{CO_2}$ production are formed when frost accumulating at shallow depths in cold conditions are released suddenly during illumination. The vertical line marks the inbound equinox.}
     \label{fig_preper_H2O_CO2}
\end{figure}

We therefore tested a number of initial depths $d_{\rm CO2}>1.94\,\mathrm{m}$ for the aphelion $\mathrm{CO_2}$ sublimation front depths in the north. We did this for a nucleus 
$\mathrm{CO_2}$ abundance of 11 per cent relative $\mathrm{H_2O}$ by number in the north (roughly half--ways between the two extreme values tested previously, remembering 
that the production rate of $\mathrm{CO_2}$ does not depend strongly on the absolute abundance at large distances). We always applied 32 per cent in the south. We found, 
that the combined contribution from the south (with $d_{\rm CO2}=1.94\,\mathrm{m}$) and the north (with $3.02\stackrel{<}{_{\sim}}d_{\rm CO2}\stackrel{<}{_{\sim}}3.77\,\mathrm{m}$), 
matched the full width of the observed $\mathrm{CO_2}$ production in August 2014 ($r_{\rm h}\approx 3.6\,\mathrm{au}$). We propagated a model with $d_{\rm CO2}=3.77\,\mathrm{m}$ in the north and 
$d_{\rm CO2}=1.94\,\mathrm{m}$ in the south to perihelion, as shown in Fig.~\ref{fig_preper_H2O_CO2} as solid curves.

This model achieves a decent simultaneous reproduction of the pre--perihelion $\mathrm{H_2O}$ and $\mathrm{CO_2}$ production rate measurements. 
It is possible that further adjustments of the aphelion $\mathrm{CO_2}$ ice depths latitude--by--latitude would make the $\mathrm{CO_2}$ curve follow the 
centre of the data cloud more closely. However, because the reason for the nearly one order of magnitude scatter in the measured data is unknown, it is 
not certain that such a solution provides physically real information about the target. It is also possible that clean $\mathrm{CO_2}$ ice is not the 
sole contributor of $\mathrm{CO_2}$ vapour -- a portion may originate from occluded $\mathrm{CO_2}$ within crystallising amorphous water ice. 
An additional source of $\mathrm{CO_2}$ might be needed to better match the rather high production rate beyond $2.5\,\mathrm{au}$ pre--perihelion.
For the time being, we settle for the model in Fig.~\ref{fig_preper_H2O_CO2} as an acceptable solution, but we may return to a more thorough investigation 
in a future publication.

The \textsc{nimbus} simulations are performed for a spherical model nucleus. Although it has the same surface area as the real irregular nucleus, it is 
possible that systematic differences in solar--exposed surface area and shape differences causes production rate deviations. To investigate the severity 
of this problem, we considered an irregular shape model obtained by degrading the $3.1\cdot 10^6$ facet SHAP5 version 1.5 shape model \citep{jordaetal16} 
to $5\cdot 10^3$ facets. The irregular model nucleus was advanced in $10^{\circ}$ rotational angle increments from $3.6\,\mathrm{au}$ pre--perihelion to perihelion. 
For each facet, the co--declination (angle between the positive spin vector and the facet outward surface normal) and the solar incidence angle were calculated, and the 
closest available proxy on the spherical model was identified. This procedure accounted for the varying spin rate of the comet, i.~e., the solar incidence angle is 
in--phase with the real nucleus. We used spin periods reported by \citet{kelleretal15b} and by H.~U.~Keller (private communication).  Because the facet and the proxy 
have identical illumination histories (apart from potential temporary shadowing taking place at 
earlier rotational phases), their instantaneous production rates ought to be similar. We apply those production rates locally, adjusted for the actual facet surface area. However, we then apply 
the model of \citet{davidssonandrickman14} to identify the facets that are shadowed by nucleus topography at any given rotational phase. For shadowed facets we re--set their surface temperatures 
and water production rates to the lowest values encountered on the nightside for that co--declination. The $\mathrm{CO_2}$ production rate is maintained during temporary dayside shadowing, because the $\mathrm{CO_2}$ 
sublimation front is located at a depth that is rather insensitive to diurnal temperature variations. However, the applied speed of $\mathrm{CO_2}$ molecules when entering the coma reflects the fact that they need to 
diffuse through a cold surface region. 

The error in the $\mathrm{H_2O}$ production rate introduced by the mapping technique is small. At shadowing onset, the real $\mathrm{H_2O}$ 
production drops gradually, while the mapping leads to an abrupt reduction to night--time production levels. Observations of jets show 
that it takes $\sim 1\,\mathrm{h}$ for the water activity to diminish to a level where it no longer can sustain a detectable dust production. Thus, the model 
temporarily has a deficit corresponding to $\sim 1/P_{\rm rot}$ or roughly 8 per cent of the daily water production. However, when the region exits from shadow, 
the mapping causes an immediate return to full activity, while the real nucleus re--activates gradually. That creates a temporary over--production for the modelled nucleus 
that partially or fully compensates for the previous deficit. Because we consider diurnally--averaged production rates, we expect the calculated rate to be off by $\ll 8$ per cent 
because of the mapping (compared to a model that would accurately consider the activity changes during shadowing).

For $\mathrm{CO_2}$ the situation is different, because it is located at a depth where diurnal temperature variations are damped out (as mentioned previously, 
$\mathrm{CO_2}$ is continuously active). An error is introduced by the mapping technique because the integrated daily energy absorption is too high when shadows are not 
accounted for. The severity of this problem can be estimated from the total amount of energy absorbed during an orbit at different locations calculated by \citet{sierksetal15}. 
The frequently shadowed Hapi valley receives $\sim 5.5\cdot 10^9\mathrm{J\,m^{-2} orbit^{-1}}$, which is $\sim 35$ per cent lower than regions with similar co--declination 
that are not being shadowed. However, this only affects an estimated 1/5 of the northern hemisphere. The estimated $\mathrm{CO_2}$ production is therefore $\stackrel{<}{_{\sim}} 7$ 
per cent too high, also considering that all the excess energy does not necessarily go to $\mathrm{CO_2}$ production. This is small compared to the factor 2--3 spread in the measured data.

We average the total nucleus production over one nucleus rotation, and plot the production rates in Fig.~\ref{fig_preper_H2O_CO2} as dashed--dotted curves. At the resolution 
of the figure, the two sets of curves are barely distinguishable. The largest differences are found near the May 11, 2015 equinox (at $r_{\rm h}=1.67\,\mathrm{au}$), where the 
production rate of the irregular model nucleus is somewhat below that of the spherical model. That is because the irregular nucleus has the smallest cross section when 
viewed from within the equatorial plane of the comet. However, the difference is small in comparison to the scatter of the measurements. Therefore, we do not find that the 
nucleus shape has a measurable influence on the production rates for 67P/C--G. The same conclusion was drawn by \citet{marshalletal19}.

\subsection{The post--perihelion branch} \label{sec_results_postper}

During the perihelion passage, most of the northern hemisphere has polar night. Solid material emanating from the south rains down in 
the north as airfall. \citet{davidssonetal21} estimated the average thickness of the airfall layer added to the north as $0.87\,\mathrm{m}$. 
They found that cm--sized chunks would retain 44 per cent of its original water ice abundance during a $12\,\mathrm{h}$ coma transfer, and that 
a dm--sized chunk would retain  94 per cent of its ice. With the average size of returning coma chunks being $\sim 1\,\mathrm{cm}$ (see Table~\ref{tab2}), and 
the intrinsic water abundance in the southern hemisphere found to be $\mu_{\rm S}\approx 1$, it is reasonable that the airfall material has $\mu\approx 2$. We therefore 
take the most successful pre--perihelion model, add a $0.87\,\mathrm{m}$ layer on the northern hemisphere, and assume it has 
$\mu_{\rm N}=2$, $\{L,\,r_{\rm p}\}=\{100,\,10\}\,\mathrm{\mu m}$, $\xi=1$, and a bulk porosity of 70 per cent (including the previously applied 40 per cent macro porosity 
plus an assumed micro porosity within chunks). We assigned an initial temperature of $T=150\,\mathrm{K}$ to the  
airfall material. It would have had $T\approx 200\,\mathrm{K}$ when exposed to the Sun in the coma \citep{davidssonetal21} but could have cooled for 
hours once entering the shadow of the nucleus before landing. The near--surface temperature of the nucleus in polar--night regions is typically $70$--$100\,\mathrm{K}$ 
at perihelion \citep{davidssonetal21}. The nucleus erosion rate was updated for outbound conditions according to equation~\ref{eq:13}.

\begin{figure}
\scalebox{0.45}{\includegraphics{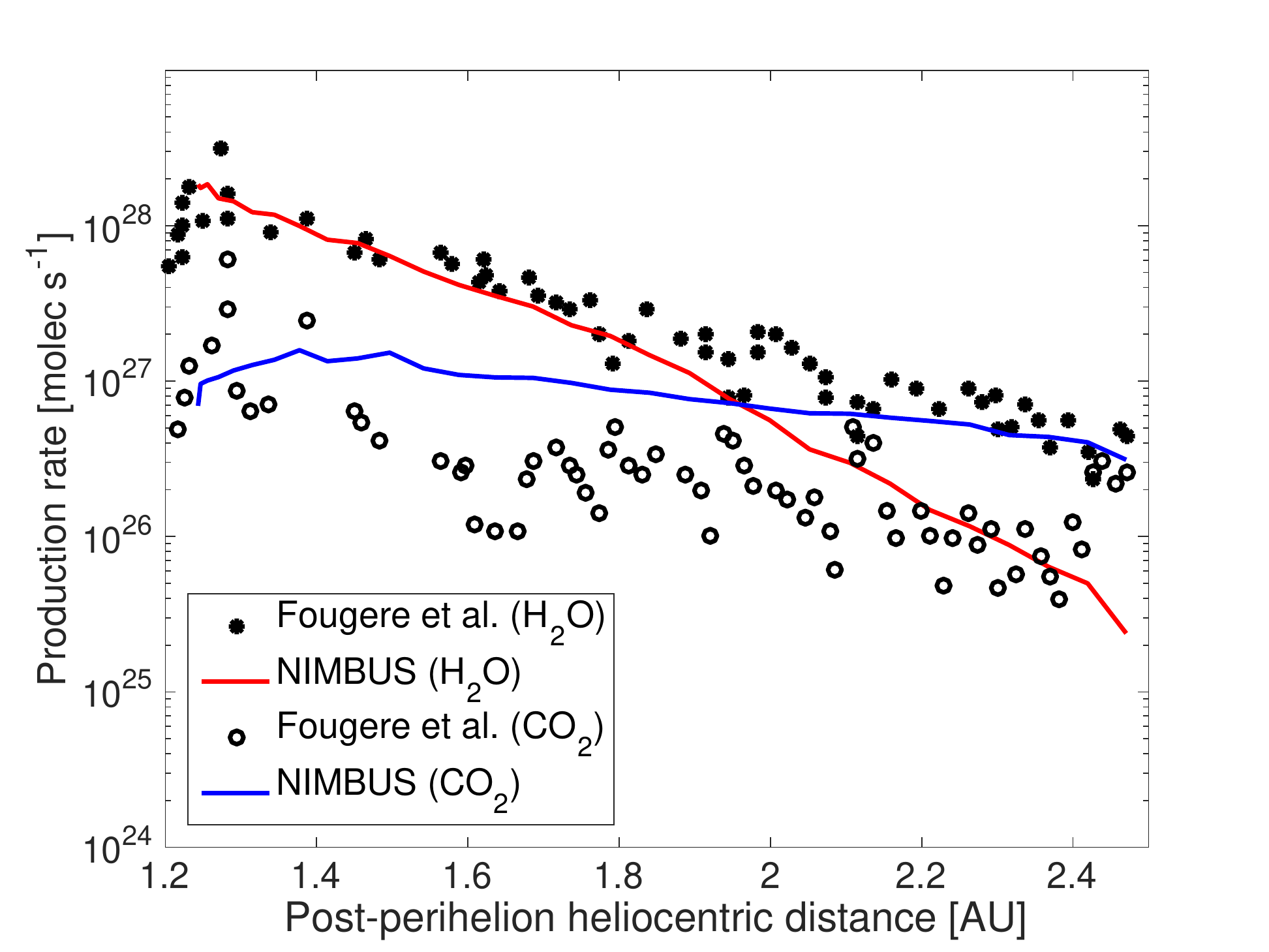}}
     \caption{The \textsc{nimbus} model that successfully reproduced the pre--perihelion $\mathrm{H_2O}$ and $\mathrm{CO_2}$ production rates (see Fig.~\ref{fig_preper_H2O_CO2} 
for details) are here propagated post--perihelion. The fits are not satisfactory neither for $\mathrm{H_2O}$, nor for $\mathrm{CO_2}$.}
     \label{fig_postper_first}
\end{figure}

The result of this model is seen in Fig.~\ref{fig_postper_first}. Interestingly, this model fails to reproduce the data for both species. 
The water production is well--reproduced within $1.8\,\mathrm{au}$, but then the model drops below the data. By the time the 
outbound equinox is reached on March 21, 2016 at $r_{\rm h}=2.6\,\mathrm{au}$, the modelled $\mathrm{H_2O}$ production is an 
order of magnitude too small. The observed $\mathrm{CO_2}$ production rate falls by more than an order of magnitude between 
perihelion and the outbound equinox, while the modelled curve is less steep. There are a few unusually high $\mathrm{CO_2}$ 
measurements near equinox that almost reach the model curve, but the bulk of the data is clearly lower. The discrepancy is a factor $\sim 3$ 
near equinox, i.~e., less severe than that for water. 

To investigate the importance of the added airfall layer (admittedly having uncertain thickness and initial temperature) we also 
propagated the best pre--perihelion model without any airfall layer. That model had $\mathrm{H_2O}$ and $\mathrm{CO_2}$ 
production rate curves that were very similar to Fig.~\ref{fig_postper_first}. This can be understood as follows. First of all, most 
water originates from the south, which is not affected by airfall. However, the north becomes increasingly exposed with time post--perihelion, yet 
there was no significant deviation in production rate between the models. The main factor influencing the near--perihelion water production rate 
is the thickness of the dust mantle. That thickness is determined by the $\mu$--value (determining how rapidly the water sublimation front is withdrawing) 
and the erosion rate (determining how quickly the surface is `catching up' with the moving water sublimation front). With the imposed erosion rates 
being identical in the two models, and the $\mu$--value being the same, the two models will only differ because they had different initial temperatures 
at perihelion ($150\,\mathrm{K}$ for the airfall model and $70$--$100\,\mathrm{K}$ for the other model) and different types of stratification (no dust mantle 
assumed for the airfall layer, versus a thin dust mantle established pre--perihelion for the other model). Apparently, these differences were equilibrated so rapidly 
once the airfall material is exposed to sunlight, that they play no practical role. A thin dust mantle is established so rapidly, and the temperature gradient over the 
near--surface region reaches a repetitive oscillatory behaviour during nucleus rotation so quickly, that the nucleus behaves similarly with or without airfall, as far as 
water production is concerned. Regarding $\mathrm{CO_2}$, the addition of airfall material quenches the contribution from the north (the $\mathrm{CO_2}$ mass flux 
to the surface is reduced by the fact that the $\mathrm{CO_2}$ ice suddenly is $0.87\,\mathrm{m}$ deeper below the surface than before). However, the vast majority 
of the $\mathrm{CO_2}$ originates from the south, which is unaffected by airfall. Therefore, the quenching of the northern contribution barely affects the 
total $\mathrm{CO_2}$ production.

We proceed to investigate what parameter changes, if any, that would lead to reproduction of the data. First, we focus on the 
water production. Our previous experience was that the near--perihelion production rate is sensitive to the water abundance 
(i.e., the dust/water--ice mass ratio) but that distant production is more sensitive to diffusivity. We therefore postulate that fresh 
airfall material has a higher diffusivity, than for the aged airfall material for which we fitted $\{L\,r_{\rm p}\}=\{100,\,10\}\,\mathrm{\mu m}$ and $\xi=1$ on the 
inbound trajectory. Accordingly, we test how much the diffusivity \emph{on the northern hemisphere} would have to increase in order to close 
the gap between the model and the observations.

\begin{figure}
\scalebox{0.45}{\includegraphics{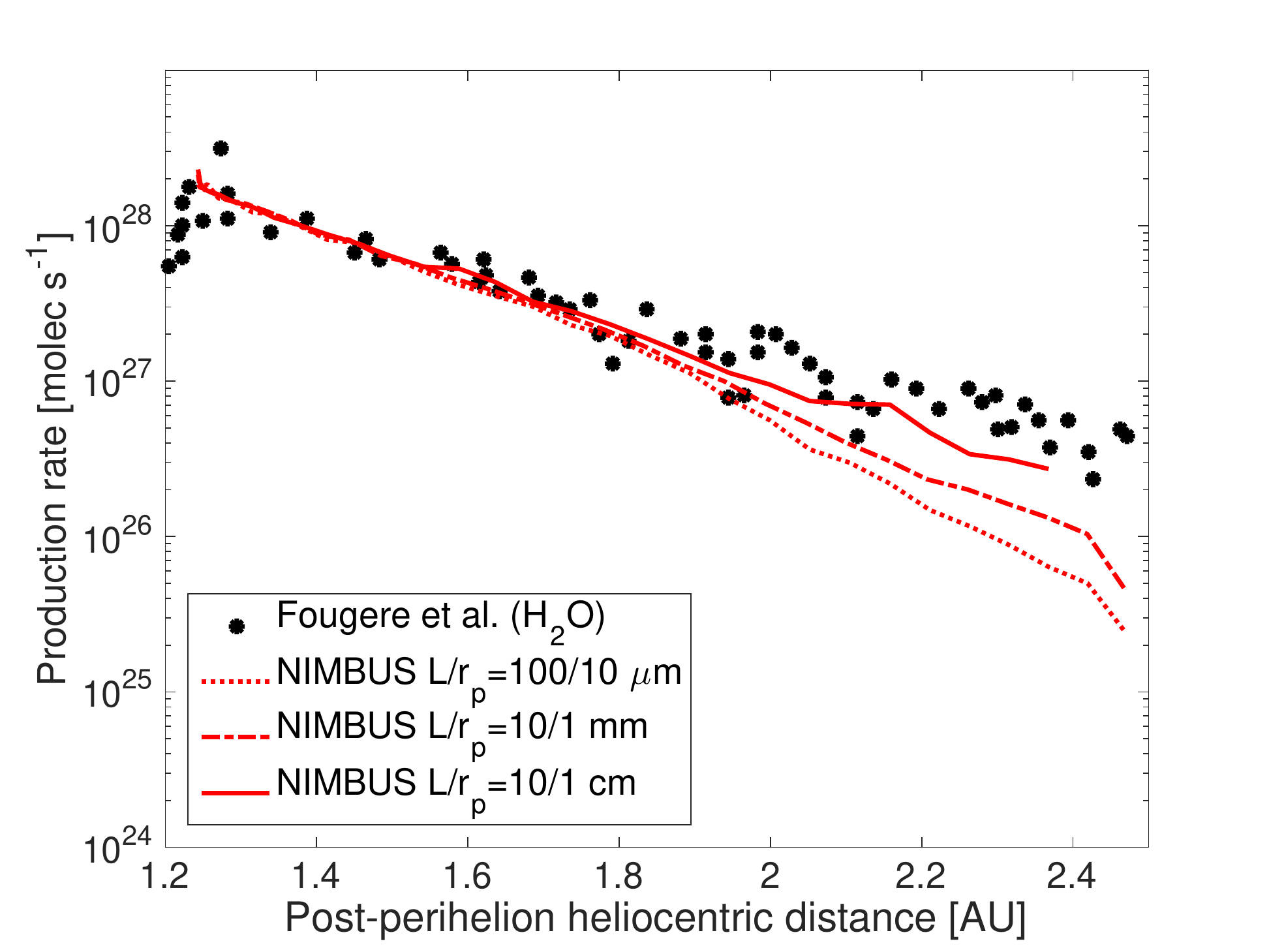}}
     \caption{\textsc{nimbus} models with different diffusivities on the northern hemisphere (realised by the tube lengths $L$ and radii $r_{\rm p}$ given in the legend), 
aiming at reproducing the measured post--perihelion water production rate. The southern hemisphere has $\{L\,r_{\rm p}\}=\{100,\,10\}\,\mathrm{\mu m}$.}
     \label{fig_postper_water}
\end{figure}

These models are shown in Fig.~\ref{fig_postper_water}. The model that has a three orders of magnitude higher diffusivity ($\{L,\,r_{\rm p}\}=\{10,\,1\}\,\mathrm{cm}$ and $\xi=1$) 
than the best pre--perihelion model ($\{L,\,r_{\rm p}\}=\{100,\,10\}\,\mathrm{\mu m}$ and $\xi=1$) is capable of increasing the modelled near--equinox water production 
rate to the level of the measurements. There are no differences between the models to speak of at $r_{\rm h}\stackrel{<}{_{\sim}}\,1.8\,\mathrm{au}$. Differences are 
only seen when sufficiently large parts of the northern hemisphere (with the high--diffusivity airfall layer) are illuminated, and start contributing measurably to the total 
water production rate.

The problem for $\mathrm{CO_2}$ is the opposite: the near--equinox production needs to be reduced. Covering the northern hemisphere by a rather 
thick airfall layer is not sufficient. We therefore proceed to explore to what extent a reduction of the diffusivity for $\mathrm{CO_2}$ on the southern hemisphere 
is capable of solving the problem. A possible mechanism for such a near--perihelion diffusivity reduction is discussed in Section~\ref{sec_discussion}.

\begin{figure}
\scalebox{0.45}{\includegraphics{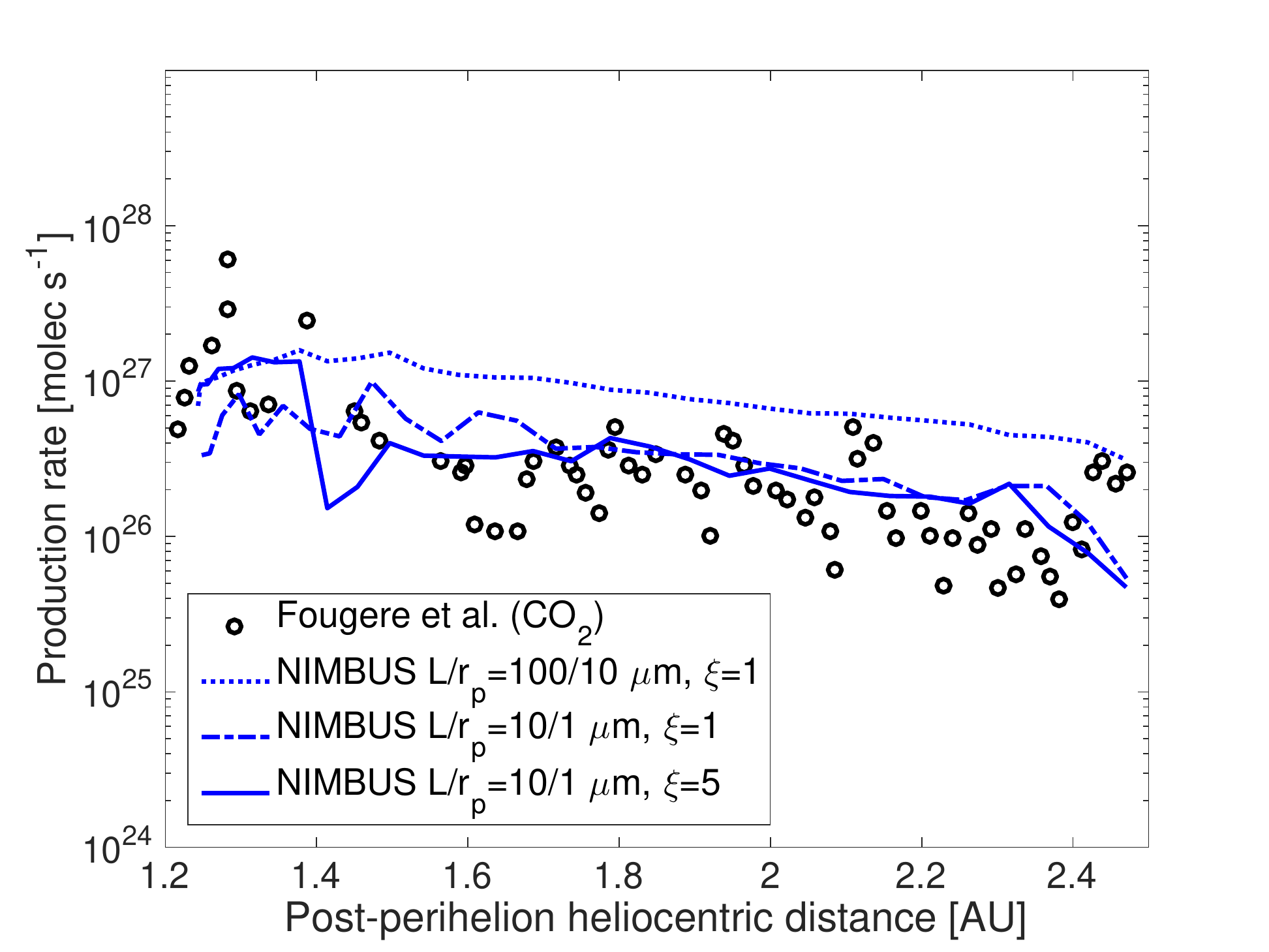}}
     \caption{\textsc{nimbus} post--perihelion $\mathrm{CO_2}$ production rates for different values of diffusivity. Note that the solid 
curve assumes $\{L,\,r_{\rm p}\}=\{100,\,10\}\,\mathrm{\mu m}$ and $\xi=1$ at $r_{\rm h}\leq 1.4\,\mathrm{au}$, before switching to 
$\{L,\,r_{\rm p}\}=\{10,\,1\}\,\mathrm{\mu m}$ and $\xi=5$ beyond that distance.}
     \label{fig_postper_CO2}
\end{figure}

Figure~\ref{fig_postper_CO2} shows the effect of reducing the diffusivity one order of magnitude, realised by changing 
$\{L,\,r_{\rm p}\}=\{100,\,10\}\,\mathrm{\mu m}$ to $\{L,\,r_{\rm p}\}=\{10,\,1\}\,\mathrm{\mu m}$ with $\xi=1$ fixed. That indeed causes a 
substantial improvement near the equinox, although the production rates now just touches the lower part of the empirical data cloud near perihelion. 
Therefore, it was decided to keep $\{L,\,r_{\rm p}\}=\{100,\,10\}\,\mathrm{\mu m}$ at $r_{\rm h}\leq 1.4\,\mathrm{au}$ and lower the 
diffusivity beyond that distance. Using tube radii and lengths much below $\{L,\,r_{\rm p}\}=\{10,\,1\}\,\mathrm{\mu m}$ is probably 
not particularly realistic, considering that the dimensions of the basic solid `monolithic' component in comet material is about one 
micrometer \citep[e.g.,][]{brownleeetal06}. However, straight tubes (tortuosity $\xi=1$) are not particularly realistic either. Therefore, 
diffusivity was lowered further by considering curvy tubes, with a length being a factor $\xi$ larger than the vertical distance travelled by 
flowing through the tube (this is the very definition of tortuosity). Figure~\ref{fig_postper_CO2} shows that an 
additional factor 25 reduction of the diffusivity does not lower the production rate further.  We conclude that $\{L,\,r_{\rm p}\}=\{10,\,1\}\,\mathrm{\mu m}$ 
and $\xi\geq 1$ is necessary for the \textsc{nimbus} curve to touch the upper part of the main data cloud near equinox.

\begin{figure}
\scalebox{0.45}{\includegraphics{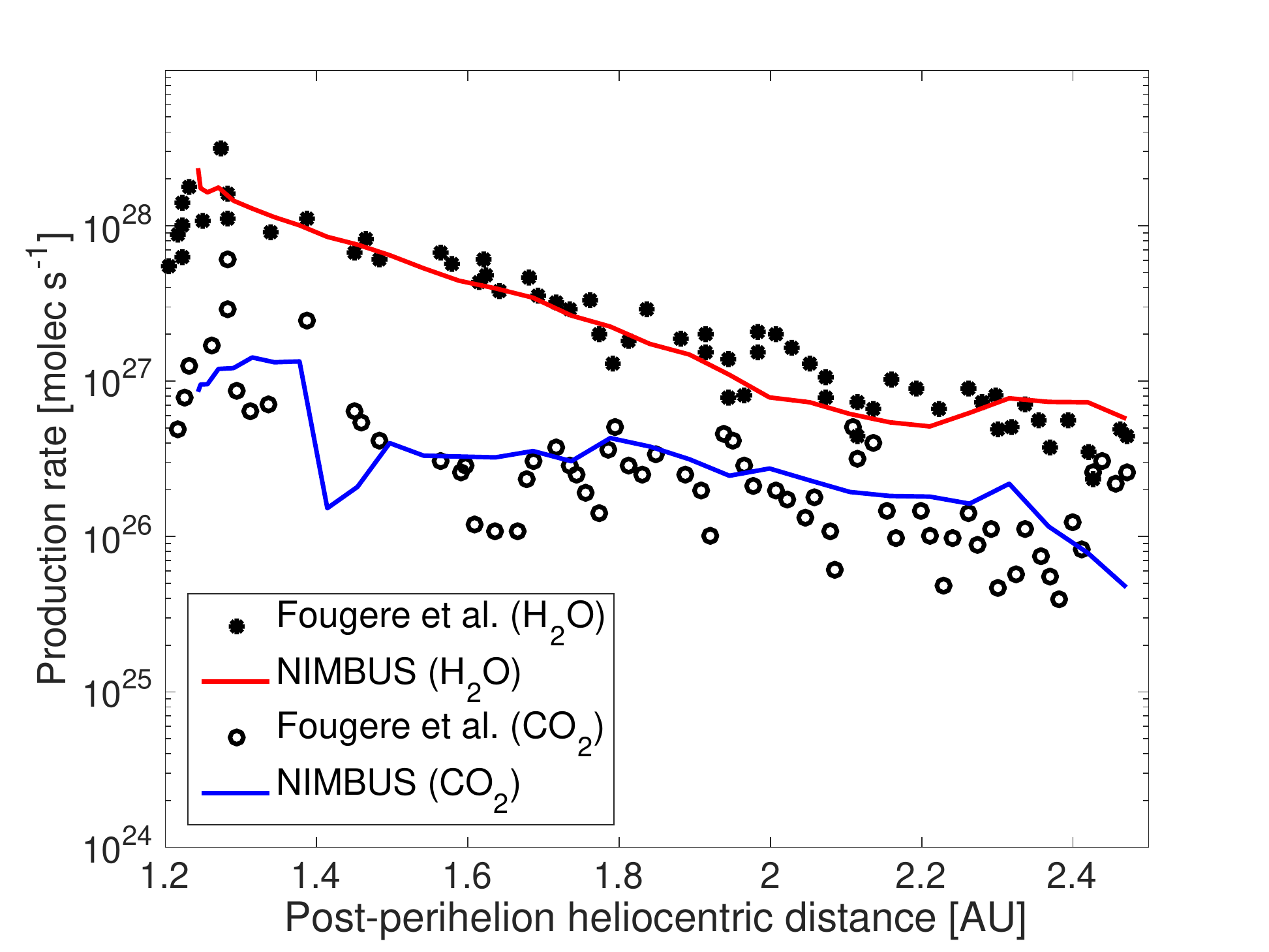}}
     \caption{\textsc{nimbus} model of the post--perihelion $\mathrm{H_2O}$ and $\mathrm{CO_2}$ production rates. The northern hemisphere 
(covered by a freshly deposited $0.87\,\mathrm{m}$ layer of $\mathrm{H_2O}$--rich material) has $\{L,\,r_{\rm p}\}=\{10,\,1\}\,\mathrm{cm}$, $\xi=1$, and $\mu_{\rm N}=2$. 
The northern $\mathrm{CO_2}$,  here located 4--$5\,\mathrm{m}$ underground has $\{L,\,r_{\rm p}\}=\{100,\,10\}\,\mathrm{\mu m}$, $\xi=1$, and abundance 11 per cent relative to water. 
The southern hemisphere has $\{L,\,r_{\rm p}\}=\{100,\,10\}\,\mathrm{\mu m}$ and $\xi=1$ at $r_{\rm h}<1.4\,\mathrm{au}$, but at $r_{\rm h}>1.4\,\mathrm{au}$ the 
$\mathrm{CO_2}$ (at 32 per cent abundance) moves into a region with $\{L,\,r_{\rm p}\}=\{10,\,1\}\,\mathrm{\mu m}$ and $\xi=5$ (hence the sudden drop and slow recovery 
to higher production rates). In the south, the water with $\mu_{\rm S}=1$ still diffuses through a $\{L,\,r_{\rm p}\}=\{100,\,10\}\,\mathrm{\mu m}$ and $\xi=1$ region near the surface). }
     \label{fig_postper_H2O_CO2}
\end{figure}

We consider this solution a reasonable reproduction of the $\mathrm{CO_2}$ production rate post--perihelion. For the southern hemisphere 
we envision a (quasi--primordial) material with $\mu_{\rm S}=1$, a top layer with $\{L,\,r_{\rm p}\}=\{100,\,10\}\,\mathrm{\mu m}$ (applied for the 
water production), and a moving $\mathrm{CO_2}$ sublimation front that transits from a region with $\{L,\,r_{\rm p}\}=\{100,\,10\}\,\mathrm{\mu m}$ and $\xi=1$ 
to one with  $\{L,\,r_{\rm p}\}=\{10,\,1\}\,\mathrm{\mu m}$ and $\xi=5$ around $r_{\rm h}=1.4\,\mathrm{au}$. For the northern hemisphere we envision the 
near--perihelion addition of a $0.87\,\mathrm{m}$ airfall layer that has $\{L,\,r_{\rm p}\}=\{10,\,1\}\,\mathrm{cm}$. For $\mathrm{CO_2}$ we applied the same 
diffusivity as for the southern hemisphere for technical reasons, but the contribution to the total $\mathrm{CO_2}$ production from the north is small ($\sim 5$ per cent) and the 
exact diffusivity value matters little. Figure~\ref{fig_postper_H2O_CO2} shows the post--perihelion $\mathrm{H_2O}$ and 
$\mathrm{CO_2}$ production rate obtained simultaneously by applying these conditions. Best--fit parameters are summarised in Table~\ref{tab4}.

\begin{figure}
\scalebox{0.45}{\includegraphics{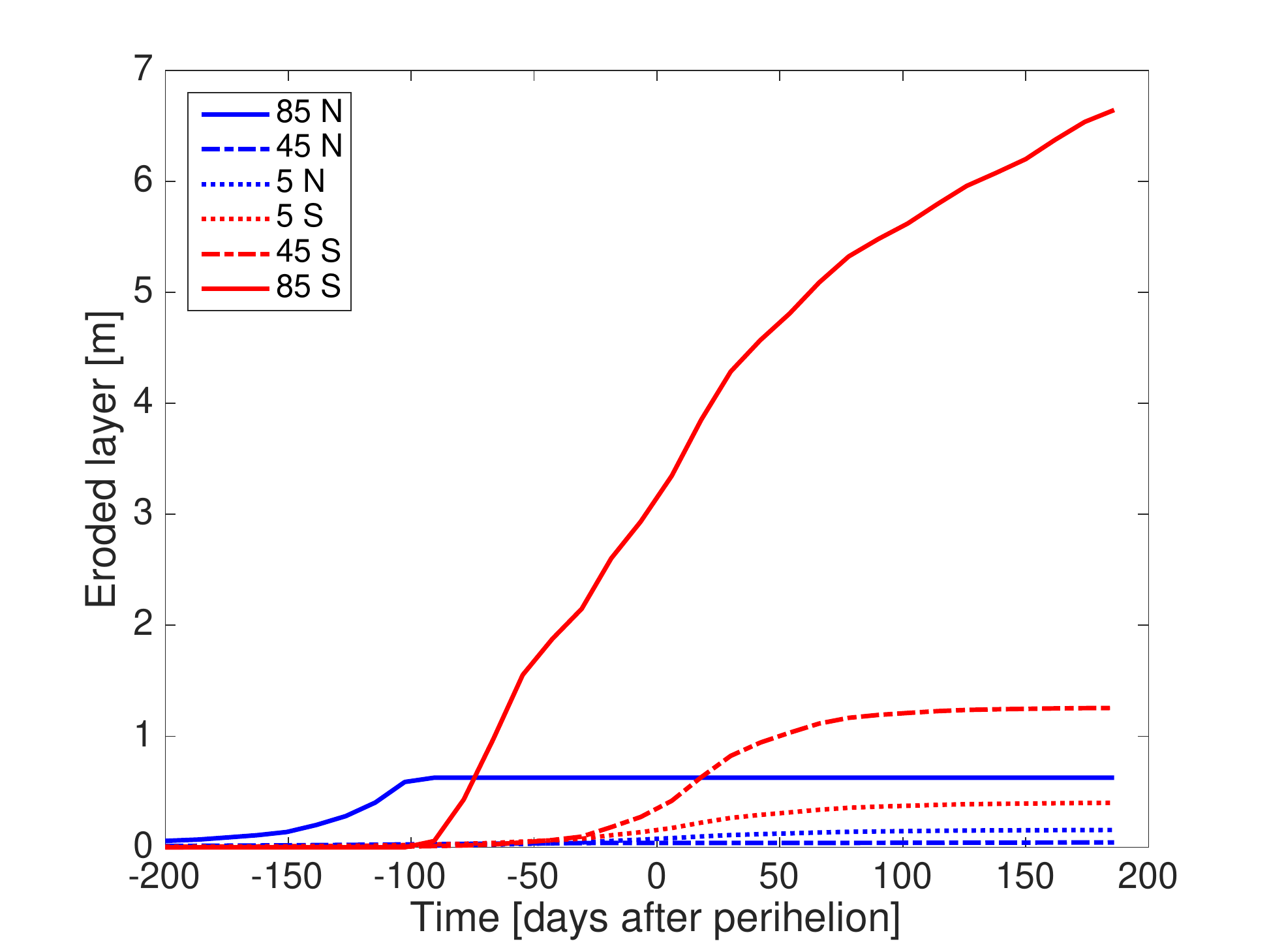}}
     \caption{The accumulated level of surface erosion resulting from the enforced dust production described in Section~\ref{sec_rates_erosion}, as function of time near perihelion, 
for selected latitudes on the northern and southern hemispheres. For reference, $\{0,\,50,\,100,\,150,\,200\}\,\mathrm{days}$ from perihelion corresponds to heliocentric 
distances of $\{1.24,\,1.38,\,1.71,\,2.09,\,2.47\}\,\mathrm{au}$.}
     \label{fig_erosion}
\end{figure}

\begin{table}
\begin{center}
\begin{tabular}{||l|l|l|l|l||}
\hline
\hline
Quantity & Pre (N) & Pre (S) & Post (N) & Post (S)\\
\hline
$\mu$ & 2 &   1 & 2 & 1\\
$\mathrm{CO_2}$ & 11 per cent & 32 per cent & 11 per cent & 32 per cent\\
abund. & & & & \\
$L$ & $100\,\mathrm{\mu m}$ & $100\,\mathrm{\mu m}$ & $10\,\mathrm{cm\,(H_2O)}$ & $100\,\mathrm{\mu m\,(H_2O)}$\\
 &  &  & $100\,\mathrm{\mu m\,(CO_2)}$ & $10\,\mathrm{\mu m\,(CO_2)}$\\
$r_{\rm p}$ & $10\,\mathrm{\mu m}$ & $10\,\mathrm{\mu m}$ & $1\,\mathrm{cm\,(H_2O)}$ & $10\,\mathrm{\mu m}$ ($\mathrm{H_2O}$)\\
 & & & $10\,\mathrm{\mu m\,(CO_2)}$  & $1\,\mathrm{\mu m}$ ($\mathrm{CO_2}$)\\
$\xi$ & 1 & 1 & 1 & 1  ($\mathrm{H_2O}$)\\
 & & & & 5 ($\mathrm{CO_2}$)\\
$\mathrm{CO_2}$ & $3.8\,\mathrm{m}$ & $1.9\,\mathrm{m}$ & -- & --\\
depth & & &\\
\hline 
\hline
\end{tabular}
\caption{Best--fit parameters obtained for Comet 67P/Churyumov--Gerasimenko, before (pre) and after (post) perihelion on the 
northern (N) and southern (S) hemispheres.}
\label{tab4}
\end{center}
\end{table}

\subsection{Erosion, front depths, and temperatures} \label{sec_results_fronts}

We here summarise some aspects of the best available pre-- and post--perihelion simulations in terms of erosion, dust mantle thickness, 
peak daily surface temperature, depth and temperature of the $\mathrm{CO_2}$ sublimation front, as functions of latitude and time. 

\begin{figure*}
\centering
\begin{tabular}{cc}
\scalebox{0.45}{\includegraphics{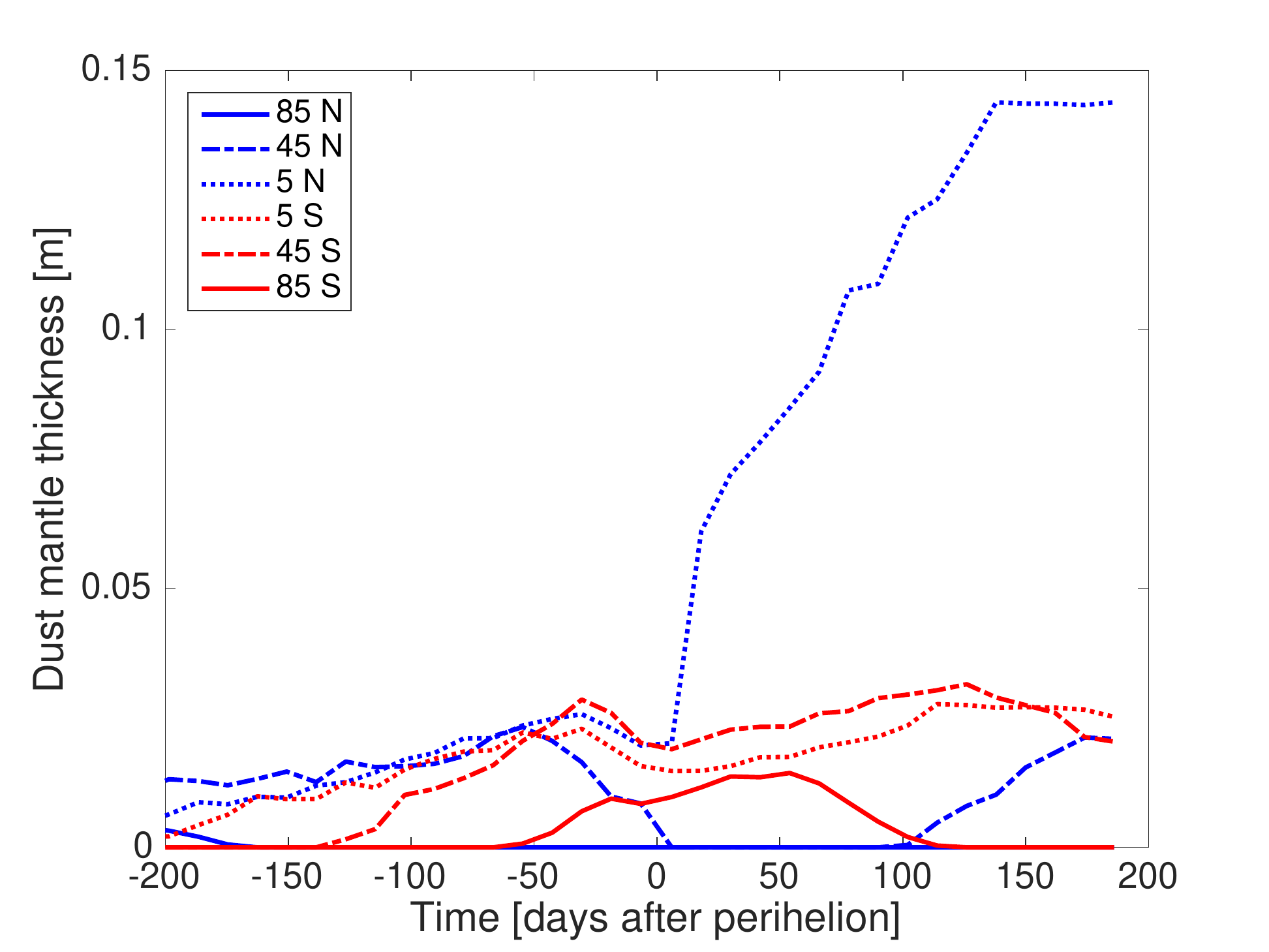}} & \scalebox{0.45}{\includegraphics{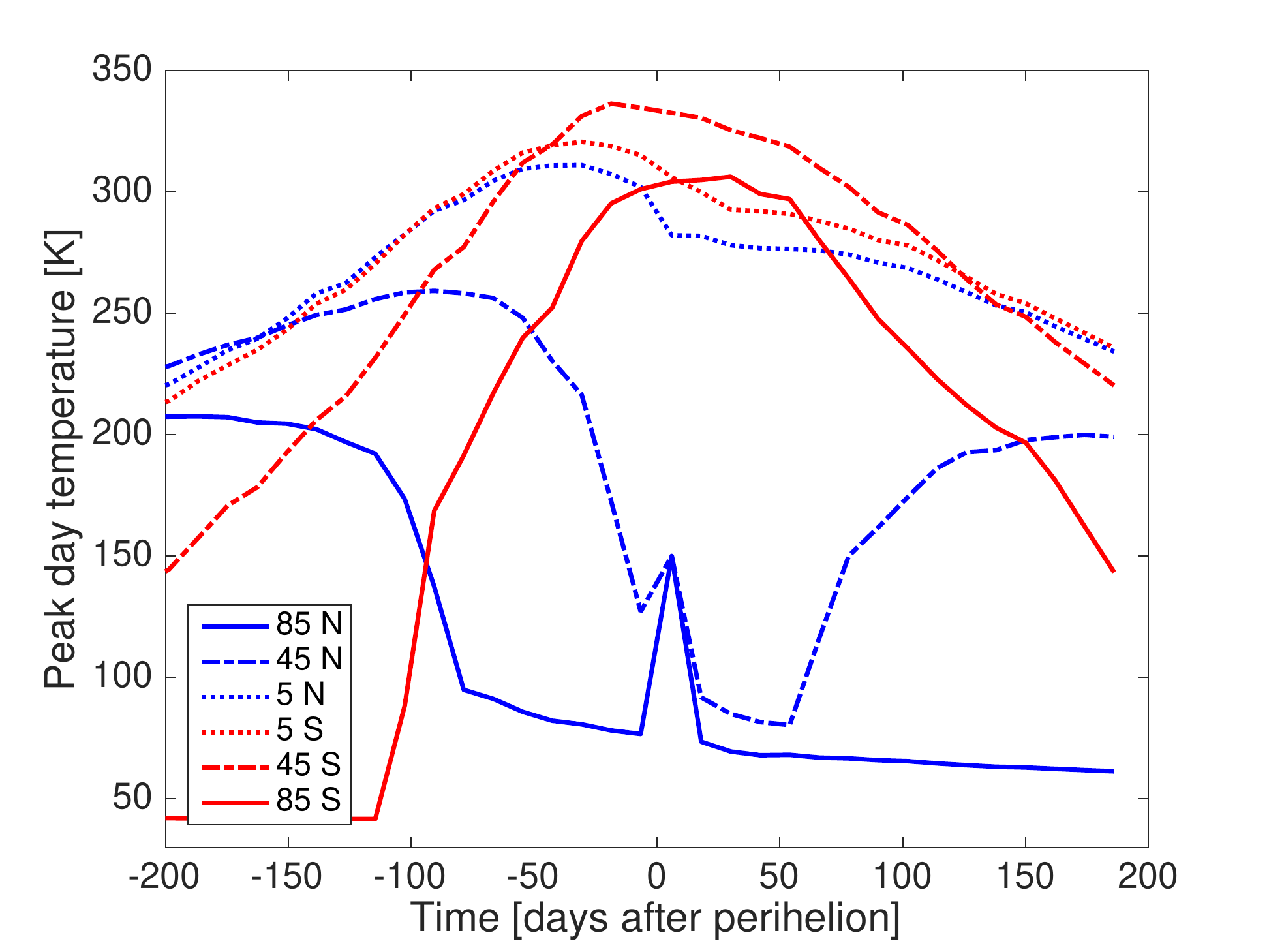}}\\
\scalebox{0.45}{\includegraphics{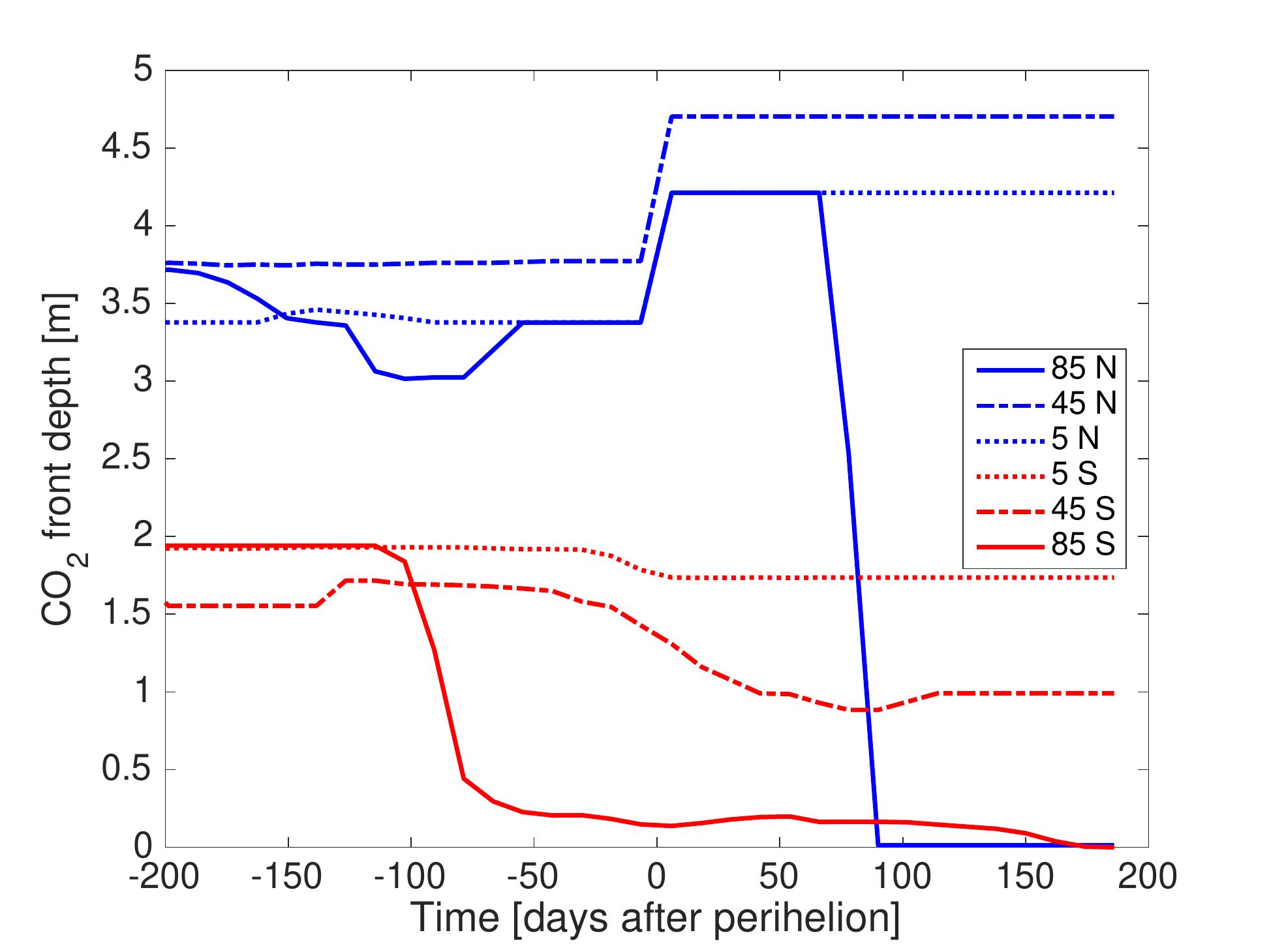}} & \scalebox{0.45}{\includegraphics{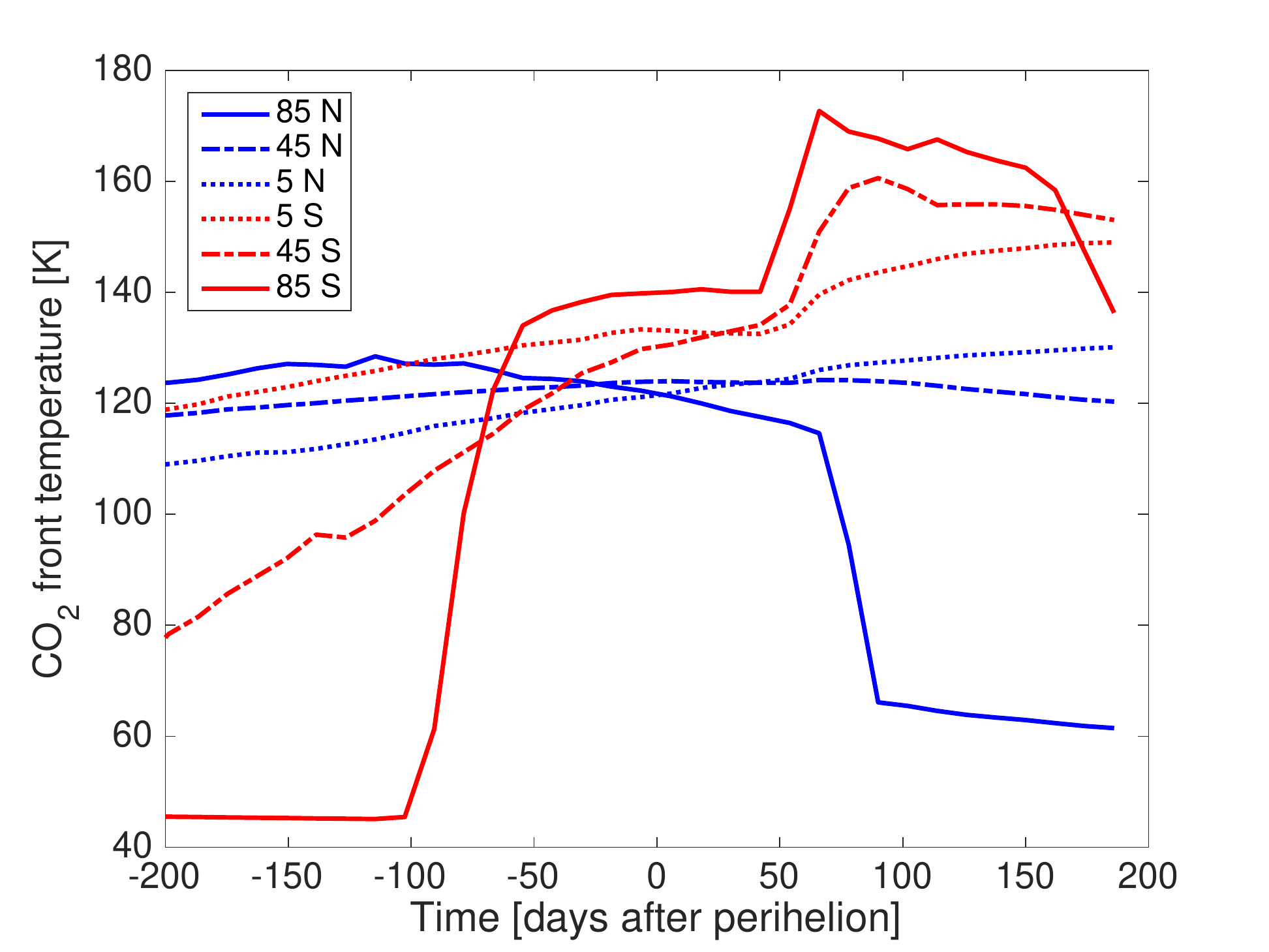}}\\
\end{tabular}
     \caption{\emph{Upper left:} Thickness of the dust mantle, equivalent to the depth of the $\mathrm{H_2O}$ sublimation front. \emph{Upper right:} The daily peak 
surface temperature. \emph{Lower left:} The depth of the $\mathrm{CO_2}$ sublimation front. \emph{Lower right:} The temperature at the $\mathrm{CO_2}$ sublimation front. 
All panels show the quantities in question as functions of time near perihelion, for selected latitudes.}
     \label{fig_fronts}
\end{figure*}

Pre--perihelion, the southern hemisphere provides 93.4 per cent of the $\mathrm{H_2O}$ and 90.4 per cent of the $\mathrm{CO_2}$. Post--perihelion, 
the southern hemisphere provides 64.0 per cent of the $\mathrm{H_2O}$ and 96.8 per cent of the $\mathrm{CO_2}$.

Figure~\ref{fig_erosion} shows the amount of erosion taking place up to, and beyond, the perihelion passage. We emphasise that this erosion 
results from an enforced dust production rate used as input to \textsc{nimbus} (see Section~\ref{sec_rates_erosion}). The amount of erosion varies 
substantially with latitude, illustrating the long--term effects of varying illumination conditions at different parts of the nucleus, combined with 
temporal variations in the total dust production rate. The northern hemisphere has an area--averaged erosion of $0.10\,\mathrm{m}$, being as small 
as $0.04\,\mathrm{m}$ at mid--northern latitudes, and peaking with $0.63\,\mathrm{m}$ at the north pole. Between aphelion and $\sim 100$ days pre--perihelion, the 
Sun is circumpolar as seen from latitudes above $\sim 50^{\circ}\,\mathrm{N}$ \citep[see Fig.~2 in ][]{kelleretal15}. The continuous heating at such latitudes causes the larger degree 
of erosion seen in the far north. The modest level of erosion in the north ($<1\,\mathrm{m}$) is consistent with the lack of observed wide--spread erosion 
in OSIRIS images (being under the resolution limit), except for small isolated patches where 
erosion rose above detection level and typically reached $\sim 1\,\mathrm{m}$ \citep{huetal17}. Taken at face value, these numbers mean that the airfall 
layer thickness increases every apparition. The southern hemisphere has an area--averaged 
erosion of $1.2\,\mathrm{m}$, peaking at the south pole with $6.6\,\mathrm{m}$. This is more substantial, but still below the estimated maximum 
possible level of erosion, that is $\sim 10\,\mathrm{m}$ according to \citet{kelleretal15}.

We now turn to the output features of the models, which result from combining the enforced dust mantle erosion rate with the usage of the 
best--fit parameters (found in Secs.~\ref{sec_results_preper}--\ref{sec_results_postper}), applied to the energy and mass conservation equations of \textsc{nimbus}, 
selected for their capability of reproducing the observed $\mathrm{H_2O}$ and $\mathrm{CO_2}$ production rates. Figure~\ref{fig_fronts} (upper left) shows the 
thickness of the dust mantle. The southern hemisphere, which is facing the Sun near perihelion, 
is covered by a mantle that typically is 1--$2\,\mathrm{cm}$ thick. The presence of water ice to within millimetres or centimetres of the 
surface is supported by the switch--off of jets 1--$2\,\mathrm{h}$ after rotating into darkness \citep{shietal16}. Note that such a dust mantle is 
maintained despite the $1.2\,\mathrm{m}$ average erosion mentioned before: water ice cannot, and will not, remain at the surface under these conditions. 
If the erosion makes the mantle too thin, the nucleus (through the energy and mass conservation equations solved by \textsc{nimbus}) will adjust its water 
production rate until the $\mathrm{H_2O}$ sublimation front has withdrawn to the depth where the energy consumption due to net sublimation balances the 
energy supply from above, and losses to the interior. Similar dust mantle thicknesses (i.~e., $\stackrel{<}{_{\sim}} 2\,\mathrm{cm}$) are seen on the northern hemisphere pre--perihelion. 
At perihelion, a $0.87\,\mathrm{m}$ thick airfall layer is added to the northern hemisphere. 
Most of the northern hemisphere has polar night, thus the water ice is inactive. As a result, water ice in the airfall material remains on the very surface 
and there is no dust mantle. However, near the equator the airfall material is illuminated and the water ice is sublimating. The reason for this rapid thickening at 
latitude $5^{\circ}\,\mathrm{N}$ can be understood as follows. The airfall material has channels in the centimetre--decimetre class, thus the diffusivity is very large. Therefore, water vapour can 
enter the coma rather effortlessly, which means that the water sublimation front moves rapidly. Consequently, the dust mantle thickness grows quickly.

The daily peak surface temperature is seen in the top right panel of Fig.~\ref{fig_fronts}. The dust mantle in the south reaches temperatures 
that are substantially higher ($300$--$340\,\mathrm{K}$) than the characteristic temperature of water sublimation ($\sim 200\,\mathrm{K}$). 
Such dust mantle temperatures are typical for comets near perihelion according to \emph{in situ} measurements \citep[e.~g.,][]{soderblometal02,groussinetal07}. Because of polar--night 
conditions on the northern hemisphere, the near--perihelion temperatures are below $100\,\mathrm{K}$ at the north pole. Note the near--perihelion 
temperature spike at northern latitudes: that marks the deposition of $150\,\mathrm{K}$ airfall originating from the southern hemisphere. 

The depth of the $\mathrm{CO_2}$ sublimation front is seen in the lower left panel of Fig.~\ref{fig_fronts}. The initial depth on the 
northern hemisphere at aphelion was $3.77\,\mathrm{m}$. At the north pole, the $\mathrm{CO_2}$ is brought to within $\sim 3\,\mathrm{m}$ 
of the surface, primarily because of a rather substantial water--driven erosion. The effect is seen at near--equatorial regions in the north as well, but 
less strong. The sudden jump in depth of northern--hemisphere $\mathrm{CO_2}$--front depths to 4--$5\,\mathrm{m}$ at perihelion is because 
of the addition of the airfall layer. The north pole has a peculiar behaviour post--perihelion: the $\mathrm{CO_2}$ sublimation front rapidly reaches 
the surface. This is because vapour, originating from the original front, diffuses upward to the surface where it condenses because of the 
extremely low post--perihelion temperature (see the upper right panel of Fig.~\ref{fig_fronts}). This $\mathrm{CO_2}$ frost formation is 
prevented at, e.~g., $45^{\circ}\,\mathrm{N}$, where the surface temperature is warmer.

On the southern hemisphere, the $\mathrm{CO_2}$ sublimation front starts at a depth of $1.94\,\mathrm{m}$ at aphelion. There is a net 
reduction of this depth that is rather modest north of $45^{\circ}\,\mathrm{S}$, because of water--driven erosion. However, at the south 
pole the erosion is so strong that the $\mathrm{CO_2}$ is brought to within decimetres of the surface near perihelion. Note that erosion 
removes several metres worth of material (Fig.~\ref{fig_erosion}), substantially more than the original depth of the $\mathrm{CO_2}$ sublimation 
front. The fact that the $\mathrm{CO_2}$ ice does not become exposed means that it finds a balance, a few decimetres below the surface, where the 
propagation speed of the $\mathrm{CO_2}$ sublimation front matches the erosion speed of the mantle. VIRTIS observed exposed $\mathrm{CO_2}$ ice in 
an $80\times 60\,\mathrm{m}$ patch in the Anhur region in March 2015 \citep{filacchioneetal16}. That is somewhat early ($\sim 145$ days pre--perihelion) 
and too far from the pole ($\sim 55^{\circ}\,\mathrm{S}$) to be readily explained by the current simulations. However, the retrieved abundance 
\citep[$<0.1$ per cent $\,\mathrm{CO_2}$ ice,][]{filacchioneetal16} is too low to be consistent with a locally and temporarily exposed $\mathrm{CO_2}$ sublimation 
front. It is therefore plausible that the $\mathrm{CO_2}$ ice observed by VIRTIS represents frost condensed at the surface, which originates from the 
actual $\mathrm{CO_2}$ sublimation front located at larger depths (located $\sim 1.5\,\mathrm{m}$ below the surface according to the current simulations). 

The lower right panel of Fig.~\ref{fig_fronts} shows the temperature at the $\mathrm{CO_2}$ sublimation front. Pre--perihelion, the latitudes that 
have actively sublimating $\mathrm{CO_2}$ fronts typically have $110 \stackrel{<}{_{\sim}}T\stackrel{<}{_{\sim}} 130\,\mathrm{K}$. This is rather 
high compared to the `sublimation temperature of $\mathrm{CO_2}$ ice', which is frequently assigned a value near $80\,\mathrm{K}$ in the 
comet literature \citep[e.~g.,][]{prialniketal04,filacchioneetal16,davidssonetal16,gascetal17,weissmanetal20}. However, such low values refer to 
the onset of sublimation of $\mathrm{CO_2}$ ice exposed to vacuum \citep[classically, the temperature at which the saturation density equals 
a specific low value, e.~g., $10^{13}\,\mathrm{cm^{-3}}$ reached at $77\,\mathrm{K}$;][]{yamamoto85}. When the $\mathrm{CO_2}$ sublimation front is located 
under ground, local temperature equilibrium means that net sublimation 
consumes the available heat flux provided by solid--state conduction. In order to have net sublimation, pre--existing vapour 
has to diffuse away from the sublimation front, to give room to new vapour. Efficient diffusion requires that sufficiently strong 
temperature and saturation pressure gradients are set up around the front. When the diffusivity is low, the front temperature may 
have to climb high in order to reach the necessary diffusion velocity. When $\{L,\,r_{\rm p}\}=\{100,\,10\}\,\mathrm{\mu m}$, $\xi=1$, and the $\mathrm{CO_2}$ sublimation 
front is located meters under the surface, the effective $\mathrm{CO_2}$ sublimation temperature is pushed into the 
$110 \stackrel{<}{_{\sim}}T\stackrel{<}{_{\sim}} 130\,\mathrm{K}$ range. Similar temperatures for sub--surface sublimation of $\mathrm{CO_2}$ were obtained by \citet{skorovandblum12}. 
The importance of diffusivity is clearly seen on the southern hemisphere 
in Fig.~\ref{fig_fronts} (lower right) 50 days post--perihelion ($r_{\rm h}\geq 1.4\,\mathrm{au}$). At that point, the diffusivity is reduced strongly by 
setting $\{L,\,r_{\rm p}\}=\{10,\,1\}\,\mathrm{\mu m}$ and $\xi=5$. Additionally, the front is located very closely to the warm surface. In such conditions, the 
temperature at the $\mathrm{CO_2}$ sublimation front is elevated to $165\,\mathrm{K}$ in order to allow for the same mass flux and net energy consumption as before.

\begin{figure}
\scalebox{0.45}{\includegraphics{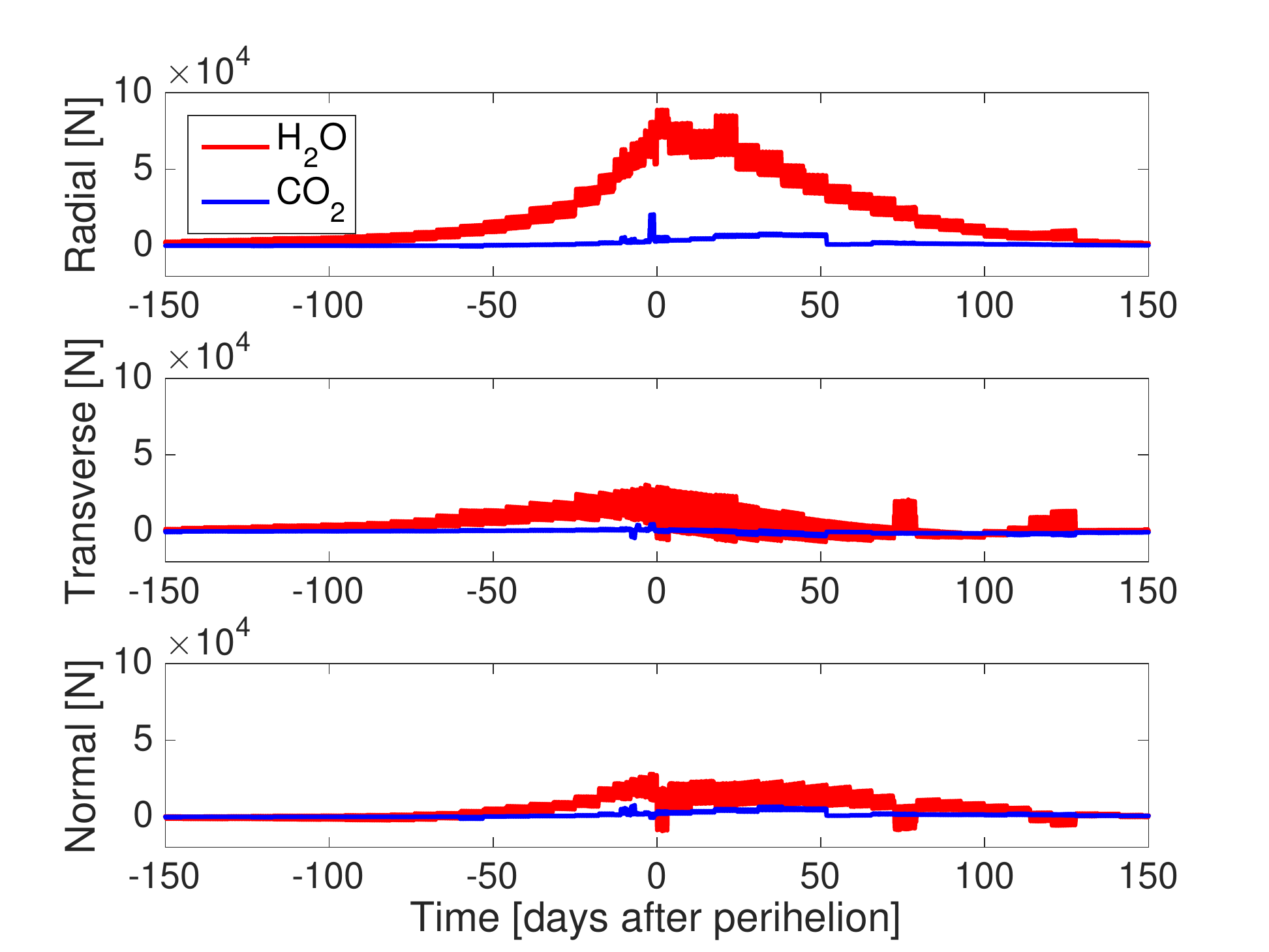}}
     \caption{The radial, transverse, and normal force components due to $\mathrm{H_2O}$ and $\mathrm{CO_2}$ outgassing, as functions 
of time during 300 days surrounding perihelion (using $\eta=0.31$). Note that force ranges of the three panels are identical, so that the 
magnitudes of $\{F_{\rm r},\,F_{\rm t},\,F_{\rm n}\}$ can be readily compared. The width of the curves shows the level of diurnal variation of the force magnitude, 
primarily due to changing cross--section of the irregular nucleus during rotation. The `block--like' structure of the curves is due to the fact 
that \textsc{nimbus} output is provided every 12$^{\rm th}$ rotation, and then applied for the following eleven revolutions. However, the calculation 
properly accounts for changes to the spin period also within such blocks. Small jumps in the production rate take place under certain conditions 
when $\mathrm{H_2O}$ frost formed in the dust mantle (or $\mathrm{CO_2}$ frost in the $\mathrm{H_2O}+$dust layer) is released vigorously at sunrise.}
     \label{fig_RTN}
\end{figure}

\subsection{Forces acting on the nucleus} \label{sec_results_forces}

The modelling in Secs.~\ref{sec_results_preper}--\ref{sec_results_postper} provides local outgassing rates $Z_{i,j}$ and surface 
temperatures $T_i$ necessary to evaluate the resultant non--gravitational force acting on the model nucleus according to equation~\ref{eq:00a}. We remind 
that the force evaluation applies the mapping from the spherical to the irregular nucleus shape model, including the rudimentary treatment 
of temporary daytime shadowing, as described in Section~\ref{sec_results_preper}. Also, note that the force evaluation accounts for the changes to 
the spin period throughout the perihelion passage, so that the nucleus has the appropriate rotational phase at any given moment. 

We decompose the force vector into its radial, transverse, and normal components $\mathbfit{F}=\{F_{\rm r},\,F_{\rm t},\,F_{\rm n}\}$ (i.~e., $F_{\rm r}$ is 
along the unit radius vector $\hat{r}$, $F_{\rm n}$ is along the angular momentum unit vector of the orbit $\hat{h}$, and $F_{\rm t}$ is along the 
vector $\hat{h}\times\hat{r}$). The force components for both $\mathrm{H_2O}$ and $\mathrm{CO_2}$ are plotted in Fig.~\ref{fig_RTN}.

\begin{figure}
\scalebox{0.45}{\includegraphics{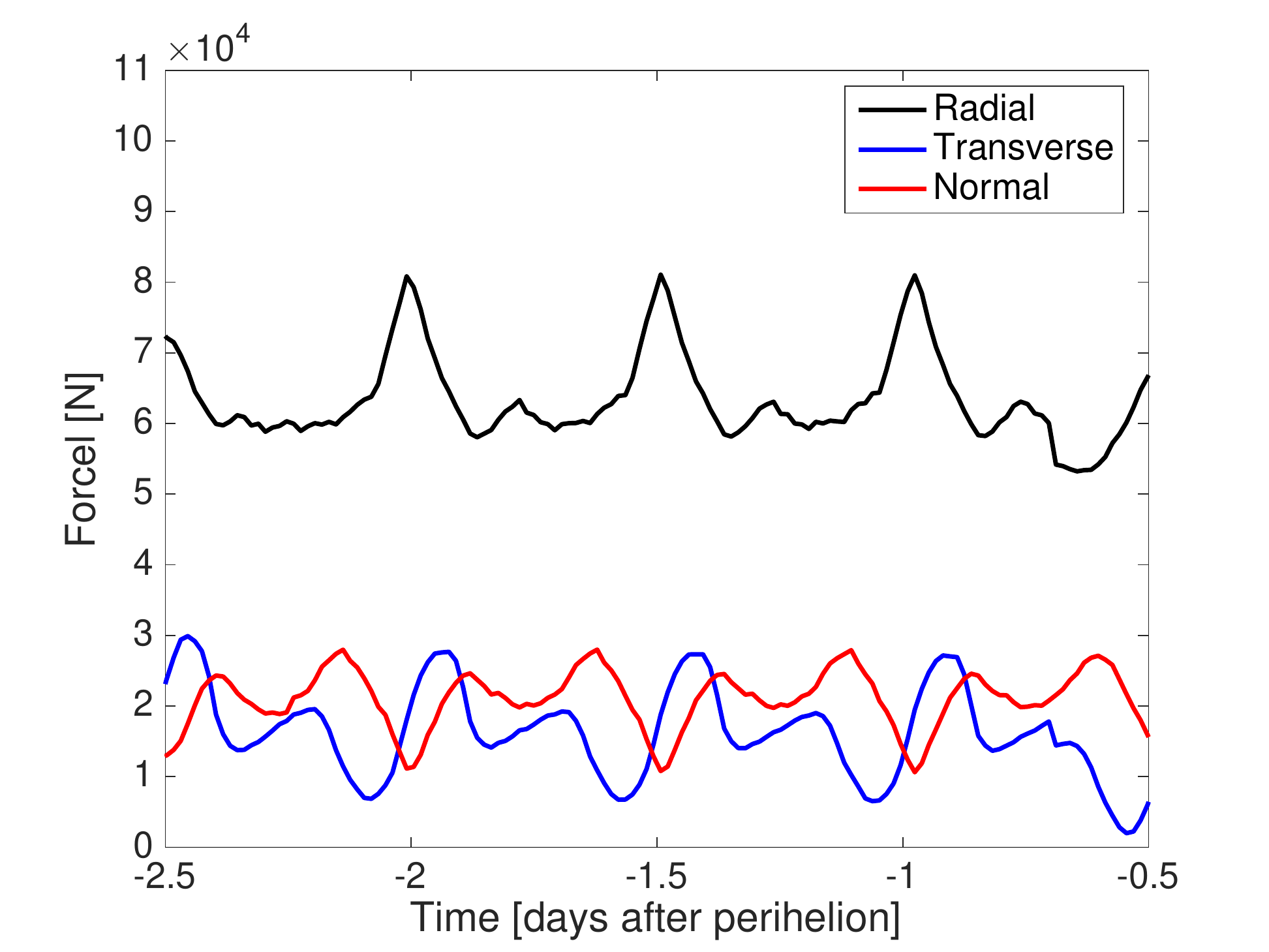}}
     \caption{Close--up of the radial, transverse, and normal force components due to $\mathrm{H_2O}$ outgassing during a few days 
prior to perihelion ($\eta=0.31$).}
     \label{fig_RTN_zoom}
\end{figure}

For water, the radial component is strongest, as expected (the highest comet outgassing rate takes place near the sub--solar point, causing a reaction 
force in the opposite direction, that roughly aligns with the radius vector). Yet, the transverse and normal components reach $\sim 1/3$ of the radial component 
magnitude near perihelion. The irregular nucleus shape, in combination with a latitude--dependent activity level, and thermal lag effects caused by the finite conduction timescale of heat to 
sub--surface ice deposits, evidently causes some significant deviations from perfectly radial alignment of the force vector. All components have a pronounced 
perihelion--asymmetry, which is a consequence of a stronger outgassing post--perihelion (see equation~\ref{eq:13}). 
The width of the curves shows the level of diurnal variations of the force components. Figure~\ref{fig_RTN_zoom} shows this more clearly by exemplifying 
a close--up of the force components due to water outgassing during a couple of days just prior to the perihelion passage, to illustrate their temporal behaviour on 
the timescale of a nucleus rotation. We emphasise that these variations are primarily caused by changes to the illuminated nucleus cross--section, and 
secondarily, due to associated differences in the topography that is being illuminated.

Figure~\ref{fig_RTN} also shows the radial, transverse, and normal force components caused by $\mathrm{CO_2}$ outgassing. They are 
much smaller than those caused by water, for two reasons. Firstly, the $\mathrm{CO_2}$ outgassing rate is an order of magnitude smaller 
than that of water (see Figs.~\ref{fig_preper_H2O_CO2} and \ref{fig_postper_H2O_CO2}). Secondly, the $\mathrm{CO_2}$ sublimation front 
is located at a depth far below the diurnal skin depth, so that day/night differences in the outgassing rate becomes very small. Consequently, 
the $\mathrm{CO_2}$ outgassing reaction force has similar strength in all directions within the orbital plane, thus cancellation effects are strong. 

The sum of the $\mathrm{H_2O}$ and $\mathrm{CO_2}$ force components can be used to calculate the non--gravitational changes of the orbit 
caused by the comet outgassing. This allows us to further calibrate our model to ensure it complies with measured data, as well as possible. 
The change in the orbital period is
\begin{equation} \label{eq:22}
  \Delta\mathcal{P}=H\frac{6\upi\sqrt{1-e^2}}{M n^2}\left(\frac{e}{q(1+e)}\int_0^{\mathcal{P}} F_{\rm r}\sin\nu\,dt+\int_0^{\mathcal{P}}\frac{F_{\rm t}}{r_{\rm h}}\,dt\right),
\end{equation}
the change in the longitude of perihelion is

\begin{equation} \label{eq:23}
\begin{array}{c}
\displaystyle \Delta\varpi =-\Delta\Omega\cos i+\frac{\sqrt{q(1+e)}H}{Mk_{\rm G}e}\left(\frac{1}{e}\int_0^{\mathcal{P}}F_{\rm r}\left(1-\frac{q(1+e)}{r_{\rm h}}\right)\,dt+\right.\\
\\
\displaystyle \left.\int_0^{\mathcal{P}}F_{\rm t}\left(1+\frac{r_{\rm h}}{q(1+e)}\right)\sin\nu\,dt\right),
\end{array}
\end{equation}
and the change in the longitude of the ascending node is
\begin{equation} \label{eq:24}
\Delta\Omega = \frac{H}{Mk_{\rm G}\sqrt{q(1+e)}\sin i}\int_0^{\mathcal{P}}F_{\rm n}r_{\rm h}\sin(\omega+\nu)\,dt,
\end{equation}
see, e.~g., \citet{sekanina93}. In these equations, $M=9.982\cdot 10^{12}\,\mathrm{kg}$ is the 67P/C--G nucleus mass \citep{patzoldetal16}, $e$ is the eccentricity, 
$n\,\mathrm{(day^{-1})}$ is the mean motion, $\nu$ is the true anomaly, $q\,\mathrm{(au)}$ is the perihelion distance, 
$\mathcal{P}=(q/(1-e))^{3/2}\,\mathrm{(yr)}$ is the orbital period (recalculated to days), $i$ is the inclination, $\omega$ is the argument of 
perihelion, $k_{\rm G}$ is the Gaussian gravitational constant, and $H=0.049900175\,\mathrm{au\,s^2\,m^{-1}\,day^{-2}}$ allows for the usage of SI units 
for $M$ and $\mathbfit{F}$, while remaining quantities have Gaussian units. According to several pre--\emph{Rosetta} orbit determinations discussed by \citet{davidssongutierrez05} the 
empirical values for 67P/C--G are $\Delta\mathcal{P}=20\pm 4\,\mathrm{min}$, $\Delta\varpi=-1.4\pm 0.6\arcsec$, and $\Delta\Omega=1.0\pm 1.0\arcsec$. As indicated by 
the error bars, the $\Delta\mathcal{P}$--value is the most reliable, while the $\Delta\Omega$--value is the least reliable. 

We find, that reproduction of the empirical $\Delta\mathcal{P}$ requires that our model force (equation~\ref{eq:00a}) is evaluated for $\eta=0.31\pm 0.06$. This 
is somewhat below the typically considered interval ($0.4 \stackrel{<}{_{\sim}}\eta\stackrel{<}{_{\sim}} 1$, see Section~\ref{sec_model}). We discuss possible reasons and implications of this in Section~\ref{sec_discussion}. 
About 36 per cent of the net change in $\Delta\mathcal{P}$ (or $7.2\,\mathrm{min}$) is established pre--perihelion, whereas the remaining 64 per cent (or $12.8\,\mathrm{min}$) of 
the change happens post--perihelion. 
When applying $\eta=0.31$ we also find $\Delta\varpi=-1.4\arcsec$ and $\Delta\Omega=0.58\arcsec$. Both parameters are consistent with the empirical counterparts. We find it reassuring that our force model simultaneously reproduces all three of $\{\Delta\mathcal{P},\,\Delta\varpi,\,\Delta\Omega\}$ for 
a low but reasonable momentum transfer coefficient. 

If the $\mathrm{CO_2}$ contribution is removed when evaluating equation~(\ref{eq:22}) we find that $\Delta\mathcal{P}$ is reduced by 
merely 0.3 per cent  (much below the empirical $\pm 20$ per cent uncertainty). It therefore seems like $\mathrm{CO_2}$ has a completely 
negligible non--gravitational effect on the orbit, at least for Comet 67P/C--G. 

If comet outgassing is symmetric about perihelion, the contribution from the radial force component in equation~(\ref{eq:22}) becomes zero when 
integrated over time, so that a non--zero $\Delta\mathcal{P}$ is entirely caused by the transverse component \citep[see e.~g., ][]{rickmanetal91}. Because 
Comet 67P/C--G has asymmetric outgassing, it is therefore interesting to understand the relative importance of the radial and 
transverse force components. Evaluating equation~(\ref{eq:22}) without the transverse force yields $\Delta\mathcal{P}=6.4\,\mathrm{min}$, 
showing that the radial component is responsible for $\sim 32$ per cent of the total change of the orbital period. Pre--perihelion, the radial 
component strives to \emph{reduce} $\mathcal{P}$ by $4.1\,\mathrm{min}$. Post--perihelion, the radial component instead \emph{increases} $\mathcal{P}$ 
by $10.5\,\mathrm{min}$, resulting in the net change $\Delta\mathcal{P}=6.4\,\mathrm{min}$ mentioned above. This directly shows that 
the asymmetric outgassing removes the complete cancellation effect. The transverse component here systematically aims at increasing $\Delta\mathcal{P}$. 
Interestingly, the pre--perihelion transverse contribution to the change ($+11.2\,\mathrm{min}$) is significantly larger than the post--perihelion one ($+2.4\,\mathrm{min}$). 
A glance at the middle panel of Fig.~\ref{fig_RTN}, and at the second integral in equation~(\ref{eq:22}), shows the reason for this behaviour. Pre--perihelion, $F_{\rm t}$ is larger 
and systematically positive, while post--perihelion,  $F_{\rm t}$ is smaller and briefly goes negative during nucleus rotation. 

\begin{figure}
\scalebox{0.45}{\includegraphics{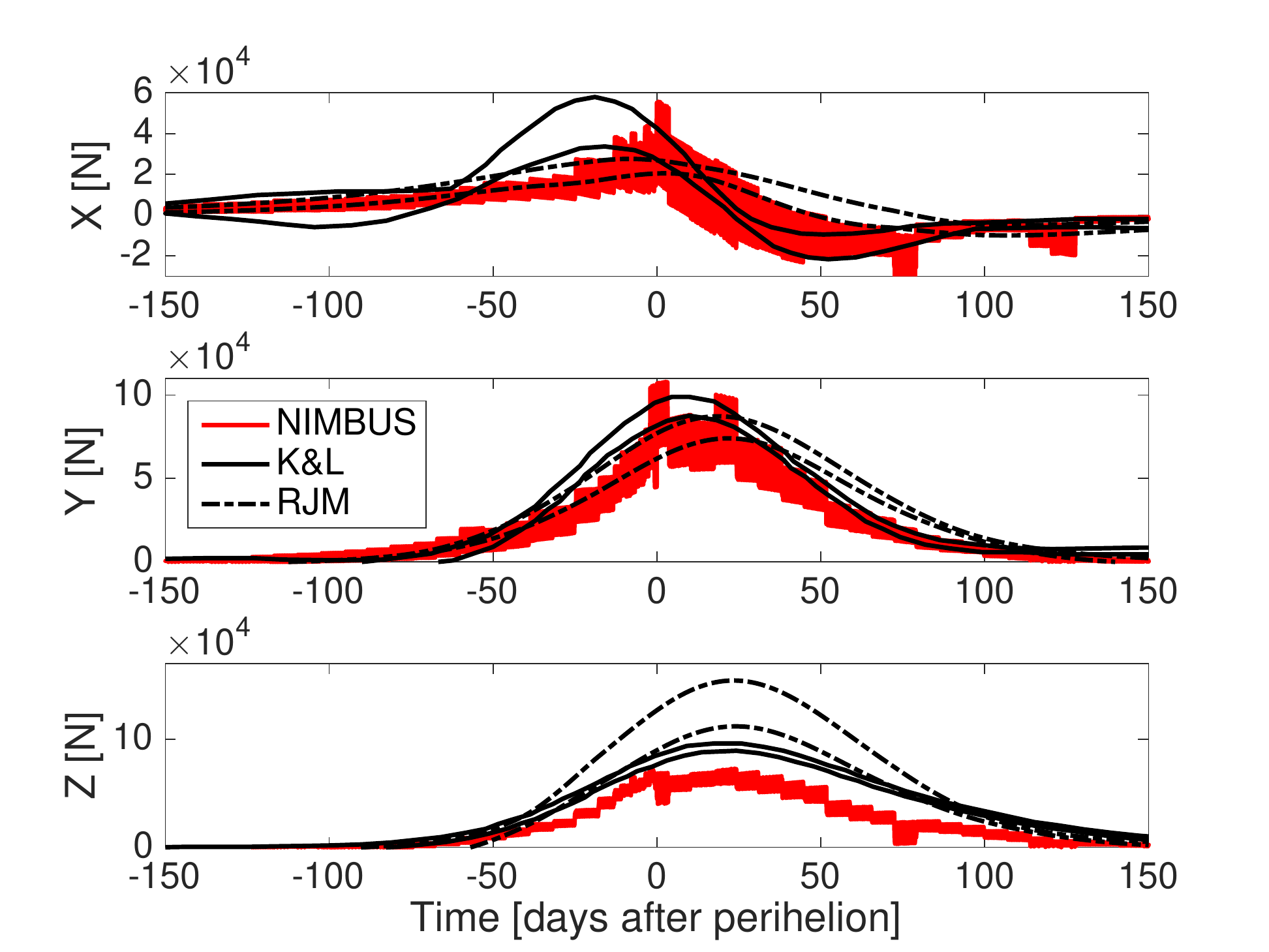}}
     \caption{The total ($\mathrm{H_2O}$ and $\mathrm{CO_2}$) non--gravitational force components in the heliocentric equatorial system according to 
\textsc{nimbus} (applying $\eta=0.39$) compared to the solution by \protect\citet{kramerandlauter19} and the Rotating Jet Model of \protect\citet{farnocchiaetal21}, labelled K\&L and RJM, respectively. 
The K\&L curves reproduces the upper and lower ranges displayed in their Fig.~2, the RJM curves trace the 1--$\sigma$ uncertainty envelope.}
     \label{fig_XYZ_comparison}
\end{figure}

Figure~\ref{fig_XYZ_comparison} shows the \textsc{nimbus} total non--gravitational force due to both $\mathrm{H_2O}$ and $\mathrm{CO_2}$ in the equatorial J2000 frame. 
This forward--modelling `first--principles' force is compared with two different empirical solutions: (1) the force derived by \citet{kramerandlauter19} 
from piece--wise orbital solutions for 67P/C--G provided by the \emph{Rosetta} flight dynamics team at ESOC; (2) the force according to the `Rotating Jet Model' \citep{chesleyyeomans05}, 
used by \citet{farnocchiaetal21} to reconstruct the 67P/C--G trajectory from \emph{Rosetta} tracking data and March 2014 to June 2018 high--precision optical astrometry from the Very Large Telescope, 
Pan--STARRS1, and the Catalina Sky Survey. We refer to those as K\&L and RJM, respectively, in the following. Note that K\&L and RJM provide acceleration, here re--calculated to force by 
multiplying with the nucleus mass $M=9.982\cdot 10^{12}\,\mathrm{kg}$ \citep{patzoldetal16} in order to be directly comparable to \textsc{nimbus} results. We first note that the RJM solution implies $\Delta\mathcal{P}=24.9\,\mathrm{min}$, 
$\Delta\varpi=-1.6\arcsec$, and $\Delta\Omega=2.6\arcsec$. The $\Delta\mathcal{P}$ value is somewhat higher than, but still consistent with, the pre--\emph{Rosetta} estimate $\Delta\mathcal{P}=20\pm 4\,\mathrm{min}$, 
which would suggest a slight increase of our momentum transfer coefficient from $\eta=0.31\pm 0.06$ to $\eta\approx 0.39$ (which would shift the changes of the perihelion and ascending 
node longitudes to $\Delta\varpi=-1.8\arcsec$ and $\Delta\Omega=0.73\arcsec$). The change in the longitude of perihelion is identical to pre--\emph{Rosetta} values ($\Delta\varpi=-1.4\pm 0.6\arcsec$) 
within error bars, while the change in the longitude of the ascending node is larger (pre--\emph{Rosetta} $\Delta\Omega=1.0\pm 1.0\arcsec$). We used $\eta=0.39$ when plotting the 
\textsc{nimbus} force in Fig.~\ref{fig_XYZ_comparison}.

We note that the three solutions all differ with respect to each other. The \textsc{nimbus} $X$-- and $Y$--components are similar to RJM pre--perihelion, 
while their post--perihelion behaviours follows more closely those of K\&L. The \textsc{nimbus} $Z$--component falls somewhat short of both RJM and K\&L. Preliminary tests indicate that 
a latitude--dependent $\eta$--value might be sufficient to nudge the \textsc{nimbus} curves towards either K\&L or RJM. A thorough investigation of the mutual discrepancies between 
\textsc{nimbus}, RJM, and K\&L is out of scope in the current paper. We intend to return to this issue when analysing the effect of \textsc{nimbus}--derived torques on the spin state.

\section{Discussion} \label{sec_discussion}

We start by discussing the outcome of our investigation of extended sources (Section~\ref{sec_rates_water}) and dust erosion (Section~\ref{sec_rates_erosion}). 
According to the ratio between equations~(\ref{eq:01}) and (\ref{eq:13}) shown in the upper panel of Fig.~\ref{fig_prod_erosion}, the nucleus contribution to 
the total water outgassing is $0.92\leq Q_{\rm nuc}/Q_{\rm H_2O}\leq 1$. It means that the extended source of water from ice--rich chunks in the 
coma is at most 8 per cent at perihelion. \citet{biveretal19} used MIRO observations of the inner $20\,\mathrm{km}$ of the coma to search for evidence of 
an extended water source. According to their analysis, an extended source contributes less than 15 per cent of the water near perihelion. We therefore 
think that our analysis in Section~\ref{sec_rates_water} is realistic, and that our decision to compare \textsc{nimbus} water production calculations directly 
with the observed $Q_{\rm H_2O}$ is justified. 

The erosion rate of the nucleus $E$ is expressed in units of $Q_{\rm H_2O}$ by the function $G(t)$. In order to evaluate $G(t)$ we needed 
the total sublimating area $A_{\rm tot}$ of nucleus and chunks, as well as the mass ratio $F$ between escaping refractories and water vapour, 
according to equation~(\ref{eq:19}). Because $A_{\rm tot}$ is directly related to $Q_{\rm nuc}$ (equations~\ref{eq:01}--\ref{eq:02}), its validity follows 
from the previous discussion of extended sources. We find $F=1.05$ with our method, suggesting that the escaping mass of refractories is very 
similar to the escaping mass of water vapour. This can be compared to measured values obtained using different instruments and 
techniques: $F=0.64$ \citep{hansenetal16}, $F=0.72$ \citep{marschalletal20}, $F=1.06$ \citep{lauteretal19}, $F=1.14$ \citep{combietal20}, 
$F=1.63$ \citep{lauteretal20}, and $F=3.4$ \citep{biveretal19}, i.~e, a mean and standard deviation of $\langle F\rangle = 1.4\pm 1$.

We have not considered neither CO nor amorphous water ice in this paper. With CO trapped in amorphous water ice, the crystallisation process could either 
produce or consume energy, depending on the CO abundance \citep[e.~g.,][]{gonzalezetal08}. Because \citet{hoangetal20} demonstrated that crystallisation 
and CO release has a negligible effect on the $\mathrm{H_2O}$ and $\mathrm{CO_2}$ production rates, we do not think our omission of CO and amorphous 
water ice has biased our analysis. 

Next, we discuss the refractories/water--ice mass ratio $\mu$ of the nucleus. The southern hemisphere is essentially airfall--free \citep{elmaarryetal16}, therefore 
its activity near perihelion is representative of the nucleus itself, as opposed to the airfall material covering the northern hemisphere, that already has been 
processed in the coma. We find that $\mu_{\rm S}=1.0$ for the southern hemisphere yields $0.89Q_{\rm H_2O}$ (Table~\ref{tab3}), which formally 
would allow for an extended source of $11$ per cent that provides the remaining vapour. As can be seen from Figs.~\ref{fig_preper_H2O_CO2} and \ref{fig_postper_first} 
(or Fig.~\ref{fig_postper_H2O_CO2}), this provides a satisfactory fit to the data at $r_{\rm h}\stackrel{<}{_{\sim}} 1.8\,\mathrm{au}$, where the southern contribution 
dominates. As our best estimate, we therefore propose that the least processed parts of the nucleus has a refractories/water--ice mass ratio close to unity. 
However, if we had applied an erosion based on another $F$ value in the $\langle F\rangle$ range, we likely would have obtained $0.4\leq\mu\leq 2.4$ for the nucleus interior.

The northern hemisphere is expected to have a higher refractories/water--ice mass ratio than the south, because of volatile loss in the airfall material 
prior to its deposition in the north. \citet{davidssonetal21} used \textsc{nimbus} simulations of coma chunks to demonstrate that a $1\,\mathrm{cm}$ diameter 
chunk is expected to lose $56$ per cent of its water ice, while a $10\,\mathrm{cm}$ diameter chunk loses $\sim 6$ per cent of the ice. If such chunks are representative of 
the airfall material of the north, we would expect $1.1\stackrel{<}{_{\sim}} \mu_{\rm N}\stackrel{<}{_{\sim}} 2.3$. Figure~\ref{fig_preper_H2O} shows that 
$\mu_{\rm N}=2$ provides a good fit to the data, whereas $\mu_{\rm N}=1$ is somewhat on the high side, suggesting that some loss indeed has taken place. 

We now compare our estimate $\mu_{\rm N}=2$ with others in the literature, which also considered the pre--perihelion branch when the northern hemisphere dominated activity.
\citet{huetal17} fitted the water production rate curve with a thermophysical model using $\mu_{\rm N}=99$ under a $0.5\,\mathrm{cm}$ dust mantle, or alternatively, 
by using $\mu_{\rm N}=9$ under a $1\,\mathrm{cm}$ dust mantle. In both cases, the dust mantle was assumed to consist of $1\,\mathrm{mm}$ diameter 
solid particles, i.~e., the mantles were 5--10 grain--layers thick. \citet{blumetal17} fitted the water production rate curve with a thermophysical model using $\mu_{\rm N}=19$ underneath a 
single monolayer consisting of $1\,\mathrm{cm}$ solid grains. These estimates ($9\stackrel{<}{_{\sim}} \mu_{\rm N}\stackrel{<}{_{\sim}}99$) are drastically different 
from ours ($\mu_{\rm N}=2$), and we now discuss the reason for this discrepancy. Both teams used the thermophysical model developed by \citet{gundlachetal11}. This model 
solves the energy conservation equation with terms for heat conduction and energy consumption due to sub--surface sublimation, but does not consider mass conservation. The magnitude of the sub--surface 
sublimation rate is given by the Hertz--Knudsen formula $Z(T)=p_{\rm sat}(T)\sqrt{m_1/2\upi k_{\rm B}T}$ \citep[e.~g.][]{fanaleandsalvail84}, where $p_{\rm sat}(T)$ is 
the saturation pressure, times a constant correction factor $\Psi$ to account for the mantle diffusivity. This approach is roughly equivalent to considering exposed surface ice 
sublimating into vacuum, which covers an area fraction $\Psi$ of any surface element of the nucleus, while an area fraction $1-\Psi$ is covered by inert dust. From the work of 
\citet{kelleretal15} we know that such a model requires $\Psi\approx 0.06$ at perihelion (and even smaller values, $\Psi\approx 0.02$, at larger heliocentric distances). It means that 67P/C--G 
produces $\stackrel{<}{_{\sim}} 6$ per cent of the amount of water that an equal--area ice--only nucleus would do. In the model of \citet{gundlachetal11}, 
the correction factor is evaluated as a function of a single parameter: the ratio between the dust mantle thickness and the diameter of the constituent grains. According to that formula, \citet{huetal17} 
applies either $\Psi=0.42$ or $\Psi=0.59$, while \citet{blumetal17} applies $\Psi=0.88$. The fact that the $\mu_{\rm N}$--value of  \citet{blumetal17} is intermediate 
between the two of \citet{huetal17}, while their $\Psi$--value is not, is a consequence of different thermal inertia applied in the models: 16--$50\,\mathrm{MKS}$ for 
$100\leq T\leq 200\,\mathrm{K}$ in \citet{blumetal17}, but a fixed $30\,\mathrm{MKS}$ in \citet{huetal17}. In all cases, the $\Psi$--values are significantly higher than 
the necessary $0.02\stackrel{<}{_{\sim}}\Psi \stackrel{<}{_{\sim}} 0.06$. This forces \citet{huetal17} and \citet{blumetal17} to introduce very low area fractions of ice underneath the mantles 
(of order 1--10 per cent) that correspond to the $\mu_{\rm N}$--values mentioned above. 

\textsc{nimbus} instead solves the coupled energy and mass conservation equations, which is the traditional and physically correct way of approaching thermophysical 
modelling \citep[e.~g.,][]{espinasseetal91, prialnik92, tancredietal94, oroseietal95, capriaetal96, enzianetal97, desanctisetal99}. Accordingly, the mass flux 
through the mantle sensitively depends on the temperature and vapour pressure gradients throughout the mantle, according to the Clausing formula \citep[equation~46 in][]{davidsson21}. 
This mechanism regulates how much vapour that should be removed from the sublimation front, thereby allowing for net production of more vapour. Such net production 
dictates the amount of energy consumed by sublimation, and therefore has a very strong effect on the overall temperature and gas pressure profiles. This delicate unbalance between 
sublimation and recondensation processes is governed by the classical volume mass production rate formula \citep[equation~22 in][]{davidsson21}. Here, the sub--surface net mass production rate 
is proportional to the difference $p_{\rm sat}(T)-p(T)$ between saturation pressure and the actual local pressure $p(T)$. Because this difference is small, outgassing becomes strongly quenched. 
Such quenching is evident when considering the `active area fraction' of comets -- they typically only produce a few per cent of the gas they would have been capable of producing if the 
ice was exposed at the surface (as previously exemplified with 67P/C--G itself). The \citet{gundlachetal11} model assumes that the volume mass production is directly proportional to the Hertz--Knudsen formula $Z(T)$, which is only 
applicable to flat icy surfaces facing a vacuum. The production rate is lowered only by the factor $\Psi$, that is large compared to the difference $p_{\rm sat}(T)-p(T)$ obtained 
when properly solving the mass conservation equation. The additional quenching that is needed to fit observed production rates is artificially obtained in the works by  
\citet{huetal17} and \citet{blumetal17} by considering extremely low ice abundances. 

Additionally, our \textsc{nimbus} simulations consider continuous erosion of the dust mantle, and the movement of the location of 
the sublimation front, as the finite reservoir of water ice is gradually being consumed. Instead, \citet{huetal17} and \citet{blumetal17} consider static models where no erosion takes 
place and the ice is treated as an infinite reservoir. It is clear, that with erosion ranging $0.04$--$0.63\,\mathrm{m}$ on the northern hemisphere (Fig.~\ref{fig_erosion}), the 
slabs considered by \citet{huetal17} and \citet{blumetal17} would be eroded and ejected into the coma many times over, giving the coma a refractories/water--vapour mass ratio equivalent to 
that assumed in their model at depth (i.~e., $9\stackrel{<}{_{\sim}} \mu_{\rm N}\stackrel{<}{_{\sim}}99$). This is at odds with even the more dustier estimates of the coma composition \citep[e.~g., $4\pm 2$;][]{rotundietal15}, 
compared to the $F$--values mentioned above. 

We find that a rather small diffusivity in combination with an ice--rich material underneath the thin and continuously evolving dust mantle on the 
northern northern hemisphere ($\mu_{\rm N}=2$ corresponds to 33 per cent ice by mass) reproduces the measured data. We consider our calculations a more realistic solution 
to the physical energy and mass conservation problem under study, and suspect that the extremely ice--poor material found in the studies by  \citet{huetal17} and \citet{blumetal17}  
is an artefact caused by an oversimplified treatment of gas diffusion in porous media. By extension, this also invalidates the work of \citet{fulleetal19}, who critically depends 
on the $\mathrm{\mu_{\rm N}}$ estimate of \citet{blumetal17} for the northern hemisphere airfall, when drawing the conclusion that the refractories/water--ice mass 
ratio of the bulk nucleus is $\sim 10$ times higher than the dust/water--vapour mass ratio of escaping material (taken as 0.7--3.1), i.~e., $7\stackrel{<}{_{\sim}} \mu\stackrel{<}{_{\sim}} 31$.

We find that a molar abundance of $\sim 30$ per cent $\mathrm{CO_2}$ relative $\mathrm{H_2O}$ is necessary to reproduce the high post--perihelion $\mathrm{CO_2}$ production, 
and that the $\mathrm{CO_2}$ sublimation front depths on average were located at $3.8\,\mathrm{m}$ depth in the north, and at $1.9\,\mathrm{m}$ in the south, at aphelion. 
Thus far, only \citet{hernyetal21} have published estimates of the nucleus $\mathrm{CO_2}$ abundance (they find a molar abundance of 7--11 per cent) and front depth (about $1\,\mathrm{m}$ in the 
north at perihelion, where we have 3.4--$3.8\,\mathrm{m}$ before the airfall, and exposed at the surface in the south, where we have 0.15--$1.8\,\mathrm{m}$). Thus, our nucleus 
$\mathrm{CO_2}$ abundance is higher, and the front depth is larger, than obtained by \citet{hernyetal21}. We attribute this difference to the usage of the `Effective Area Fraction' of 20 per cent 
that \citet{hernyetal21} introduce to quench their high model water production rate down to the observed one. The same quenching is needed for $\mathrm{CO_2}$. 
We believe that their $\mathrm{CO_2}$ production is too high because their $\mathrm{CO_2}$ front is too shallow, and that the abundance is biased to a too low value for the same reason.

We now discuss the interpretation of our model fits in Section~\ref{sec_results_preper}--\ref{sec_results_postper} in terms 
of the diffusivity of 67P/C--G and how it evolved with time and depth. We first focus on $\mathrm{H_2O}$ and then $\mathrm{CO_2}$, 
in both cases starting at perihelion and marching through the orbit (assuming that the comet behaviour is fairly repetitive from one 
apparition to the next). 

One of the biggest surprises in this work is that the comet activity of both hemispheres are strongly asymmetric, 
in the sense that significantly different model parameters must be used before and after perihelion (see Fig.~\ref{fig_postper_first}). 
We first concentrate on the northern hemisphere, which is asymmetric in terms of the $\mathrm{H_2O}$ production. Close to perihelion, 
it is the target of substantial airfall. Because of the polar night conditions, it provides a negligible contribution to the water production 
rate during the first $\sim 100\,\mathrm{days}$ after perihelion ($\stackrel{<}{_{\sim}} 1.7\,\mathrm{au}$). It then gradually receives more 
illumination, and by the time the comet reaches $\sim 2.5\,\mathrm{au}$ ($\sim 200\,\mathrm{days}$ post--perihelion) 
a substantial fraction of the water activity must be provided by the northern hemisphere. Figure~\ref{fig_fronts} (upper right) shows that most of the 
northern hemisphere is substantially colder at this time, compared to similar distances pre--perihelion. At mid--northern latitudes the difference 
is several times $10\,\mathrm{K}$. Despite being colder, it produces a $\sim 30$ per cent higher total water production rate than at the same heliocentric 
distance pre--perihelion. The only way for a relatively cold hemisphere to produce large amounts of water vapour, at a distance 
where the sensitivity to the water ice abundance is low, is if the diffusivity is very high.

We find that we need to use a diffusivity corresponding to $\{L,\,r_{\rm p}\}=\{10,\,1\}\,\mathrm{cm}$ and $\xi=1$ (Fig.~\ref{fig_postper_water}) on the northern 
hemisphere, post--perihelion. This means that there must be plenty of cavities, cracks, and channels on centimetre--scale or larger,  through which water 
vapour may flow from depth very freely. This is entirely consistent with the presence of a thick layer of loosely stacked chunks with diameters in the centimetre--decimetre 
scale. Thanks to the last \emph{Rosetta} images acquired before landing at end--of--mission (414 days post--perihelion at $r_{\rm h}=3.8\,\mathrm{au}$), we have close--up images 
of airfall material at Sais in the Ma'at region on the northern hemisphere (latitude $35^{\circ}\,\mathrm{N}$). The best images, with a resolution of $1.4\,\mathrm{cm\,px^{-1}}$ and a chunk--size 
completion limit at $\stackrel{>}{_{\sim}} 7\,\mathrm{cm}$ shows very coarse debris in the $7$--$70\,\mathrm{cm}$ diameter range \citep{pajolaetal17b}. Therefore, 
we think that our post--perihelion diffusivity parameter for the northern hemisphere is realistic.

Assuming a repetitive apparition behaviour, the same type of material is responsible for the northern water activity on the inbound orbit, that was 
observed by \emph{Rosetta} early in the mission. For that segment of the orbit, a fundamentally different diffusivity is required,  $\{L,\,r_{\rm p}\}=\{100,\,10\}\,\mathrm{\mu m}$ 
and $\xi=1$. If real, the three orders of magnitude drop in diffusivity must be caused by a significant change of the size distribution of near--surface material 
taking place during the aphelion passage. We postulate that the chunks in a top layer crumbled, fragmented, and pulverised into a fine--grained dust cover, having 
typical cavities and channels of sub--millimetre size. The mechanism for such fragmentation likely starts with the removal of the water ice that binds the chunks 
together. Figure~\ref{fig_fronts} (upper left) shows that dust mantle growth is significant in the north during post--perihelion. The generally low temperatures at aphelion may 
have caused remaining cohesive material (primarily organics) to become brittle. In combination with a substantial diurnal cycling 
\citep[typically in the $120$--$160\,\mathrm{K}$ range;][]{davidssonetal21} this may have led to thermal fatigue and fragmentation, analogous to regolith--formation 
on asteroids \citep{delboetal14}. We note, however, that the timescales of fragmentation differ drastically: the fragmentation of competent rock 
taking place on asteroids is slow compared to the fragmentation of a cometary low--cohesion grain assemblage that is sublimating. The process we propose would be 
more akin to the rapid fragmentation observed in comet comae \citep[e.~g.][]{hadamcikandlevasseurregourd03,hoetal07,jonesetal08,rosenbushetal17}.

\emph{Rosetta} observations support such crumbling of the top part of the airfall layer. The ROLIS camera on the \emph{Philae} lander documented the airfall material at 
Agilkia, also in the Ma'at region at latitude $12^{\circ}\,\mathrm{N}$, (at $r_{\rm h}=3.0\,\mathrm{au}$ inbound) with $1\,\mathrm{cm\,px^{-1}}$ resolution \citep{mottolaetal15}. Above a chunk 
diameter of $\sim 0.26\,\mathrm{m}$, the differential size--frequency distribution power--law slopes are very similar for the two sites: $-4.2_{-0.8}^{+0.4}$ at 
Sais and $-4.3\pm 0.5$ at Agilkia \citep{pajolaetal17b}. Whereas the fresher airfall at Sais has a distinct break in the distribution at smaller sizes, being $-1.7\pm 0.1$ in the 
$0.07$--$0.26\,\mathrm{m}$ region, the aged airfall at Agilkia has a much steeper slope (ranging $-3.8$ to $-3.4$ for two different sub--units) that is more similar 
to the one at larger sizes \citep{pajolaetal17b}. Evidently, smaller chunks are relatively more common at Agilkia than at Sais, which we attribute primarily to 
fragmentation processes near aphelion \citep[although transport of small chunks from Hapi to Agilkia during activity switch--on may contribute as well;][]{pajolaetal17b}.

We now shift to the southern hemisphere, which is asymmetric in terms of the model parameters needed to explain the $\mathrm{CO_2}$ production. First note 
that the $\mathrm{CO_2}$ sublimation front depth between latitude $45^{\circ}\,\mathrm{S}$ and the south pole starts to decrease rapidly  around 
$100\,\mathrm{days}$ ($r_{\rm h}\approx 1.7\,\mathrm{au}$) pre--perihelion, according to Fig.~\ref{fig_fronts} (lower left). This is caused by strong water--driven erosion 
of the surface material. This causes a tenfold increase in the modelled $\mathrm{CO_2}$ production rate, from $10^{26}\,\mathrm{molec\,s^{-1}}$ to $10^{27}\,\mathrm{molec\,s^{-1}}$ 
between  $r_{\rm h}\approx 1.7\,\mathrm{au}$ and perihelion according to Fig.~\ref{fig_preper_H2O_CO2}. The nucleus $\mathrm{CO_2}$ outgassing may briefly have reached 
$6\cdot 10^{27}\,\mathrm{molec\,s^{-1}}$ just after perihelion (measurements in Fig.~\ref{fig_postper_first}). The measured and modelled surges suggest that the drastic 
reduction of the $\mathrm{CO_2}$ sublimation front depth is real. As already discussed in Section~\ref{sec_results_preper} this level of outgassing requires a combination of 
high intrinsic nucleus $\mathrm{CO_2}$ abundance, as well as a shallow $\mathrm{CO_2}$ front. 

The model that reproduces the high near--perihelion $\mathrm{CO_2}$ production rate does not fall off as rapidly as the observations farther from perihelion (Fig.~\ref{fig_postper_first}). 
We found that a substantial reduction of the diffusivity, by a factor $\sim 100$--$250$, realised by setting $\{L,\,r_{\rm p}\}=\{10,\,1\}\,\mathrm{\mu m}$ and $\xi=1$--$5$, was required to reproduce the 
measurements. We now discuss the possible cause of such a change.

In the current simulations, we find that the $\mathrm{CO_2}$ vapour pressure reaches $0.7\,\mathrm{kPa}$ pre--perihelion at the south pole, 
but almost a factor 30 higher post--perihelion (because of the imposed strong reduction of diffusivity). The $\mathrm{CO_2}$ pressure peaks at the sublimation front and 
falls off to nearly zero towards the surface, as well as towards regions somewhat below the front. Such gradients would be capable of ejecting refractories and water ice outwards into the coma, as well as 
compressing material at depth. A pressure difference of $19.1\,\mathrm{kPa}$ is sufficient to compress granular 
water--ice media to 77 per cent porosity according to laboratory measurements by \citet{loreketal16}. Pure silica dust would compress to $62$ per cent porosity according 
to laboratory measurements by \citet{guettlereta09}, suggesting that compression could be substantial if the material is weakened by the presence of dust. Such compacted 
terrain would have lower porosity, smaller pores, having less connectivity, thus lower diffusivity and higher tortuosity than uncompressed comet material.

We therefore propose that the intense sublimation near perihelion compressed the material underneath the $\mathrm{CO_2}$ sublimation front. When the front gradually moved into this 
pre--compressed material post--perihelion, as the $\mathrm{CO_2}$ was being consumed, the local diffusivity dropped drastically, and the vapour could not flow as freely. We modelled this 
as an abrupt change at $r_{\rm h}=1.4\,\mathrm{au}$, but in reality it would have been gradual. This caused a reduction of the $\mathrm{CO_2}$ outgassing (with respect to Fig.~\ref{fig_postper_first}), 
thus explaining the observed data (Fig.~\ref{fig_postper_H2O_CO2}). As the $\mathrm{CO_2}$ front moved into deeper and colder region, its sublimation rate waned, and further compression 
ceased (see the $\mathrm{CO_2}$ front temperature in Fig.~\ref{fig_fronts}, lower right, day 100--200 post--perihelion). At the same time, water activity was still substantial, leading to several meters 
worth of erosion (Fig.~\ref{fig_erosion}). That erosion may have consumed most of the compressed layer in the near--surface region. That could explain why $\mathrm{CO_2}$ outgassing from 
the southern hemisphere on the inbound trajectory, after the aphelion passage, was characterised by a relatively high diffusivity anew ($\{L,\,r_{\rm p}\}=\{100,\,10\}\,\mathrm{\mu m}$ and $\xi=1$).

We therefore conclude that the idea of fixed nucleus thermophysical parameters may have to be abandoned. Comet activity changes the physical properties of the surface layer, 
and is therefore self--regulatory. Although this certainly makes the task of modelling comet activity more challenging, it also provides novel opportunities to better understand 
comet behaviour. The perihelion asymmetry in the water production of 67P/C--G is apparently caused by the large difference in diffusivity between fresh airfall and `regular' nucleus 
material (that is more similar to aged and pulverised airfall). Nucleus erosion may locally bring the $\mathrm{CO_2}$ sublimation front to a steady--state depth near perihelion, 
where this depth depends on the absolute $\mathrm{CO_2}$ abundance. Strong perihelion asymmetries in the $\mathrm{CO_2}$ production rate are prohibited by low post--perihelion diffusivity caused by $\mathrm{CO_2}$ 
activity itself. 

We note that the $\mathrm{H_2O}$ production in this scenario is repetitive. After perihelion, the dust mantle thickness returns to its pre--perihelion values at most 
latitudes (Fig.~\ref{fig_fronts}, upper left). The changes in diffusivity caused by fresh airfall deposition and near--aphelion fragmentation would take place each orbit. The low--diffusivity 
surface layer formed by fragmentation around aphelion is either removed when approaching the Sun, or simply covered with new airfall, in both cases leading to a high--diffusivity 
surface coverage on the outbound orbital arc. The resulting similarities in the water production rate from one orbit to the next seems to be consistent with 
observations over several apparitions \citep{bertaux15}. The repetitive water outgassing is also evident from the long--term stability of the non--gravitational force: the Jet Propulsion Laboratory 
solution\footnote{https://ssd.jpl.nasa.gov/tools/sbdb\_lookup.html\#/?sstr=67P} K215/10 spans 26 years with low RMS residuals.

Concerning $\mathrm{CO_2}$, the situation is not as clear. We predict that the $\mathrm{CO_2}$ sublimation front is receding from the surface 
in the north (due to net airfall deposition) and approaching it in the south, i.~e., there might be changes in the $\mathrm{CO_2}$ production from one apparition to the next. 
However, because $\mathrm{CO_2}$ can only be observed from space, future observations will have to determine whether or not the $\mathrm{CO_2}$ production pattern is repetitive.

Finally, we discuss the significance of the momentum transfer coefficient. We found that $\eta=0.31\pm 0.06$ or $\eta=0.39$ provided the best match (for slightly different orbital solutions), while classically applied values 
typically have ranged $0.4\leq\eta\leq 1$ \citep[see][and references therein]{rickman89,davidssonandskorov04}. The classical approach assumes that the sublimating surface 
is flat, faces the vacuum of space, and consists of pure water ice. The expression for the force (equation~\ref{eq:00a}) uses the mean molecular speed 
$\langle V\rangle=\langle V(T_{\rm surf})\rangle$ of a non--drifting Maxwellian gas as a normalisation factor. Fundamental gas kinetic theory leaves no doubt that the 
actual mean velocity component along the surface normal differs from $\langle V\rangle$, hence the need for the momentum transfer coefficient. If one only considers 
molecules travelling away from the surface, there is no doubt that $\eta=0.5$ \citep[e.~g.,][]{huebnerandmarkiewicz00}. If one considers the fact that downstream molecules will 
collide with each other, and that some of those are re--directed towards the surface, there is also an additional contribution to the momentum transfer from backflux molecules 
that hit the surface. Analytical solutions to the conservation equations for this type of Knudsen layer outflow exist, but they do not have unique solutions \citep{ytrehus77}. 
For reasonable assumptions about the downstream Mach number, one finds $0.53 \stackrel{<}{_{\sim}}\eta\stackrel{<}{_{\sim}}0.67$ \citep{crifo87,rickman89}.

There are several important differences between \textsc{nimbus} and the idealised sublimating surface described above: 1) the sublimating water is not exposed, but is located 
under a dust mantle; 2) molecules diffusing through the mantle thermalise and may depart with a temperature substantially higher/lower than the water ice sublimation 
temperature, if the mantle surface is hotter/cooler than the ice below \citep{skorovandrickman95,christouetal18}; 3) the dust mantle may collimate or de--collimate the flow depending on the sign of the temperature 
gradient across the mantle \citep{davidssonandskorov04}. With respect to a real comet nucleus there are additional differences: 1) the comet surface is not flat on the 
size scale of a shape--model facet, but macroscopically rough; 2) water molecules in the Knudsen layer diffuse through heavier and thus slower $\mathrm{CO}$ and $\mathrm{CO_2}$ 
vapour, modifying their backflux properties; 3) solid chunks in the coma may disturb the water flow by re--directing or emitting molecules, and they may impact the surface themselves. 

We expect surface roughness to have the largest and most systematic effect on the momentum transfer coefficient. As soon as a surface element (large enough to act as 
a local flat surface in the context of outgassing) is tilted with respect to the regional mean outward surface normal, that will reduce $\eta$ for the region. It is also noteworthy 
that we evaluate  $\langle V\rangle$ for the surface temperature. A hot dust mantle may approach $340\,\mathrm{K}$ (Fig.~\ref{fig_erosion}, upper left), 
making $\langle V\rangle$ some $\sim 1.3$ times higher than expected for sublimating water ice ($\sim 200\,\mathrm{K}$). If water molecules do not thermalise at the surface 
temperature (e.~g., by having their last interaction with the solid medium at some cooler depth below the surface) that would lower $\eta$ as well (to compensate for 
a $\langle V\rangle$ that may be too large). We therefore consider $\eta=0.31\pm 0.06$ or $\eta=0.39$ reasonable estimates of the momentum transfer coefficient. Our force model will be further 
tested in a forthcoming publication focusing on the spin state evolution due to outgassing torques, and the time--resolved non--gravitational  changes of the orbit of Comet 67P/C--G.

\section{Conclusions} \label{sec_conclusions}

We have modelled the outgassing of $\mathrm{H_2O}$ and $\mathrm{CO_2}$ from Comet 67P/C--G using the thermophysical 
code \textsc{nimbus} (Numerical Icy Minor Body evolUtion Simulator) by \citet{davidsson21}. We have adjusted key model 
parameters until we simultaneously reproduce the pre-- and post--perihelion production rates of both species according to 
ROSINA measurements \citep{fougereetal16b}. The model uses an enforced dust production rate, which has a functional form similar to 
the observed brightness of the comet and is scaled to be compatible with the observed total mass loss. The goal of this 
effort is to place constraints on the $\mathrm{H_2O}$ and $\mathrm{CO_2}$ nucleus abundances, retrieve information about 
the depths of the $\mathrm{H_2O}$ and $\mathrm{CO_2}$ sublimation fronts and how they may have changed with time and latitude, 
and estimate the diffusivity of the porous near--surface material, which informs about the size scale of macro porosity. We apply the outgassing 
model in order to calculate the non--gravitational changes of the orbit, and identify the conditions under which the observed 
changes are reproduced. The longer--term goal of this work is to apply the outgassing model to investigate the evolution of 
the nucleus spin state. 

Our main conclusions can be summarised as follows:

\begin{enumerate}
\item The contribution of an extended coma source to the total comet water production rate is $\sim 8$ per cent at perihelion.
\item The refractories/water--ice mass ratio of relatively pristine material on the southern hemisphere is $\mu_{\rm S}\approx 1$. 
The airfall material on the northern hemisphere has lost some water ice during transfer through the coma and is better characterised by $\mu_{\rm N}\approx 2$.
\item The thickness of the dust mantle is typically $\stackrel{<}{_{\sim}}2\,\mathrm{cm}$, suggesting that water ice is ubiquitous near (but not on) the surface. 
\item The observed $\mathrm{CO_2/H_2O}$ abundance ratio in the coma \citep[e.~g., 2.2--5.6 per cent at $1.8$--$2.2\,\mathrm{au}$ pre--perihelion;][]{finketal16} is 
not necessarily a reliable measure of the corresponding ratio within the nucleus. We find that nucleus models with $\mathrm{CO_2/H_2O}=32$ per cent reproduces the 
data. The coma  $\mathrm{CO_2/H_2O}$ is sensitive to the depth of the $\mathrm{CO_2}$ sublimation front in addition to the intrinsic nucleus 
$\mathrm{CO_2/H_2O}$ abundance.
\item We find that the depth of the $\mathrm{CO_2}$ sublimation front on the northern hemisphere was $\sim 3.8\,\mathrm{m}$ on average at aphelion. 
A combination of $\mathrm{CO_2}$ loss, nucleus erosion, and airfall deposition may result in a net increase of the front depth with time. 
The depth of the $\mathrm{CO_2}$ sublimation front on the southern hemisphere was $\sim 1.9\,\mathrm{m}$ on average at aphelion. Locally, the 
depth may have been reduced to as little at $\sim 0.15\,\mathrm{m}$ near the south pole. The $\mathrm{CO_2}$ sublimation front depth vary significantly 
with time and latitude.
\item Fresh airfall material has a high diffusivity, suggesting macro porosity on centimetre scales. Aged airfall material has a three orders of magnitude 
lower diffusivity, suggesting sub--millimetre macro porosity. We propose that a combination of desiccation, temperature--dependent brittleness, and 
thermal fatigue causes substantial fragmentation of a surface layer during the aphelion passage. 
\item The model with the best available reproduction of the $\mathrm{H_2O}$ and $\mathrm{CO_2}$ outgassing, results in a net force 
acting on the nucleus that reproduces the observed non--gravitational changes of the orbital period, longitude of perihelion, and longitude of 
the ascending node of 67P/C--G, as long as the momentum transfer efficiency is $\eta=0.31\pm 0.06$ or $\eta=0.39$ (depending on orbital solution). 
We interpret the relatively low $\eta$--value as a consequence of small--scale surface roughness, potentially in combination with incomplete thermalisation 
of molecules diffusing through the hot dust mantle. 
\end{enumerate}

\section*{Acknowledgements}

Parts of the research was carried out at the Jet Propulsion Laboratory, California Institute of Technology, under a 
contract with the National Aeronautics and Space Administration. BJRD and NHS acknowledge funding from 
NASA grant 80NSSC18K1272 awarded by the \emph{Rosetta} Data Analysis Program. PJG acknowledges financial support 
from PGC2018--099425--B--I00 (MCI/AEI/FEDER, UE) and from the State Agency for Research of the
Spanish MCIU through the `Center of Excellence Severo Ochoa' award to 
the Instituto de Astrof\'{i}sica de Andaluc\'{i}a (SEV--2017--0709). The authors thank the reviewer Raphael Marschall for 
numerous suggestions which made the manuscript better.\\

\noindent
\emph{COPYRIGHT}.  \textcopyright\,2021. All rights reserved.

\section*{Data Availability}

The data underlying this article will be shared on reasonable request to the corresponding author.

\bibliography{MN-21-2573-MJ.R1.bbl}

\bsp	
\label{lastpage}
\end{document}